\documentclass[sn-mathphys-num]{sn-jnl}% Math and Physical Sciences Numbered Reference Style 
\usepackage{ulem}
\usepackage{subfigure}
\usepackage{graphicx}%
\usepackage{multirow}%
\usepackage{amsmath,amssymb,amsfonts}%
\usepackage{amsthm}%
\usepackage{mathrsfs}%
\usepackage[title]{appendix}%
\usepackage{xcolor}%
\usepackage{textcomp}%
\usepackage{manyfoot}%
\usepackage{booktabs}%
\usepackage{stmaryrd}
\usepackage{soul}   
\usepackage{epstopdf}
\usepackage{setspace,booktabs}
\usepackage{multirow}
\usepackage{bm}
\usepackage{graphicx}
\usepackage{parskip}
\usepackage{listings}%
\usepackage{adjustbox}
\usepackage{float} 
\usepackage[ruled,lined,linesnumbered]{algorithm2e}
\SetKwFunction{KwFn}{Fn}
\SetKwInOut{Input}{input}
\SetKwInOut{Output}{output}
\usepackage{placeins} 

\newcommand{\Dcal}{{\mathcal D}}

\newcommand{\Rcal}{{\mathcal R}}

\newcommand{\Tcal}{{\mathcal T}}

\theoremstyle{thmstyleone}%
\theoremstyle{thmstyletwo}%

\theoremstyle{thmstylethree}%
\newtheorem{definition}{Definition}%

\begin{document}

\title[Article Title]{Class of topological portfolios: Are they better than classical portfolios?}


 \author[1]{\fnm{Anubha} \sur{Goel}}\email{anubha.goel@tuni.fi}

 \author*[2]{\fnm{Amita} \sur{Sharma}}\email{amita.sharma@nsut.ac.in}
 \equalcont{These authors contributed equally to this work.}

 \author[1]{\fnm{Juho} \sur{Kanniainen}}\email{juho.kanniainen@tuni.fi}
 \equalcont{These authors contributed equally to this work.\\ \textbf{Pre-Accepted in Financial Innovation}}

 \affil[1]{\orgdiv{Computing Sciences/Financial Computing and Data Analytics Group}, \orgname{Tampere University},\orgaddress{ \country{Finland}}}
 \affil*[2]{\orgdiv{Department of Mathematics}, \orgname{Netaji Subhas University of Technology}, \orgaddress{ \country{India}}}


\abstract{Topological Data Analysis (TDA), an emerging field in investment sciences, harnesses mathematical methods to extract data features based on shape, offering a promising alternative to classical portfolio selection methodologies. We utilize persistence landscapes, a type of summary statistics for persistent homology, to capture the topological variation of returns, blossoming a novel concept of ``Topological Risk". Our proposed topological risk then quantifies portfolio risk by tracking time-varying topological properties of assets through the $L_p$ norm of the persistence landscape. Through optimization, we derive an optimal portfolio that minimizes this topological risk. Numerical experiments conducted using nearly a decade long S\&P 500 data demonstrate the superior performance of our TDA-based portfolios in comparison to the seven popular portfolio optimization models and two benchmark portfolio strategies, the naive $1/N$ portfolio and the S\&P 500 market index, in terms of excess mean return, and several financial ratios. The outcome remains consistent through out the computational analysis conducted for the varying size of holding and investment time horizon. These results underscore the potential of our TDA-based topological risk metric in providing a more comprehensive understanding of portfolio dynamics than traditional statistical measures. As such, it holds significant relevance for modern portfolio management practices.}

%%================================%%
%% Sample for structured abstract %%
%%================================%%




\maketitle

\section{Introduction}\label{sec1}

The Markowitz model or mean-variance model \cite{marko1952}, which accounts for variance as its underlying risk measure, relies heavily on moment estimation, often leading to poor and unstable out-of-sample performance %and weights that fluctuate substantially over time due to estimation error 
(\cite{bonnans2000}, \cite{JagannathanMa2003}, \cite{ledoitwolf2003}, \cite{ledoitwolf2004}, \cite{demigueletal2009}). Indeed, some practitioners consider the concept of optimality proposed by modern portfolio theory questionable within the volatile and non-stationary environment of financial security returns \cite{kolm201460}. The degradation in out-of-sample performance of mean-variance portfolios due to estimation errors is so significant that it struggles to consistently outperform even the simplest naive $1/N$ strategy, which allocates equal weights to all components of the portfolio \cite{demiguel2009optimal}. %Given its ease of implementation and freedom from estimation errors, the $1/N$ naive portfolio strategy remains an attractive allocation rule among investors.
Further, if the covariance matrix of the asset returns is rank-deficient, the mean-variance model does not have a unique solution. Even when the number of assets is close to the number of observations, the sample covariance matrix approaches a singularity and hence not a consistent estimator to the population covariance matrix. %, leading to many uncertainties in the form of estimation errors. 

Several methods have been suggested to reduce estimation errors in the mean-variance model. For instance, Bayesian and shrinkage techniques (\cite{Jobson1981}, \cite{Jorion1991}, \cite{ledoitwolf2003}, \cite{ledoitwolf2004}) aim to mitigate errors in estimating parameters. Some studies opt for a different approach altogether, by excluding the mean return function from the model (known as the global minimum variance model), thus eliminating mean estimation errors (\cite{chopra1993}, \cite{sparsestable22}, \cite{Clarke2010MinimumVP}) or imposing weight norm constraint \cite{demiguel2009optimal}. Alternatively, researchers have explored translating the mean-variance model into its robust version by defining uncertainty sets for the mean vector and covariance matrix (\cite{ccmv2020}, \cite{leeetal_2020}). These sets encapsulate a range of possible values for these parameters, making resulting models more resilient. Despite their statistical underpinnings, all these approaches, whether directly or indirectly, grapple with estimation errors. Moreover, as \cite{scherer2007can} argues, robust optimization methods can lead to portfolios with inferior out-of-sample results.

This paper presents a novel approach to portfolio diversification that smartly avoids model-based estimation errors, i.e., the estimation error linked to distributional assumptions or statistical inputs like mean and covariance, by focusing on minimizing the portfolio's topological risk. Our approach holds significant potential in scenarios characterized by a high degree of model uncertainty accompanied by substantial estimation errors. Topological Data Analysis (TDA) is employed to process high-dimensional complex data and produce a simplified, low-dimensional representation that preserves critical topological attributes, such as shape and connectivity. This allows for the extraction of \textit{quantitative homological features, reflecting the topological variability of returns}. In this paper, to capture the topological variation of returns, we utilize the norms of persistence landscapes, which are summary statistics for persistent homology introduced by \cite{bubenik2015statistical}. An advantage of using persistence landscape is that it has a function form, and thus the theory of random variables can be applied in vector spaces defined by it.

The basic idea of persistent homology is to connect points step-wise that are ``relatively close together" and thereby examining the homology of the resulting shape. The number of connections added at each step depends on a subjective parameter, dictating the desirable closeness. %how much the points should be closed enough to get connected. 
Such network created at each step is called a simplicial complex. %The two graphs that record the homology are Persistent Barcodes and Persistent Diagrams, conveying the same information of feature appearance and disappearance. 
The persistent diagram is a 2-dimensional graph of birth (appearance) and death (disappearance) of a feature in the Euclidean plane where ($b$, $d$) = (birth, death). %\textcolor{red}{allowing more convenience in visualization. }
%In result, persistent homology allow us to draw the conclusion using algebraic tools alone- a useful method when data is too complex to examine visually. 
Due to the incompleteness of the metric space induced by the persistent diagram, an alternative topological summary called persistent landscape, is developed to convey similar information as the persistence diagram. An important technical hallmark of the persistent landscape is that it is a function and its function space forms a separable Banach space, therefore, the theory of
random variables can be well defined in such space (\cite{bubenik2018persistence}, \cite{bubenik2015statistical}). Furthermore, persistence landscape is a sequence of piecewise-linear functions which makes them more computationally efficient in comparison to the persistence barcodes (or diagrams) in analyzing the data  (\cite{bubenik2018persistence}, \cite{bubenik2015statistical}). 

Staging the background of TDA, we intend to design an optimal portfolio via persistent landscape to capture the insight of each asset return dynamic in a better way than the traditional risk measures. We believe that as long as the portfolio selection process depends on a statistical-based risk measure to capture uncertainty in asset's return, be it variance, mean-absolute deviation \cite{konno93}, or quantiles \cite{bellklaretal14}, it continues to suffer from estimation error resulting in non-robust and unstable outcomes (\cite{chopra1993}, \cite{pearson98}, \cite{goldfarb2003robust} \cite{moon2011}). Additionally, none of the individual risk measures can capture the hidden qualitative properties of asset return time series data such as shape and structure. To achieve reliability with respect to estimation error and feature extraction, we define a new TDA-based risk measure of a portfolio named as \textit{``Topological Risk"} that aims to track the dynamic of topological properties for each asset. More precisely, the aim is to dynamically monitor the topology of each asset and the risk of an asset is assessed based on the changes in its topology over time.

%We propose an innovative approach to improve asset risk evaluation in investment portfolios. 
Our scheme focuses on assessing and quantifying the topological risk of individual assets, a critical factor in portfolio management. Specifically, we introduce the concept of topological risk for each asset, defined as the squared error between its persistence landscape norms and a reference point, the mean persistence landscape norm. Under the vector space structure of the persistence landscape, its mean which is a point-wise mean, is a well-defined function \cite{bubenik2015statistical}. The method involves dividing each training period into overlapping sub-windows and employing Takens embedding to convert asset return data into a multivariate matrix. These matrices form input point clouds for persistent homology analysis. For each asset and a sub-window, we extract topological features using persistence diagrams and derive persistence landscapes, resulting in a series of landscapes for each window. The topological risk for each asset is then calculated as the variation of the persistence landscapes from the mean landscape, using the $L_p$ norm. This comprehensive approach provides a nuanced understanding of asset topological risk, enabling investors to make informed decisions regarding portfolio composition and risk management strategies.

Upon calculating the topological risk $\Lambda_{i} \geq 0$ of each asset $i; i=1,\ldots,n$, %we form an $n \times n$ diagonal matrix $Q = \{\Lambda_{i}\}_{i=1}^n$, called a topological risk matrix with diagonal entries as each asset's topological risk $\Lambda_{i}$. 
the topological risk of a portfolio $\mathbf{w}$ = $(w_1,\ldots,w_n)$ comprising $n$ assets (where $w_i$ denotes the proportion of total budget to be invested in $i$-th asset) is then quantified via a quadratic function $\sum_{i=1}^n \Lambda_{i} w^2_i$. Therefore, for a polyhedron set of feasible portfolios, minimizing the topological risk of a portfolio $\mathbf{w}$ is a convex program and attains its global optimal solution. An empirical analysis to investigate the financial benefit of the proposed scheme is carried on the sample data of daily closing prices of the constituents from S\&P 500 (U.S) with the sample period of nearly 10 years from December 10, 2012, to August 11, 2022. The S\&P 500, a stock market index comprising 500 large-capitalization companies listed on US stock exchanges, serves as a global benchmark for the majority of economies. For the comparative analysis purpose, we consider seven famous traditional portfolio optimization (PO) models, namely the global minimum variance model, the mean-variance model, the mean-value at risk model \cite{lofti2016}, the mean-conditional value at risk model \cite{Rockafellar2000}, the reward-risk PO model maximizing Sharpe ratio \cite{sharpe94}, the reward-risk PO model maximizing STARR ratio \cite{starr_martin}, and the reward-risk PO model maximizing Omega ratio \cite{kapsosetal14} along with the two benchmark portfolios, the $1/N$ naive portfolio \cite{demigueletal2009} and the market index S\&P 500.

We consider several financial ranking measures including excess mean return, standard deviation, downside deviations, Sharpe ratio, Sortino ratio, STARR ratio, and Rachev ratio to assess the performance of all considered models. Under the current setting, we notice excel performance of the proposed scheme in comparison to all other models in almost all financial measures considered, showing the reliable outcomes when our scheme is implemented in practice. Furthermore, our findings are consistent with those presented in \cite{demiguel2009optimal}, wherein it is observed that the out-of-sample Sharpe ratio of the sample-based mean-variance strategy is lower than that of the $1/N$ strategy. This suggests that errors in estimating means and covariances negate the benefits of optimal diversification compared to naive diversification. However, we have devised a strategy that improves portfolio performance by incorporating topological risk associated with each asset, rather than simply evenly distributing wealth among them (Naive Strategy).

The main contributions of the paper are summarized as follows:

\begin{itemize}
\item \textit{Conceptual contribution:} While the use of TDA in analyzing financial time series is not entirely novel, its application in portfolio construction remains a relatively underexplored area. Contributing to the growing literature on TDA in finance, we introduce the concept of \textit{Topological Risk} for the first time—a risk measure that is entirely independent of traditional statistical metrics. We utilize persistence landscapes, a summary statistic derived from persistent homology, to capture the topological variation in asset returns. The proposed optimization model minimizes this topological risk and is formulated as a quadratic program, thus preserving computational efficiency.

\item \textit{Empirical contribution:} We evaluate the performance of our proposed TDA-based model using nearly a decade of S\&P 500 data. The financial benefits of the TDA approach are compared against seven classical PO models (ranging from mean-risk to return-reward ratio models), as well as two benchmark portfolios: the naive $1/N$ portfolio and the S\&P 500 index. An extensive empirical analysis is conducted using a rolling window scheme with varying window sizes to assess robustness. Specifically, we consider two in-sample periods (1 year and 2 years) and four out-of-sample horizons (1 month, 3 months, 6 months, and 1 year), resulting in a total of eight different evaluation settings. Across all scenarios, the TDA-based portfolios consistently outperform the benchmarks in terms of excess mean returns and risk-adjusted performance ratios, demonstrating the significance of TDA in modern portfolio management practices.
\end{itemize}

The rest of the paper is organized as follows: Section 2 presents the literature reviews; Section 3 gives the basics of TDA tools; Section 4 explains the proposed scheme;  Section 5 and Section 6 respectively, explain the empirical analysis and robust analysis of the computational study; and the paper concludes in Section 7. 

\section {Literature Review}
The core of modern portfolio theory is to make an optimal trade-off between return and risk, where return is generally estimated by the first order moment of underlying distribution (mean return), risk can be defined by different measures in order to capture most of the uncertainty in the data. Statistical risk measures such as variance \cite{marko1952}, mean-absolute deviation \cite{konno93}, Gini mean difference \cite{yitzhaki}, and central semi-deviations \cite{ogryczak01}, aim to deal with the deviations around mean return, whereas the tail risk measures namely, quantiles \cite{linsmeier} and conditional-value-at-risk \cite{rockafeller02}, focus on the risk of large losses. One can choose the risk measure that matches with his objective of portfolio or after a brief discussion with his portfolio manager. 

Traditionally, while optimizing the trade-off between mean return and underlying risk function in practice, the unknown model parameters (or distribution) are replaced by their sample estimations (or some empirical distribution such as uniform distribution) \cite{roman09}.  This results into unstable, and non-robust portfolio weights, for example, it is well known that the simplest naive $1/N$ strategy outperforms the mean-variance portfolios (\cite{demiguel2009optimal}, \cite{jobson81}, \cite{kolma14}). Further, the portfolio optimizers are often called as ``error-maximizers" \cite{michaud}, and the mean-variance model can produce extreme weights (\cite{black91}, \cite{demiguel2009optimal}). Though, it is commonly believed that estimation errors due to mean terms are of much greater significance than covariance terms (\cite{chopra1993}; \cite{best15}) yet errors due to covariances can also have hefty impact. As it is comparatively easier to estimate covariances than means, the presence of heavy tails in distributions can result in substantial errors in the covariance estimates too \cite{martin09}. The estimation problem is even more severe for high-dimensional data when the size of portfolio exceeds the sample size. In such case, the sample covariance matrix is no longer a consistent estimator for the true population matrix, causing deviation of traditional portfolio weights from the population ones \cite{bodnar19}. Neverthless, the mean-variance PO model has a significant impact on academic research and the financial industry as a whole \cite{kolma14}.

In light of the inherent limitations of conventional PO frameworks, this work employs Topological Data Analysis (TDA) to formulate a risk measure that not only eliminates dependence on distributional assumptions but also effectively captures the dynamic structure of asset returns, with robustness to noise. TDA is a set of mathematical tools from algebraic topology for data visualization and feature extraction (\cite{carlsson2009topology}, \cite{carlsson2020topological}, \cite{wasserman2018topological}, \cite{bubenik2015statistical}, \cite{chazal2017introduction}). 
An excellent overview and literature on TDA may be referred to (\cite{carlsson2009topology}, \cite{carlsson2020topological}) and a recent article on biomedicine by \cite{skaf20221}. 
The applications of TDA are wide, ranging from biomedical \cite{skaf20221}, signal processing \cite{wang2018}, image recognition \cite{moraleda2020},  neuroscience \cite{chung2021}, time series analysis \cite{ravishankar2019} to social sciences \cite{lum.et.al.2013}. TDA tools take high-dimensional complex data as input and result in a simpler low-dimensional representation while retaining significant topological features related to its shape and connectivity. 

The two fundamental TDA methods are Persistent Homology and Mapper, where the former is useful for generating quantitative representation of homological features such as connected components or higher-dimensional holes whereas the latter constructs a compact optical summary, suitable for exploratory data analysis. 

As an application to persistent homology, work of classifying patients with chronic obstructive pulmonary disease can be seen in \cite{belchi2018}, analyzing protein folding in \cite{ichinomiya2020} and examining the epidemiology of specific diseases can be referred to \cite{lo2018}. 
%Bio medical paper
Continuing with the charm of TDA, it witnesses its wide range of applications in the shape analysis (\cite{carlsson2005};  \cite{liet.al.2014}), sensor networks (\cite{silva2007}; \cite{adams2015}), dynamic systems and signal processing (\cite{perea2015}), and amalgamation of machine learning tools with TDA \cite{pun2018}.

Tasting a great success in the above several listed fields, TDA extends its beauty in the area of time series analysis which was earlier relatively underdeveloped. In particular, equipped with the notions of birth and death for features appearing and disappearing, persistent homology is a more logical way to analyze time-changing factors of data. Early research on applying persistent homology to time series analysis includes (\cite{berwald2014}, \cite{pereira2015persistent}, \cite{wu2022topological}, \cite{perea2015sw1pers}), with specific attention to financial time series seen in (\cite{sato2016},  \cite{gidea2018topological}, \cite{saengduean2020topological}, \cite{gidea2017topological}, \cite{vandewalle2001non}, %\cite{phoa2013portfolio}, 
\cite{majumdar2020},  \cite{gidea2020topological}, \cite{aromi2021topological}, \cite{AKINGBADE2024107665}). However, applications in PO are much less common, with only a few studies such as (\cite{goel2023sparse}, \cite{goel2020topological}) exploring this area. In the field of financial time series analysis, an important class of applications of TDA concerns critical transitions in financial time series, for instance, \cite{gidea2017topological} develop a TDA-based method to detect early signs for critical transitions in financial time series data. They use persistent homology to investigate stocks during a period prior to the US financial crisis of 2007-2008, and find the presence of early signs of the critical transition whereas authors in \cite{AKINGBADE2024107665} present a heuristic argument for the tendency of TDA to detect financial bubbles. As an illustration, \cite{AKINGBADE2024107665} use their proposed approach on a sample of positive and negative bubbles in the Bitcoin historical price. \cite{sato2016} investigate the relationship of TDA's barcodes with the financial risk measures, such as growth rate, volatility, and correlation coefficients over the time series data of NIKKEI 225 in 2014 and 2015. Their results support the effectiveness of TDA as a risk measure, utilizing barcodes to detect the rapid change in a short time.

\cite{RUDKIN2023119894} demonstrate a link between persistence norms and uncertainty (observed and unobserved uncertainty). Clues from the dynamic analysis of the persistence norm and uncertainty relationship supports that the persistence norms provide a signal of impending market crashes, therefore, persistence norms have potential as a further tool in asset pricing. Next, the authors in \cite{gidea2020topological} propose to combine persistent homology with k-means clustering to detect early warning signals of the January 2018 digital asset market crash, studying the movements of four major cryptocurrencies (Bitcoin, Ethereum, Litecoin, and Ripple) whereas \cite{saengduean2020topological}
apply TDA techniques to investigate financial crashes for the two cryptocurrencies, Bitcoin and Ethereum for the same duration of digital asset market crash, 2018. \cite{saengduean2020topological} demonstrate good early warning signals before the crashes and show that the $L_1$ and $C_1$-norms of persistent landscapes peak before the occurrence of crashes. Continuing with studying the transitions in digital market systems, \cite{SONG2025130194} employ persistent homology method, based on the high-frequency price data of the ten major cryptocurrencies, to study the evolution of the topological structural features of the cryptocurrency system over time.  \cite{majumdar2020} propose to use persistent homology and time delay embedding for time series classification and clustering. Their findings show that the topological features of the time series of stock prices over different sectors are not same and have distinctive features that can be make out effectively via TDA.

In the area of PO, \cite{goel2020topological} propose to use persistent homology with an application of enhanced indexing, an investment strategy that seeks to outperform the benchmark index. The proposed method executes in two stages: filtering of assets based on $L_p$ norms of the persistence landscapes values and thereafter, solving an optimization problem to generate an optimal portfolio. The authors test the efficiency of the proposed algorithm over ten data sets from financial markets across the globe and reported favorable outcomes. Very recently, \cite{goel2023sparse} utilize persistence homology to create a sparse index tracking portfolio. They consider minimizing the objective function with the weighted Elastic-Net terms and propose to learn its regularization coefficients using $L_p$ norms of persistence landscape for robust outcomes. They also verify their results using a data set that covers 23 years of the S\&P 500 index. Related studies are also applied persistent homology to analyze market structure and improve portfolio strategies. \cite{baitinger2021better} construct a turbulence index by computing Wasserstein distances between persistence diagrams at different time points to detect regime shifts in financial markets, and then use this index to inform portfolio allocation. \cite{ruiz2022tda} develop a similar regime-detection framework based on the normed differences between persistence landscapes over time, identifying periods of heightened topological change that signal turbulent market conditions. While these studies are developed with reference to specific benchmarks or regime-detection contexts, the current study introduces the novel concept of topological risk to optimize portfolio in a more general context. We employ persistence landscapes to capture the topological variation of asset's returns, 
giving parameter free risk measure. The study also extends to incorporate the effect of cardinatility constraint in the proposed PO model minimizing topological risk. An extensive empirical analysis over varying window size confirms the outperformance of the proposed scheme in comparison to several well established PO models.

\section{Overview of Topological Data Analysis}

Under the discrete time framework of total $T$ scenarios, let $r_i(t)$ represents the return of the $i$-th asset at time point $t; ~t=1,2,\ldots, T$. Then the time series $X_i = \{r_i(t)\}_{t=1}^T$ of window size $T$ represents a univariate time series for each $i$-th asset; $i=1,2,\ldots,n$. Since TDA works on point clouds %in more than one-dimension 
and the one-dimensional time series like $X_i$, fails to have point cloud representations (\cite{gidea2018topological}, \cite{goel2020topological}) in general, therefore, we first embed the time series into its corresponding high-dimensional space (point clouds) using the time-delay coordinate embedding or Takens’ embedding (\cite{takens1981detecting}). 

As a specimen, a time series $X = \{r_1, r_2,\ldots,r_T\}$, can be reconstructed in phase space in time as follows:
	\begin{align}
	{R} & =\left[
	\begin{array}{c}
	R_{1}\\
	R_{2}\\
	\vdots\\
	R_{T-(d-1)\tau}\\
	\end{array}
	\right]
 = \left[
	\begin{array}{cccc}
	r_{1}& r_{1+\tau} \dots  &r_{1+(d-1)\tau}\\
	r_{2}& r_{2+\tau}  \dots   &r_{2+(d-1)\tau}\\
	\vdots & \vdots & \vdots \\
	r_{T-(d-1)\tau} & r_{T-(d-2)\tau} \dots  &r_{T}\\
	\end{array}
	\right],\label{eqn3}
	\end{align}
where $\tau$ is the time delay, $d$ is the dimension of reconstructed space (embedding dimension), $T - (d - 1)\tau$ is the number of points (states) in the phase space where each point in the space is represented by a row of the matrix $R$. So, conceptually, the association between the time series and the constructed point clouds is
known as the Takens’ embedding. The Takens’ embedding has now become a useful tool to connect a variety of time series with persistent
homology by converting the former into a meaningful depiction of the point cloud (\cite{horak2003deterministicky}, \cite{khasawneh2017utilizing}, \cite{kim2019}) while 
preserving the topology during the process \cite{takens1981detecting}.
%Anubha experts paper

We then propose to obtain a simplicial complex for a given point cloud using the Vietoris-Rips method. A simplicial complex, $S$, is a collection of finite sets (called simplices) that satisfy two essential conditions:
(i) Any face of a simplex from $S$ is also in $S$. In other words, if $\alpha \in S$ and $\beta \subset \alpha$, then $\beta \in S$. Here, if $k = |\alpha| - 1$, $\alpha$ is called a $k$-simplex, and $\beta \subset \alpha$ is known as a face of $\alpha$.
(ii) The intersection of any two simplices in $S$ is either empty or shares faces. Vietoris-Rips complex is among some of the popular procedures to construct a simplicial complex 
by connecting the pairs of points (vertices) in a given point cloud that are sufficiently close 
as described by the following definition:

\begin{definition}
    The Vietoris-Rips complex of $R$ with a parameter $\epsilon > 0$ is defined to be the simplicial complex denoted by $\mathbb{R}_{\epsilon}(X)$ satisfying $\{R_1, R_2, \ldots, R_l\} \in \mathbb{R}_{\epsilon}(X)$ if and only if Diam $(R_1, R_2, \ldots, R_l) < 
 \epsilon $ where diam is the largest distance between any two points in the set.
\end{definition}

The general idea of this procedure is to connect points that are relatively close together and then examine the homology of the resulting 
shape. A natural question is how to select the value for $\epsilon$ in the Vietoris-Rips complex procedure while connecting the points as it can greatly affect the topological features in the complex. For example, for smaller values of $\epsilon$ we might see no edges whereas for large values, every point is connected to every other point leaving just one connected component and therefore, has no topological
features of interest in both of the cases. Thankfully, TDA provides an efficient way for the right selection of $\epsilon$ by computing the shape of a point cloud over its entire range and studying the topological structure as a function of $\epsilon$ rather than taking any of its random value. Consequently, we get a sequence of Vietoris-Rips simplicial complexes corresponding to different values of $\epsilon$, named as Vietoris–
Rips filtration, and is denoted by $\{\mathbb{R}_{\epsilon_k}(X)\}_{k \in \mathbb{N}}$, for a non-decreasing sequence $\{\epsilon_k\} \in \mathbb{R}^+\cup\{0\}$ with $\epsilon_0= 0$. 
The intuition behind this procedure is that via a ``discrete" filtration of simplicial complexes associated with a finite number of parameters
$0<\epsilon_1<\epsilon_1<\ldots<\epsilon_k$, one can track the topological features (connected components, holes, etc.) that persist throughout the filtration. To achieve the same, each topological feature is given a ‘birth’ (appear) and ‘death’ value (disappear) during the filtration process and the difference between the birth and death values represents the feature’s persistence in the corresponding filtration. The topological features that persist over a wider range of scales ($\epsilon$)
are considered the most significant and representative of
the shape of the point cloud. This process of tracking and analyzing the changes in the topological features of complex data 
across multiple resolutions is known as persistence homology (see \cite{goel2023sparse} for a visual illustration of this process).

Once we obtain the Vietoris-Rips filtration, it is time to summarize the persistence of topological features (or persistent homology). The two most common topological summaries of the data are:
persistent barcodes and the persistence diagram. The persistence barcode contains horizontal lines, called bars, in which each bar begins at some feature’s birth and ends at its death. The same information can also be represented by the persistence diagram which is a 
graph in the Euclidean plane where $(b, d)$ = (birth, death). Mathematically, persistence diagram is a multi-set of points in $W \times \{0, 1,\dots, q-1\}$, where $W:= \{(b, d) \in \mathbb{R}^2 : d \geq b \geq 0\}$ and each element
$(b, d, f)$ represents a homological feature of dimension $f$ that appears at scale $b$ during a Vietoris–Rips filtration and
disappears at scale $d$. Intuitively speaking, the feature $(b, d, f)$ is a $f$-dimensional hole lasting for duration $d - b$, called persistence. Namely, features with $f = 0$ correspond to connected components, $f = 1$ to loops or holes, and $f = 2$ to voids. 

In other words, a persistence diagram (or barcode) offers more than a traditional summary statistic by showing not only the number of persistent features but also when they appear and overlap along the filtered simplicial complex. Having said that, the persistence diagram is of limited use when equipped with Wasserstein distance as it forms an incomplete metric space and therefore, is not appropriate to apply tools from statistics. An alternative topology summary called the persistence landscape, proposed by \cite{bubenik2015statistical}, has a function form. Thus, the vector space structure can be used for its underlying function space. Additionally, computations with a persistence landscape are much faster than the persistence diagram, removing a second obstruction to the use of topological methods in data analysis. The other methods to embed the persistence diagrams are persistence image, or kernel-based methods such as persistence scale space kernel, to name a few. One can refer to (\cite{carlsson2009topology}, \cite{gidea2018topological}, \cite{adams2017}) for more description.

The persistence landscapes are sequences of piecewise-linear functions, defined on a re-scaled birth-death coordinate with the peaks representing the significant topological features. Mathematically, with each birth-death pair $p(a, b) \in D$, where $D$ is the persistence
	diagram, a piece-wise linear function $\Lambda _p : \mathbb{R}\rightarrow [0,\infty)$ is associated as follows:
	\begin{eqnarray}
	\Lambda _p(t) = \left\{\begin{array}{ll} t-a &\quad t \in [a,\frac{a+b}{2}] \\[0.5em]
	b-t &\quad t \in [\frac{a+b}{2},b]\\[0.5em]
	0 &\quad \mbox{otherwise}.
	\end{array}\right. \label{pl}
	\end{eqnarray}
	
A persistence landscape of the birth-death pairs $p_i(a_i, b_i),\; i=1,\ldots,m,$ is the sequence of functions $\eta: \mathbb{N}\times\mathbb{R}\rightarrow [0,\infty)$, as $\eta(k,t) = \eta_k(t)$ where $\eta_k(t)$ denotes the $k$-th largest value of $ \{\Lambda _{p_i}(t),\; i=1,\ldots,m\}.$ We set $\eta_k(x) = 0$ if
the $k$-th largest value does not exist; so, $\eta_k (t)= 0$ for $k > m.$ The persistence landscapes form a subset of the Banach space 
$L^p(\mathbb{N} \times \mathbb{R})$ consisting of sequences $\eta= (\eta_k)_{k \in \mathbb{N}}$. This set has an obvious vector space structure (\cite{gidea2018topological}), and it becomes a Banach space when endowed with the norm

 \begin{align}
	||\eta||_p=\left( \displaystyle \sum_{k=1}^\infty ||\eta_k||^p_p \right)^{\frac{1}{p}},
	\end{align}

 where $||\cdot||_p$ is the $L^p$-norm.  Further, it is shown in \cite{bubenik2015statistical} that the persistence landscape is stable with respect to the $L^p$ norm for $1 \leq p \leq \infty $. 

Forming a separable Banach space, the mean of the landscape has a well-defined function form as proposed by \cite{bubenik2015statistical}. Formally, for a random variable $Z$ defined on some underlying probability space $(\Omega, F, P)$, with corresponding persistence landscape $\eta$, with values in the separable Banach space $L^p(S);\, 1 \leq p \leq \infty,\, S = \mathbb{R}$ i.e., for $w \in \Omega,~ X(w) \, \textrm{is the data and } \, \eta(w) = \eta(X(w)) =: \eta$ is the
corresponding topological summary statistic. Further let $X_1, \ldots, X_M$ be independent and identically distributed random variables with corresponding persistence landscapes $\eta^1, \ldots, \eta^M$. Then the mean landscape $\bar{\eta}$ is given by the following pointwise mean:

\begin{align}
    \bar{\eta}(k,t) = \frac{1}{M}\displaystyle\sum_{i=1}^M \eta^i(k,t) \label{tdamean}
\end{align}

and the norm of the mean landscape is as follows:

\begin{align}
	||\bar{\eta}||_p=\left( \displaystyle \sum_{k=1}^\infty ||\bar{\eta}(k,t)||^p_p \right)^{\frac{1}{p}}, \label{meannorm}
	\end{align}

where $||\cdot||_p$ is the $L^p$-norm.

With the above-mentioned crucial framework of TDA, we have now come to the point to explain our proposed strategy particularly, incorporating the persistence landscape to define the risk of a portfolio comprising assets with uncertain returns. 

\section{Proposed TDA-based Scheme}
Persistence homology captures the shape of data across multiple resolutions. On applying a sliding window approach to a time series, resulting point cloud encodes the dynamic behaviors of topological features such as connected components, cycles and other structural changes. The stability of persistent homology of the point cloud under small perturbations makes this approach suitable for analyzing financial time series. The method can be used, for example, for detection of critical transitions (tipping points). Persistence landscape, a summary statistics for persistent homology introduced by \cite{bubenik2018persistence}, is a powerful tool to quantify “topological activities”, i.e., the behavior of shape of data over time. For instance, higher fluctuations in time series often result in more complex topological structures resulting in higher persistence norms. Further, the higher the concentration (scattering) of the point cloud, and thus the more (less) stable the returns over the subintervals from a topological point of view. 

The amount of scatteredness among return observations over time can be thus effectively quantified using $L_p$ norm of persistence landscape called as TDA norm, with its lower (higher) value corresponding to a higher concentration (scattering) of the point cloud \cite{goel2023sparse}. Therefore, the TDA norm can effectively track changes in the state of stock return dynamics without using any prior distribution assumption. Moreover, the persistence landscape \footnote{Persistence landscape forms a Banach space and therefore, the theory of random variables can be applied in such spaces which is a natural choice when dealing with the financial asset selection.} by definition involves no parameter and thus it is free from parameter tuning and over-fitting risk 
\cite{bubenik2015statistical}. The relevance of persistence norms as risk indicators mentioned above is also supported by numerous empirical studies (see \cite{guo2020empirical}, \cite{gidea2018topological}, \cite{ruiz2022tda}, \cite{saengduean2020topological}, \cite{goel2020topological}, \cite{goel2023sparse}
). 

We now explain the procedure of calculating the topological risk contribution by each asset followed by the necessary notation and optimization model minimizing the topological risk of the portfolio.

\textbf{Notation:}
Consider a set of $n$ feasible assets $A_1, A_2,\ldots, A_n$ available for investment with uncertain respective future returns $r_1, r_2, \ldots, r_n$. Let $\mathbf{w}=(w_1, w_2, \ldots, w_n)^{'} \in \mathbb{R}^n$ denotes a decision vector corresponding to  portfolio allocation where $w_i$ denotes the proportion of total budget to be investment in $i$-th asset $i=1,\ldots, n$.
\textbf{Asset topological risk:} 
For a portfolio $\mathbf{w}$ consisting of $n$ assets, the topological risk contribution of the $i$th asset, where $i = 1, \ldots, n$, is computed through the following steps:
% {\color{red}{We compute the topological risk contribution of each asset in a portfolio $\mathbf{w}$ of $n$ assets using the following steps:}}
\begin{itemize}
    \item Given the return time series of the $i$th asset, denoted as $X_i = {r_i(t)}_{t=1}^T$ with a window size of $T$, the process of generating a sequence of persistence landscapes involves the following sequential steps:    
%    {\eta_{i}^{(j)}(k)\}$
 \begin{itemize}
    \item [(i)] Obtain a set of point clouds by employing sub windowing technique and applying Takens embedding to each of the sub windows. We choose $\tau=1$ and $d=3$ following (\cite{tran2023detecting}, \cite{ruiz2022tda}, \cite{goel2020topological}, \cite{seversky2016time}, %\cite{small2005applied}, 
    \cite{pereira2015persistent}).
    \item [(ii)] Obtain Rips filtration corresponding to each point cloud.
    \item [(iii)] Generate a persistence diagram based on the Rips filtration carried out in the previous step.
    \item [(iv)] Finally transform the persistence diagrams to obtain persistence landscapes. 
   \end{itemize}

\item We then obtain the corresponding series of persistence landscape $p$-norms for $p$ =1 from the persistence landscapes corresponding to each of the sub windows.  The choice of $p=1$ is motivated by (\cite{akingbade2024topological}, \cite{gidea2017topological}, \cite{goel2020topological}, \cite{goel2023sparse}). %Empirical validation for $p=1$ and $p=2$ also yielded consistent results.
Empirical validation for $p=1$ and $p=2$ are generally found to yield similar results which is also consistent with our findings.

\item The mean persistence landscape of $i$th asset and henceforth its norm is calculated using the equations (\ref{tdamean}) and (\ref{meannorm}), respectively.  The mean persistence landscape serves as the reference persistence landscape %\footnote{There are other measures which can serve as the reference point like .... [][] but we select mean of persistence landspace due to its ... } 
and its norm, called as expected norm serves as the reference landscape norm.
%, serves the reference landscape for us. It represents the average persistence landscape over time.

\item The asset topological risk for $i$th asset $\Lambda_{i}; i=1\dots,n$, is then measured as the average squared difference of its series of persistence landscape norm from its reference landscape norm{\color{blue}{\footnote{In this sequel, we define topological risk by quantifying the deviation of the norm of individual persistence landscapes from the norm of the mean persistence landscape. This construction builds upon key properties of persistence landscapes: they are stable under perturbations, admit a well-defined notion of expectation in Banach space, and allow for direct computation of $L^p$ norms. These properties collectively support their use in defining a principled measure of variability across topological summaries. While our empirical application centers on 0-dimensional persistence diagrams ($H_0$), this is a modeling choice made for reasons of interpretability and computational efficiency. Crucially, the definition of topological risk is not restricted to $H_0$: it extends naturally to persistence landscapes derived from higher-dimensional features (e.g., $H_1$, $H_2$).  More broadly, the definition remains valid when applied to any suitable topological summary as long as a meaningful notion of mean and deviation can be established akin to the approach outlined in this context. The framework is thus broadly applicable, with the flexibility to incorporate richer topological signals where appropriate.}.}} 
 \end{itemize}

Follow the detailed Algorithm \ref{algo:main_algo} to get $\Lambda_{i}; ~i=1,\ldots,n$.\footnote{
Please refer to \cite{goel2023sparse} for detailed algorithms on obtaining the Rips filtration (step 4), persistence diagrams (step 5), and persistence landscapes (step 6).} The topological risk is then defined as follows:

\begin{algorithm}[!ht]
  %\SetAlgoLined
    \KwData{Time-series data $ X_i = \{r_{i}(t)\}_{t=1}^T$ of $M$ months, i.e., $T=21\times M$, for a given constituent $i \in \{1, 2, \dots, n\}$ of a portfolio $w$ and a sequence of resolutions $\epsilon_0 < \epsilon_1 < \ldots < \epsilon_N$.}
    
%     \KwResult{The series of persistence landscape $\{\eta_{i}^{(j)}(k)\}
% $; $i \in \{1, 2, \dots, n\}$.}
  \KwResult{The series of deviation of the persistence landscape norm from the norm of mean persistence landscape}
  %$\{\eta_{i}^{(j)}(k)\} $; $i \in \{1, 2, \dots, n\}$.}

Split time-series data into $\Tcal$ overlapping sub-series of length $\tilde{T}$ days with a shift of $h<\tilde{T}$ days such that $\Tcal=\frac{T-\tilde{T}}{h}+1$ (we use $M=12,~\tilde{T}=126$ and $h = 21$)\;
        \For{$j = 0,1, 2, \dots, \Tcal-1$}{
        %\For{each sub-series $\{R_{i,t}\}_{t=1}^{\tilde{T}}$}{
            Extract the sub-series $\{R_{i,t}\}_{t=jh+1}^{jh+\tilde{T}}$
            Apply Takens' time-delay embedding with time delay $\tau$ and embedding dimension $d$ to obtain point cloud $X_{i}^{(j)}$ (we use $\tau=1$ and $d=3$)
{\scriptsize        
    \[
	X_{i}^{(j)}=
    \left[
	\begin{array}{cccc}
	R_{i,jh+1}& \dots  &R_{i,jh+1+(d-1)\tau}\\
	R_{i,jh+2}& \dots   &R_{i,jh+2+(d-1)\tau}\\
	\vdots & \vdots & \vdots \\
	R_{i,jh+\tilde{T}-(d-1)\tau} &\dots  &R_{i,jh+\tilde{T}}\\
	\end{array}
	\right]
	\]
 }

            Compute Rips filtration $\{\Rcal_{\epsilon_n}(X_{i}^{(j)})\}_{n\in\mathbb{N}}$ for the point cloud $X_{i}^{(j)}$. 
            
            Compute persistence diagram $\Dcal_{X_i^{(j)}}$ for the filtration $\{\Rcal_{\epsilon_n}(X_{i}^{(j)})\}_{n\in\mathbb{N}}$. 
            
            Compute persistence landscape $\{\eta_{i}^{(j)}(k)\}$ from the birth-death pairs $\{(b_m, d_m)\}_{m=1}^{r}$ corresponding to 0- dimensional components extracted from the persistence diagram $\Dcal_{X_i^{(j)}}$.

      }

Compute the mean persistence landscapes $\{\overline{\eta}_{i}(k)\}$ as follows: 
\begin{equation*}
   \overline{\eta}_{i}(k)= \frac{1}{\Tcal} \displaystyle \sum_{j=0}^{\Tcal-1} \eta_{i}^{(j)}(k).
\end{equation*} 

Compute the corresponding $p$-norm (we use $p=1$ and $k=1$) 
\begin{equation*}\begin{split}
||\overline{\eta}_{i}||_p = \left( \displaystyle \sum_{k=1}^\infty ||\overline{\eta}_{i}(k)||^p_p \right)^{\frac{1}{p}}, \;\; \text{and}\;\;\; ||\eta_{i}^{(j)}||_p = \left( \displaystyle \sum_{k=1}^\infty ||\eta_{i}^{(j)}(k)||^p_p \right)^{\frac{1}{p}},j=0,\ldots, \Tcal-1\\ \end{split}\end{equation*}
 
Return $\Lambda_{i} \leftarrow 
\displaystyle\sum_{j=0}^{\Tcal-1}\left(||\eta_{i}^{(j)}||_p - ||\bar{\eta_i}^n||_p\right)^2$, 
%$\Lambda_{i,1} \leftarrow 
%\displaystyle\sum_{i=1}^M(||\eta^i(t)||_p - ||\bar{\eta_i}^n||_p)^2$
\caption{Algorithm to obtain the topological risk for all assets $i=1,\ldots,n$.} 
\label{algo:main_algo}
\end{algorithm}

\textbf{Portfolio topological risk (PTR):} For a portfolio allocation vector $\mathbf{w}=(w_1, w_2, \ldots, w_n)^{'} \in \mathbb{R}^n$, its topological risk is defined as: 
\begin{equation}\label{eq:tdap}
\mathbf{w}'Q\mathbf{w} 
\end{equation}

where $Q=[\Lambda_{i}]_{i=1}^n$ is a diagonal matrix with $i$th diagonal entries corresponds to the $i$th asset topological risk $\Lambda_{i}$; $i=1,\ldots,n$, named as topological risk matrix. As by definition $\Lambda_{i} \geq 0, \, \forall \, i=1,\ldots,n$, the topological risk matrix $Q$ is a positive definite matrix.

For a polyhedron set of feasible portfolios $\mathbf{W} = \{(w_1, w_2, \ldots, w_n)^{'} \in \mathbb{R}^n; \displaystyle\sum_{i=1}^{n}w_i = 1, w_i \geq 0\}$ containing budget and no-short selling restrictions, the proposed optimization model minimizing the topological risk is then given by the following Quadratic program:

\begin{center}
    \textbf{(TDA-PO)} \quad min $ \mathbf{w}^T Q \mathbf{w} $\\
  \;\;s.t. \hspace{0.7cm}$\mathbf{w} \in \mathbf{W}$ \\
\end{center}

The problem TDA-PO is a convex program and hence computationally efficient. For a desired cardinality $\mathbf{k}$ of assets to select, an investor solves the following 0-1 form of the proposed model:  

\begin{center}
    \textbf{(TDA-IPO)}  \quad min $ \mathbf{w}^T Q \mathbf{w} $\\
  %s.t.\hspace{0.7cm} $\mathbf{w} \in \mathbf{Z} $ \\
  %  \hspace{0.7cm}   $\;\;\;\mathbf{z}^{'} \mathbf{e} = k$\\
\hspace{2cm}  $\displaystyle\sum_{i=1}^{n}z_i = \mathbf{k}$ \\
\hspace{4.5cm}  $0 \leq w_i \leq z_i; \; z_i \in \{0,1\},  \; i=1,2,\ldots,n.$

\end{center}

Cardinality constraint allows an investor to enforce a specific number of stocks to invest, avoiding both the cases of excessively diversified portfolios that may dilute potential returns and under-diversified portfolios pertaining to higher market risk.

The proposed model TDA-PO (or TDA-IPO for cardinality condition) aims to minimize the topological deviation or instability of an asset's persistence landscape compared to its reference landscape as quantified by its norm. It is designed to capture the inherent topological properties of an asset's behavior and quantify how far its persistence landscapes deviate from the average landscape. A lower value of asset's topological risk indicates that its time-varied persistence landscape norms are closer to the reference landscape norm, suggesting a more stable and predictable behavior. Conversely, its higher value signifies a higher deviation from the reference landscape norm, indicating potential topological instabilities or structural changes in the asset's behavior. By incorporating topological information, PTR provides additional insights into an asset's risk profile that might not be captured by traditional risk measures based on statistical moments. Furthermore, by its definition, PTR is free from any model-based estimation error and hence able to generate robust and stable out-of-sample results which we also confirm numerically in our empirical section. Overall, PTR can be a valuable risk measure in the context of portfolio management and asset allocation.

\FloatBarrier 
\subsection{Proposed Methodology}
We now elucidate the methodology to carry out an empirical investigation of the proposed study by detailing the sample data, sample period, rolling window scheme and the seven benchmarked PO models considered for the comparative analysis.
%of our proposed strategy v/s the traditional PO models.

\subsubsection{Sample Data and Sample Period}
The sample data contains daily closing prices of the S\&P500 and its constituents over a sample period spanning nearly 10 years from December 10, 2012, to August 11, 2022. The data is collected from Thomson Reuters Datastream \footnote{Thomson Reuters Datastream is a licensed financial database and may require institutional access.}. Since the index's composition is time-varying, we follow the standard approach from the literature to select the constituents. More precisely, we choose only those constituents for which the data is available for the whole period and drop the stocks with missing observations from our analysis. This results in 462 constituents. 

The daily return for $i$-th stock at $t$-th day is calculated as $r_{it}=\frac{p_{it}^{c}-p_{it-1}^{c}}{p_{it-1}^{c}};$ $i=1,\ldots, n, \ t=1,\ldots, T,$  where $p_{jt}^{c}$ and $p_{jt-1}^{c}$ are respectively, the closing prices at $t$-th and $(t-1)$-th day. 

\subsubsection{Rolling Window Scheme}
To investigate the performance of all the portfolios, we utilize a rolling window approach consisting of solving a sequence of problems, each taking into account an in-sample period and an out-of-sample period, both of them measured in days. Each optimal portfolio obtained by solving a model over the in-sample data is then evaluated in the corresponding out-of-sample data. The resulting 
out-of-sample return is observed and recorded. The in-sample time horizon is then shifted ahead by the number of weeks corresponding to the out-of-sample period to get the new in-sample window, and the procedure is reiterated. We primarly employ %\textcolor{red}{.. varying case of window size to check the consistency of numerical outcomes:}
a rolling window of 273 trading days (13 months) with an in-sample period of 252 trading days (12 months) and an out-of-sample period of 21 trading days (one month) in line with \cite{fastrich2014cardinality,goel2024sparse}. We shift the window by 21 trading days (one month) at each rolling step, leading to a total of 108 windows. To check the consistency of numerical outcomes from this window size setting, we re-investigate computational analysis for the varying window size as briefly described in the section 6 under the heading of robust analysis.

\subsubsection{Benchmarked PO Models for the Comparative Analysis}
To check the efficacy of the proposed model TDA-PO, we compare its out-of-sample results with the following seven famous traditional PO models from literature and two benchmark portfolios, the naive $1/N$ portfolio and the benchmark index S\&P 500: 

\begin{itemize}
\item [1.] Mean-variance (MP) model \cite{marko1952}

\item [2.] Global minimum variance model (GMV) \cite{Clarke2010MinimumVP}: 

\item [3.] Mean-value at risk model (MVaR) \cite{lofti2016}

%Somayyeh Lofti, Stauros Andrea Zenios, Equivalence of Robust Var and Cvar Optimization, Financial Institutions Center, 2016.

\item [4.] Mean-conditional value at risk model (MCVaR) \cite{Rockafellar2000}

\item [5.] Sharpe model \cite{sharpe94}%: 

\item [6.] STARR model \cite{starr_martin}%: STARR model maximizes Sharpe-CVaR ratio over the feasible set $\mathbf{W}$. 

\item [7.] Omega model \cite{mausser2006}%: Omega model 
%maximizes the Omega ratio which is a ratio of upper deviation $E(R_w - L)^{+}$ to the lower deviation $E(L- R_w)^{+}$ over the set $\mathbf{W}$ where $L \in \mathbb{R}$ is a benchmark point. We set $L$ as the average return of the benchmark index S\&P 500 for computation purpose. 
\end{itemize}

The mathematical detail of all the above seven benchmarked PO models is given in the Appendix \ref{sec:Appendix_II}. All the optimization problems including our proposed model 
(TDA-PO) have been solved using $\mathbf{R}$ software with gurobi interface for integer programming (see Appendix \ref{sec:Appendix_II} for details on packages and libraries used). %We use Windows 10, 64-bit operating system with 64 GB RAM and Intel(R) Core 2.10 GHz processor. 

%}

\section{Results and Analysis}
% Table generated by Excel2LaTeX from sheet 'sum_ind'
\subsection{Descriptive Statistics}

\begin{figure}[ht!]
		\begin{center}
  			\subfigure{%
				\includegraphics[scale=.5]{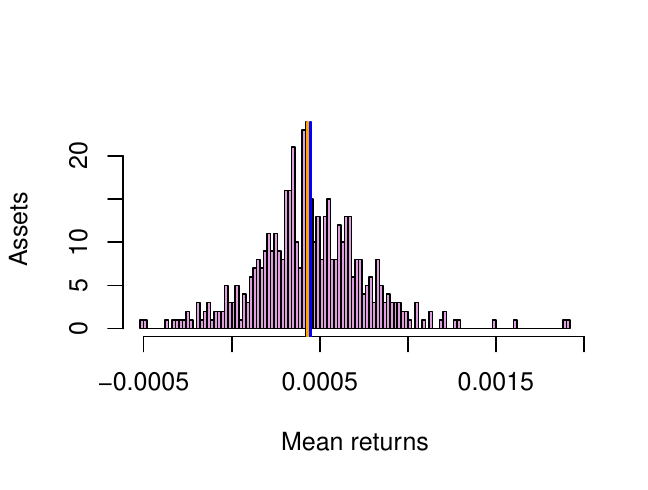}
				\label{Returns}}
    \quad
			\subfigure{%
				\includegraphics[scale=.5]{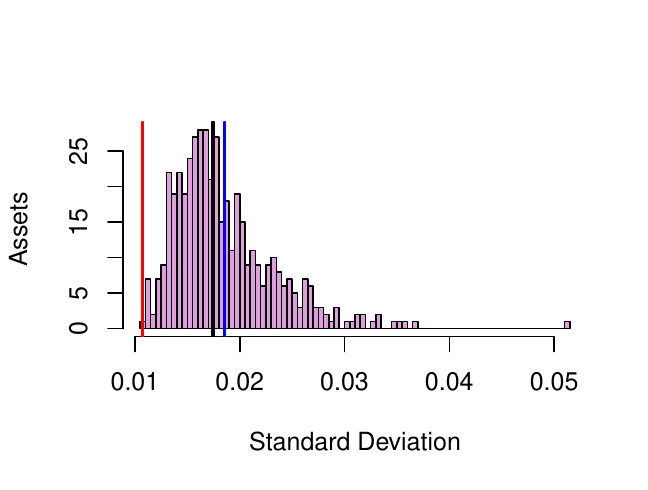}
				\label{Standard Deviation}}
    \quad
			\subfigure{%
				\includegraphics[scale=.5]{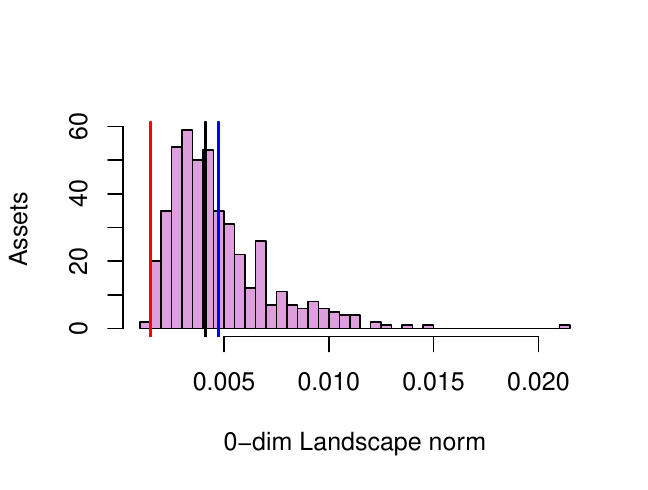}
				\label{Landscape norm for 0-dimensional features}}
		\end{center}
		\caption{Mean returns, standard deviation, and 0-dim Landscape norm for all the constituents in the period of study. The blue (black) line in each of the histograms represents the corresponding mean (median) value of the statistic across all assets. The red line depicts the statistics value corresponding to the Index} \label{summary_assets}
	\end{figure}

Table \ref{tab:des_index} presents the descriptive statistics for the benchmark index S\&P 500 for the considered period of 10 years from December 10, 2012, to August 11, 2022, confirming the typical stylized facts of financial time series, such as negative skewness (asymmetry) and high kurtosis (heavy tails). Due to the large number of constituents, presenting a complete table for descriptive statistics of each of the constituents is impractical. Therefore, we show the histograms in Figures \ref{summary_assets} summarizing the mean return, the standard deviation, and the 0-dim landscape norm for all the 462 constituents. The blue (black) line in each of the histograms represents the corresponding mean (median) value of the statistic across all assets. The red line depicts the statistics value corresponding to the Index S\&P 500.

\begin{table}[ht]
  \centering
  \caption{Descriptive statistics for the S\&P500 index of 10 years from December 10, 2012, to August 11, 2022}
    \begin{tabular}{c c}
     \toprule

    {Mean} & 4.31E-04 \\
   {Max} & 8.97E-02 \\
    {Min} & -1.28E-01 \\
   {Std Dev} & 1.07E-02 \\
   {Skewness} & -9.60E-01 \\
   {Kurtosis} & 1.92E+01 \\
    10th percentile   & -9.54E-03 \\
    50th percentile      & 3.74E-04 \\
    90th percentile      & 1.09E-02 \\
\bottomrule
    \end{tabular}%
  \label{tab:des_index}%
\end{table}%

\subsection{Out-of-Sample Analysis}

Table \ref{tab:compare}\footnote{As we have a considerable number of windows for the considered data period, to provide comprehensive details, we present the results based on the series created by concatenating the out-of-sample returns from each optimal in-sample portfolio.} records the out-of-sample results of the portfolios from the six models namely, the proposed model TDA-PO and seven benchmarked PO models, GMV, MP, MVaR, MCVaR, Sharpe, STARR, and Omega, and the two benchmark portfolio strategies, the naive $1/N$ and the market index S\&P 500, in terms of several performance measures (see Appendix \ref{sec:Appendix_I}) calculated over the out-of-sample period. % \textcolor{red}{All the optimization models are 

\begin{table}[htbp]
\footnotesize
\setlength{\tabcolsep}{3pt}
  \centering
  \caption{Comparison analysis: Out-of-sample performance matrices obtained over rolling window scheme from the eight optimization models and two benchmark portfolio strategies. The best values (second best values) are highlighted in bold (italics) for the reader's convenience. The $*$ is used to mark the significant values in the statistical tests for the Sharpe ratio at $90\%$ confidence level for the TDA PO model vs others. Values highlighted in red represent performance metrics that are statistically inferior to those of the naive portfolio at the 95\% confidence level, with respect to risk measures: variance, VaR$_{0.05}$, and CVaR$_{0.95}$. Acronyms of the metrics are expanded and defined in the Appendix \ref{sec:Appendix_I}. The Assets are averages of total assets with non-zero weights across different windows.}
  \label{tab:compare}   
\begin{tabular}{lrrrrrrrrrr}
\toprule
Metric & TDA PO & GMV & MP & Sharpe & STARR & Omega & MCVaR & MVaR & Naive & Index \\
\midrule
\multicolumn{11}{c}{\textbf{In-sample period = 1 year, Out-of-Sample period = 1 month}} \\
\midrule
EMR & \textbf{6.330E-04} & 2.036E-04 & 2.343E-04 & 2.101E-04 & 4.658E-04 & \textit{4.696E-04} & 2.693E-04 & 2.087E-04 & 3.601E-04 & 3.655E-04 \\
Min & -1.447E-01 & \textbf{-8.943E-02} & \textit{-9.442E-02} & -1.171E-01 & -1.004E-01 & -1.367E-01 & -9.348E-02 & -1.038E-01 & -1.392E-01 & -1.277E-01 \\
Stdev & 1.143E-02 & \textbf{8.296E-03} & \textit{8.348E-03} & 9.998E-03 & \textcolor{red}{1.234E-02} & \textcolor{red}{1.456E-02} & 9.169E-03 & 9.593E-03 & 1.154E-02 & 1.104E-02 \\
DD & 8.433E-03 & \textbf{6.208E-03} & \textit{6.230E-03} & 7.398E-03 & 9.064E-03 & 1.082E-02 & 6.785E-03 & 7.234E-03 & 8.584E-03 & 8.145E-03 \\
VaR$_{0.95}$ & 1.565E-02 & \textbf{1.140E-02} & \textit{1.143E-02} & 1.399E-02 & \textcolor{red}{1.809E-02} & \textcolor{red}{2.234E-02} & 1.298E-02 & 1.346E-02 & 1.640E-02 & 1.663E-02 \\
CVaR$_{0.95}$ & 2.713E-02 & \textit{2.014E-02} & \textbf{2.002E-02} & 2.421E-02 & \textcolor{red}{3.112E-02} & \textcolor{red}{3.680E-02} & 2.218E-02 & 2.365E-02 & 2.882E-02 & 2.796E-02 \\
SR & \textbf{5.540E-02} & 2.454E-02* & 2.806E-02* & 2.101E-02* & \textit{3.775E-02} & 3.225E-02 & 2.937E-02* & 2.175E-02* & 3.119E-02* & 3.310E-02* \\
SVR$_{0.95}$ & \textbf{4.044E-02} & 1.786E-02 & 2.050E-02 & 1.501E-02 & \textit{2.575E-02} & 2.102E-02 & 2.075E-02 & 1.551E-02 & 2.196E-02 & 2.197E-02 \\
SCR$_{0.95}$ & \textbf{2.333E-02} & 1.011E-02 & 1.170E-02 & 8.679E-03 & \textit{1.497E-02} & 1.276E-02 & 1.214E-02 & 8.821E-03 & 1.249E-02 & 1.307E-02 \\
Sortino & \textbf{7.506E-02} & 3.279E-02 & 3.760E-02 & 2.840E-02 & \textit{5.139E-02} & 4.339E-02 & 3.970E-02 & 2.884E-02 & 4.195E-02 & 4.487E-02 \\
Rachev$_{0.95}$ & \textit{6.699E-04} & 3.454E-04 & 3.485E-04 & 5.016E-04 & \textbf{8.349E-04} & 1.175E-03 & 4.249E-04 & 4.556E-04 & 7.063E-04 & 6.529E-04 \\
Assets & 168 & 35 & 35 & 61 & 13 & 14 & 39 & 34 & 462 & 462 \\
PTR & \textit{5.682E-04} & 5.745E-04 & 5.787E-04 & 6.316E-04 & 5.897E-04 & 7.998E-04 & 6.084E-04 & 7.385E-04 & 5.708E-04 & \textbf{4.810E-04} \\
\bottomrule
\end{tabular}
\end{table}

We summarize the observation from the Table \ref{tab:compare} in below: 
 
\begin{itemize}
    \item \textbf{Return performance in terms of EMR (Excess Mean Return):} It is evident from Table \ref{tab:compare} that the proposed model, TDA-PO attains the highest value of mean return in comparison to all the other seven models and both the benchmark portfolio strategies. The model TDA-PO attains almost double mean return in comparison to all the three variance-based models, i.e. GMV, MP, \& Sharpe model, and the two tail-risk measure based models i.e. MVaR \& MCVaR and from both the benchmark portfolio strategies. Interestingly, the proposed model TDA-PO which does not account for the mean return function explicitly, manages to get the highest value of EMR in comparison to the all other PO models. The outperformance of the proposed model TDA-PO in terms of EMR shows its financial gains over other popular investment strategies. 

A key takeaway from these results is that TDA-PO structural advantage does not stem from chasing returns but rather from efficiently adapting to market dynamics. By constructing portfolios without imposing rigid return-based constraints, TDA-PO enhances return stability while mitigating reliance on historical mean return estimates, which are often unreliable in real-world investment settings.

\item \textbf{Risk analysis in terms of stdev, DD, VaR$_{0.95}$, CVaR$_{0.95}$, PTR:} The risk measure standard deviation (stdev) captures the variation of portfolio return around its mean values while the other risk measures, DD, VaR$_{0.95}$, and CVaR$_{0.95}$ focus on the downside risk performance of the model. On the other hands, PTR tells the presence of topological risk (or variation in the topological features over the time) in the out-of-sample returns. We can notice from Table \ref{tab:compare} that the proposed model TDA-PO achieves less amount of risks in terms of risk measures (stdev, DD, VaR$_{0.95}$, and CVaR$_{0.95}$) in comparison to the STARR model, Omega model, and the naive $1/N$ portfolio. While comparing with the variance-based models, i.e. the GMV, MP, and Sharpe, and and the two tail-risk measure based models i.e. MVaR \& MCVaR, the TDA-PO model generates only marginally high risk. The lowest (i.e., best) PTR is reported by the S\&P 500 index and the second lowest is given by the TDA-PO model. In contrast, the highest (i.e., worst) PTR is observed for the Omega model. 

This indicates that TDA-PO does not aggressively sacrifice risk control for return maximization - a key advantage in real-world applications, where excessive risk exposure can lead to substantial drawdowns during adverse market conditions.

\item \textbf{Risk adjusted return performance in terms of Sharpe, Sortino, Sharpe-CVaR (SCR$_{0.95}$), Sharpe-VaR (SVR$_{0.95}$), and Rachev ratios:} Financial ratio measures the overall performance of a portfolio by combining return and risk in one single formula. It gives risk-adjusted return (or return per unit) for the underlying risk measure. We notice that the model TDA-PO outperforms all other models in terms of all financial ratios (except that the STARR model generates the best Rachev ratio followed by the performance by TDA-PO), depicting that it generates a high level of return without making much comprise with the risk levels. Therefore, the proposed model ranks as the first choice among all rational investors. 

The main reason for having such good out-of-sample performance in all aspects is that the proposed model captures the time-varying market features in a much better way than the traditional PO models. 
%TDA1 has lesser value of Rachev ration than Omega

\item \textbf{Performance analysis against the naive $1/N$ strategy and the market index S\&P 500:} The naive $1/N$ strategy and the market index S\&P 500 serve as important benchmarks in assessing whether the optimization-based investment strategies deliver superior performance. From \ref{tab:compare}, we observe that the proposed TDA-PO model does not only achieve almost double EMR value in comparison to both the benchmark strategies, $1/N$ and S\&P 500, but also improves the values of all financial ratios.  This confirms that proposed TDA based strategy provides noteworthy advantages over passive allocation strategies.

On the other hands, though the variance-based models i.e. GMV, MP, and Sharpe, generate relatively lesser values of risks than the $1/N$ and S\&P 500, %all other models improve in terms of EMR.  where the proposed TDA-PO model almost doubled its value than the 1/N and S\&P 500. 
they fail to deliver competitive values for EMR and risk-adjusted returns, reinforcing the inefficiency of variance minimization without adaptive structural adjustments.

\end{itemize}

In short from Table \ref{tab:compare}, we conclude the best out-of-sample performance from the model TDA-PO in comparison to several popular PO models. The proposed model achieves first rank in terms of out-of-sample mean return and financial ratios such as Sharpe, Sharpe-CVaR (SCR$_{0.95}$), Sharpe-VaR (SVR$_{0.95}$), Sortino, and Rachev ratios. It shows that the TDA-based proposed model able to produce a good amount of return without generating high risk resulting in the best risk-adjusted return performance.  These results underline the out-of-sample efficacy of TDA-based models in delivering high returns while
managing risk. Though, the model GMV generates least values of risks, it also suffers with the least return, lesser than the simplest Naive $1/N$ and the benchmark index. Therefore, our numerical findings recommend the GMV and MP based portfolios only to risk-averse investors.

\begin{figure*}[ht!]
    \centering
    \subfigure[]{%
        \includegraphics[height=6cm,width=12cm]{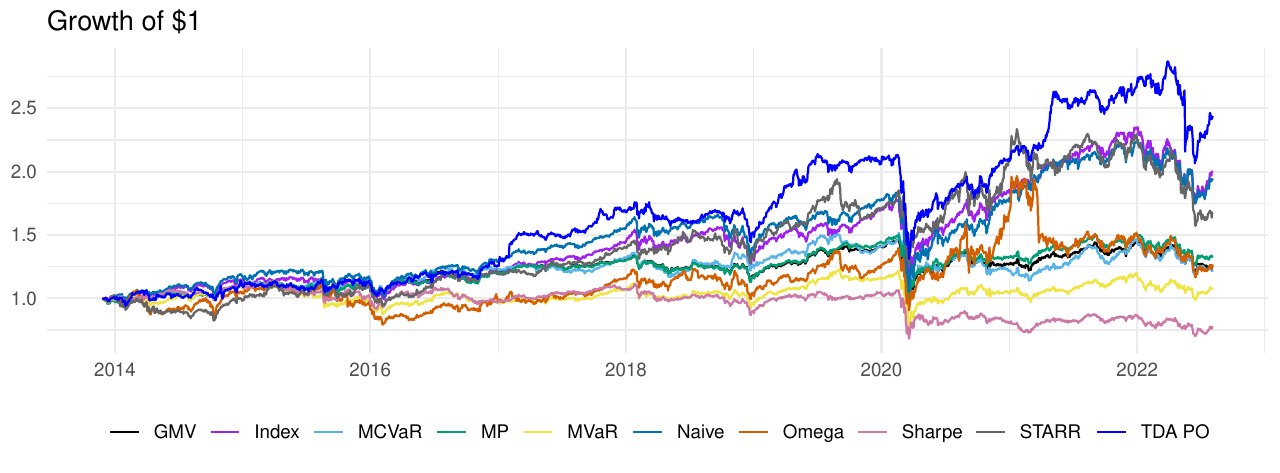}
        }
    \quad
    \subfigure[]{%
        \includegraphics[height=6cm,width=12cm]{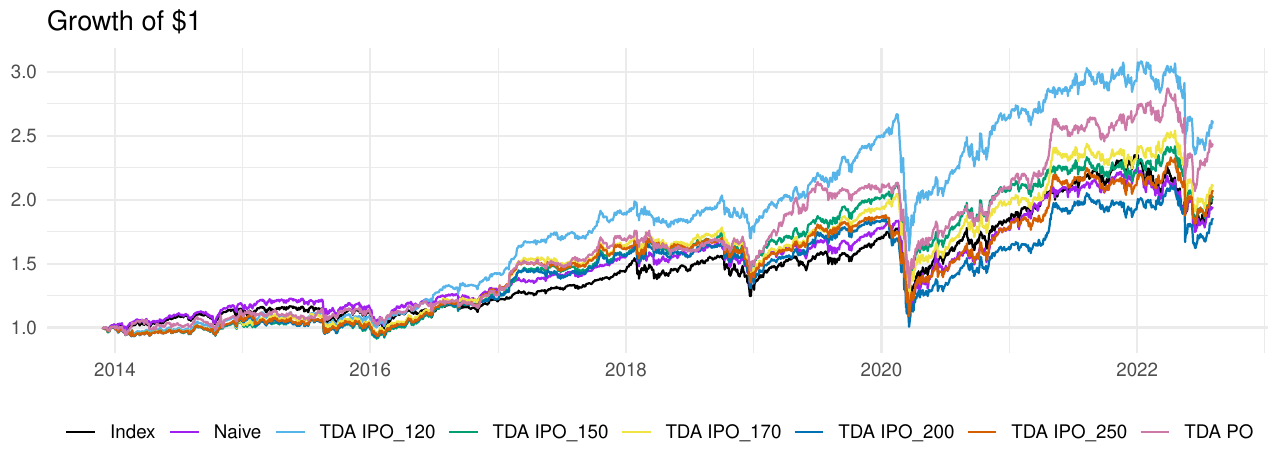}
        }
		\caption{The wealth of the portfolio starting from \$1 for (a) TDA-PO and the models under comparison and for (b) the integer constraint TDA portfolios. The returns have been adjusted to account for a transaction cost rate of 0.3\%. The plot illustrates the performance of each model and the index vs the TDA-based models over the given period. }\label{return_cumu}
\end{figure*}

For pictorial illustration purposes, we report the growth of \$1 investment adjusted with the transaction cost (TC) at rate 0.3\%\footnote{The growth of \$1 investment of the portfolio $\mathbf{w}$ on $t$-day is calculated as $1(1+A_{w1})(1+A_{w2})\ldots(1+A_{wt})$ with $A_{wt}$ be the $t$-th adjusted return realization with TC of portfolio return $R_w$ i.e. the cumulative return for each model is calculated by subtracting the TC from the daily returns and then compounding these adjusted returns over time.
For each day \(t\), the adjusted return is given by:
\[
{AR}_{wt} = R_{wt}  - TC_d
\]
where the $TC_d$ is the transaction cost at the rebalancing day $d$ is computed as:
\[
TC_d = 
\begin{cases} 
\text{turnover} \times 0.0003 & \text{if } d \text{ is the rebalancing day} \\
0 & \text{otherwise}
\end{cases}
\]
with turnover is defined as the sum of the absolute differences between the current ($mth$ window) and previous window ($mth$ window):
\[
\text{turnover} = \sum_{i} |\text{weights}_{m,i} - \text{weights}_{m-1,i}|
\]
%We took cost rate=2\%.
} %and drawdown risk () 
(wealth graph) by all the models 
in the %respective graphs \ref{returnDD}(a) and 
Figure \ref{return_cumu}. Transaction costs, which mainly emerge during portfolio adjustments or re-balancing, are crucial aspect in examining the performance of different portfolios strategies and can influence decision-making %(\cite{amihud1986liquidity}, 
(\cite{arnott1990measurement}, \cite{yu2022dynamic}). To plot the wealth graph from each model, we first concatenate their out-of-sample returns from each rolling window to get a single return series and thereby, obtaining a single wealth graph. 
%\textcolor{red}{.... days from December 2016 to December 2019} out-of-sample returns. 
Plots of these wealth graphs help us to illustrate the clear time-varying behavior from all the models. Figure \ref{return_cumu} depicts the dominance of the proposed model TDA-PO throughout the considered period in terms of wealth. %(Figure \ref{returnDD}(a)) and drawdown risk (Figure \ref{returnDD}(b)). 
A sharp drop can be seen near the point of April 2020, where the least and largest fallout is given by the PO models STARR and Sharpe, respectively. The proposed model bypasses all portfolios by generating large returns while saving an investor from the sharp fall.

\textbf{Effect of cardinality in the model TDA-PO} 

Next, in order to see the effect of cardinality in the model TDA-PO, we propose to check the numerical performance of the model TDA-IPO for five different values of cardinality $\mathbf{k}$ = 120, 150, 170, 200, 250 in Table \ref{tab:integer}. %\textcolor{red}{The model TDA-IPO is solved using R software with Gurobi interface when it takes a average time of ... per window. 
Table \ref{tab:integer} depicts that the TDA-based model selects 168 assets when no cardinality is introduced (i.e. the model TDA-PO) while it goes only up to 64 assets when $\mathbf{k} = 250$.  On comparing the out-of-sample values for performance matrices, we find the best values of mean return and all the financial ratios from the model TDA-PO in comparison to the portfolios generated by TDA-IPO for five different values of $\mathbf{k}$. Nevertheless, irrespective of the values for $\mathbf{k}$ in TDA-IPO, we observe the model generates higher values of mean and ratios in comparison to all other traditional models as prescribed in Table \ref{tab:integer}, confirming that its asset constraint does not significantly weaken its financial efficiency. Lastly, we view the wealth graphs of TDA-IPO for all the five values of $\mathbf{k}$ = 120, 150, 170, 200, 250 along with the TDA-PO in Figure \ref{return_cumu}. As evident, the model TDA-PO produces higher wealth throughout the period with the least return at TDA-IPO for $\mathbf{k}$= 200. %However, no statistical evidences were found verify that the TDA PO model consistently generates a higher Sharpe ratio than the TDA IPO models across all values of $\mathbf{k}$.

\begin{table}[htbp]
\footnotesize
\setlength{\tabcolsep}{3pt}
  \centering
  \caption{Sensitivity analysis: Out-of-sample performance metrics for the model (TDA-IPO) for five different cardinality values, $k$= $120$, $150$, $170$, $200$ and $250$. Acronyms of the metrics are expanded and defined in the Appendix \ref{sec:Appendix_I}. The Assets are averages of total assets with non-zero weights across different windows.}
  \begin{tabular}{lrrrrrr}
    \toprule
       & \multicolumn{1}{l}{TDA PO} & \multicolumn{1}{l}{TDA IPO$_{120}$} & \multicolumn{1}{l}{TDA IPO$_{150}$} & \multicolumn{1}{l}{TDA IPO$_{170}$} & \multicolumn{1}{l}{TDA IPO$_{200}$} & \multicolumn{1}{l}{TDA IPO$_{250}$} \\
    \midrule
    EMR   & \textbf{6.330E-04} & 4.845E-04 & 5.078E-04 & 5.301E-04 & 4.787E-04 & 5.307E-04 \\
    Min   & -1.447E-01 & -1.252E-01 & -1.247E-01 & -1.239E-01 & -1.134E-01 &\textbf{ -1.055E-01} \\
 Stdev & 1.143E-02 & 1.099E-02 & 1.102E-02 & 1.118E-02 & 1.103E-02 &\textbf{ 1.081E-02} \\
    DD    & 8.433E-03 & 8.106E-03 & 8.078E-03 & 8.214E-03 & 8.134E-03 & \textbf{7.879E-03} \\
  VaR$_{0.95}$   & 1.565E-02 & 1.588E-02 & 1.614E-02 & 1.560E-02 & 1.554E-02 & \textbf{1.533E-02} \\
       CVaR$_{0.95}$  & 2.713E-02 & 2.668E-02 & 2.673E-02 & 2.724E-02 & 2.694E-02 & \textbf{2.625E-02} \\
    SR & \textbf{5.540E-02} & 4.410E-02 & 4.609E-02 & 4.741E-02 & 4.339E-02 & 4.910E-02 \\
    SVR$_{0.95}$& \textbf{4.044E-02} & 3.050E-02 & 3.147E-02 & 3.397E-02 & 3.080E-02 & 3.462E-02 \\
    SCR$_{0.95}$ &\textbf{ 2.333E-02} & 1.816E-02 & 1.900E-02 & 1.946E-02 & 1.777E-02 & 2.022E-02 \\
    Sortino & \textbf{7.506E-02} & 5.977E-02 & 6.286E-02 & 6.453E-02 & 5.885E-02 & 6.735E-02 \\
    Rachev$_{0.95}$ & \textbf{6.699E-04} & 6.296E-04 & 6.338E-04 & 6.589E-04 & 6.428E-04 & 6.182E-04 \\
    Assets & 168 & 51 & 57& 59& 60& 64\\
PTR &   5.682E-04  & 4.997E-04 & 5.084E-04 & 5.074E-04 & 5.018E-04 & 4.748E-04\\
 %   Sharpe test & \multicolumn{1}{l}{NA} & 3.101E-01 & 3.721E-01 & 4.099E-01 & 2.124E-01 & 5.000E-01 \\
 %   Assets & 462   & 50.898 & 57.444 & 58.611 & 59.926 & 63.593 \\
    \bottomrule
    \end{tabular}%
  \label{tab:integer}%
\end{table}%

\subsection{Portfolio Stability}

\subsubsection{Portfolio Stability in terms of Turnover}
\begin{figure*}[ht!]
    \centering
    \includegraphics[width=\textwidth]{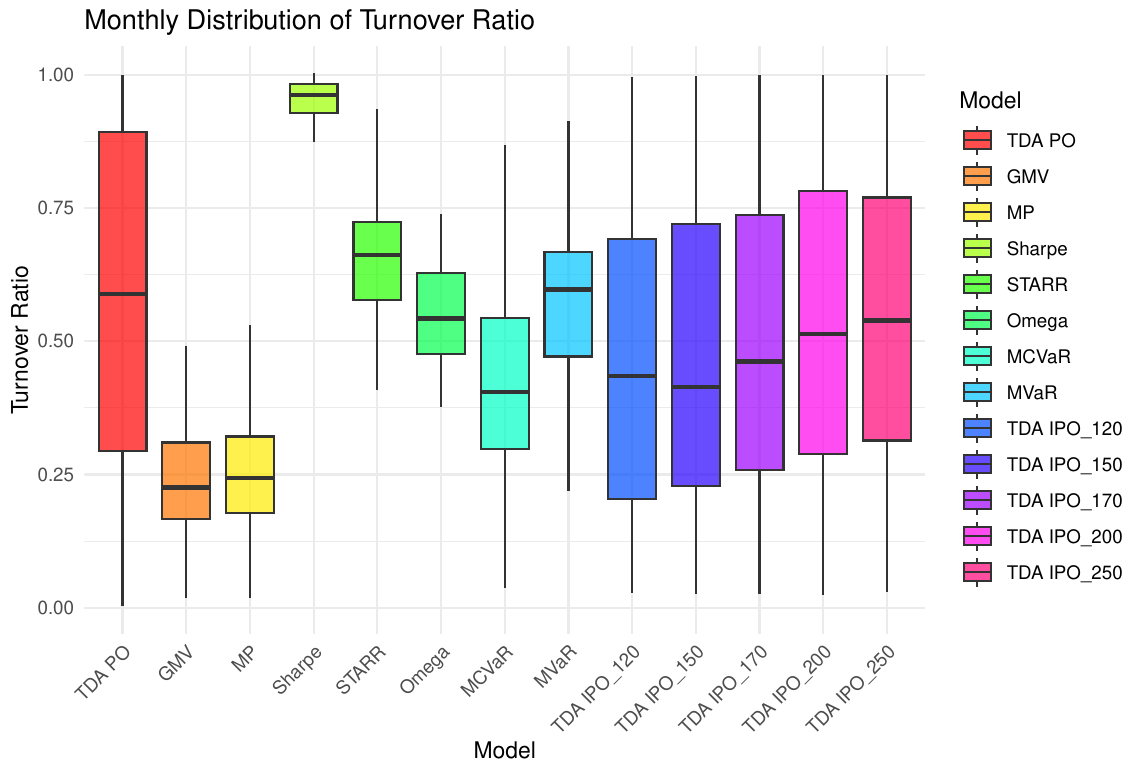}
    \caption{Monthly distribution of turnover ratio across thirteen portfolio optimization models.
%This figure presents the turnover ratio distributions for different portfolio optimization models. 
The proposed TDA-PO model and its constrained variations (TDA-IPO with different asset limits $k$ = 120, 150, 170, 200, 250) exhibit relatively lower turnover compared to benchmarked PO models such as Sharpe, STARR, MVaR and Omega (except for the GMV, and MP models while having comparable results with MCVaR model). The lower turnover suggests that TDA-based models generate stable allocations with fewer rebalancing costs while maintaining strong performance.}
    \label{fig:turnover_distribution}
\end{figure*}

Turnover is a key measure of portfolio stability and transaction costs, where lower turnover values indicate reduced trading activity and greater cost efficiency. Figure~\ref{fig:turnover_distribution} presents the turnover distribution across all considered PO models, highlighting the efficiency of the proposed TDA-PO and TDA-IPO  strategies. The length of each box-plot depicts the variation of turnover ratios obtained over 108 windows in the span of nearly 10 years from December 10, 2012, to August 11, 2022. 

A key observation from Figure~\ref{fig:turnover_distribution} is the stark contrast in turnover behavior across traditional PO models. The Sharpe PO model exhibits the \textit{highest} turnover among all the models, as it aggressively reallocates assets in response to fluctuations in mean returns or model parameters (mean and covariance terms) with the dominance of mean returns. Since mean returns are notoriously unstable and subject to estimation error, the Sharpe model reacts with frequent and often excessive rebalancing, leading to high transaction costs. In contrast, the GMV and MP portfolios maintain significantly \textit{lower} turnover, as their allocations are primarily driven by risk minimization, which is inherently more stable over time. The GMV portfolio, which optimizes for the lowest portfolio volatility, exhibits the most stable allocation adjustments, while the MP model, despite incorporating mean returns, still remains dominated by the risk term, preventing drastic portfolio changes.

The turnover distributions from GMV and MP, are entirely 
falling below to all other turnover distributions which indicating of having very stable portfolios over the span of 10 years, giving serious concern of not accomodating the new market information.

%This could be attributed to structural shifts in asset weights within the constrained universe or market conditions that necessitate frequent adjustments even with a predefined asset limit.
The TDA-PO model exhibits a lower median turnover ratio compared to the Sharpe, STARR, \& MVaR and comparable to Omega PO model, reflecting its ability to construct robust and stable allocations that do not require frequent rebalancing.
%This characteristic is particularly desirable for institutional investors and fund managers seeking to minimize transaction costs while maintaining strong out-of-sample performance.
For the TDA-IPO models, which introduce explicit constraints on the number of assets held, the expected trend of decreasing turnover with increasing $k$ (the asset constraint) is not clearly visible in the figure. Instead, turnover distributions for different TDA-IPO constraints remain relatively close to each other, suggesting that while restricting the number of assets may fix diversification, it does not necessarily lead to significantly lower turnover levels in the observed dataset. 

Nonetheless, the TDA-IPO models maintain lower turnover compared to the high-turnover strategies such as Sharpe, STARR, MVaR \& Omega portfolios (and having comparable results with MCVaR), while preserving strong risk-adjusted performance. The ability to generate stable yet adaptive allocations with lower reliance on expected return estimation gives TDA-PO and TDA-IPO a distinct advantage over traditional PO models. These results underline the financial viability of TDA-based models as practical alternatives to standard portfolio optimization approaches, achieving an optimal balance between return maximization, risk control, and cost efficiency.

\subsubsection{Transaction Cost Sensitivity Analysis}

We next investigate the sensitivity of portfolio performance to transaction costs, motivated by the variation in turnover observed across different allocation strategies. As shown in Figure~\ref{fig:turnover_distribution}, turnover levels differ significantly across models, with Sharpe and STARR portfolios exhibiting the highest rebalancing intensity, TDA-PO demonstrating a more stable turnover profile while the GMV and MP being the ones with the lowest turnover.

\begin{figure*}[ht!]
    \centering
    \subfigure[Transaction Cost = 0.1\%]{%
        \includegraphics[height=4.7cm,width=6cm]{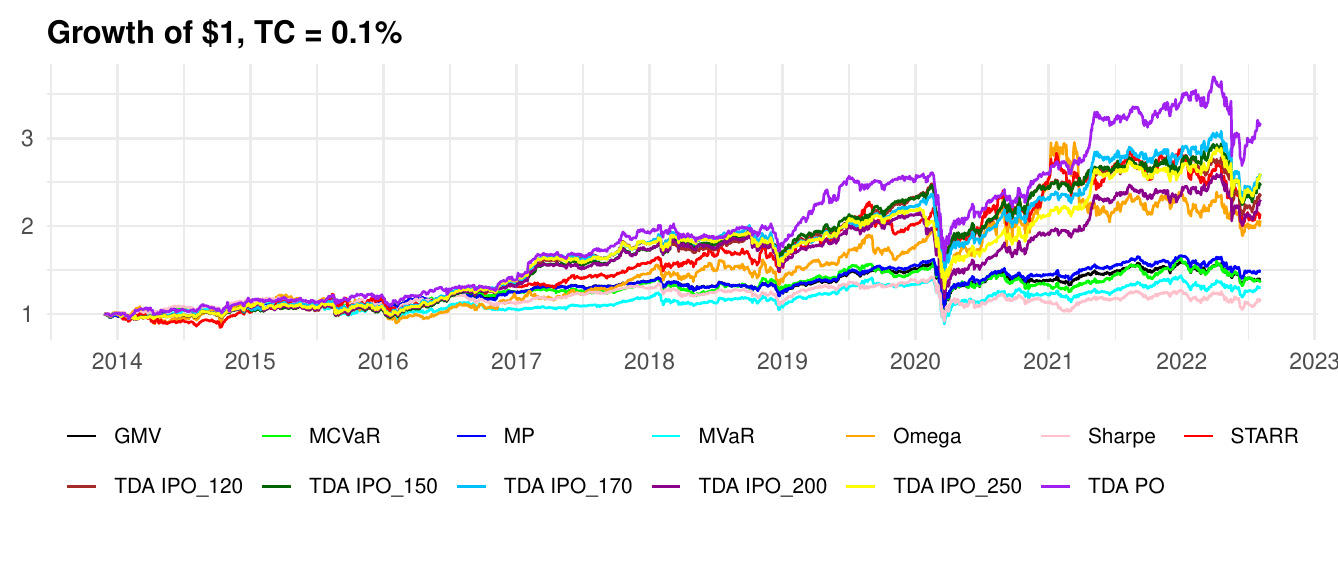}
        \label{fig:wealth_tc1}}
    \quad
    \subfigure[Transaction Cost = 0.2\%]{%
        \includegraphics[height=4.7cm,width=6cm]{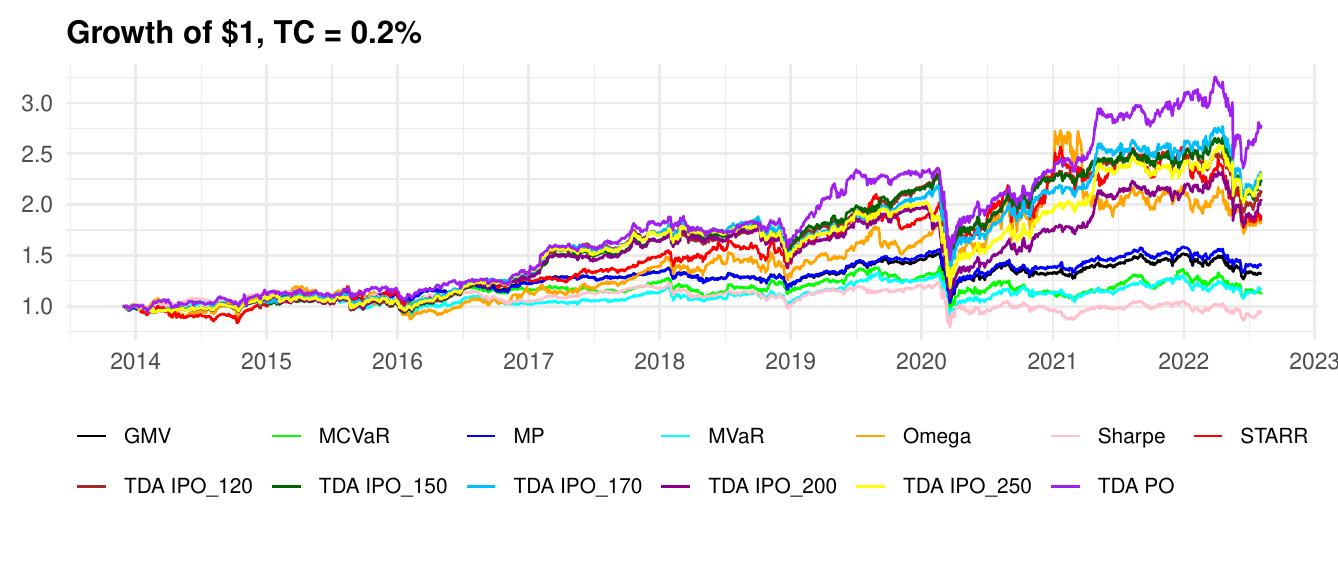}
        \label{fig:wealth_tc2}}
    \quad
    \subfigure[Transaction Cost = 0.3\%]{%
        \includegraphics[height=4.7cm,width=6cm]{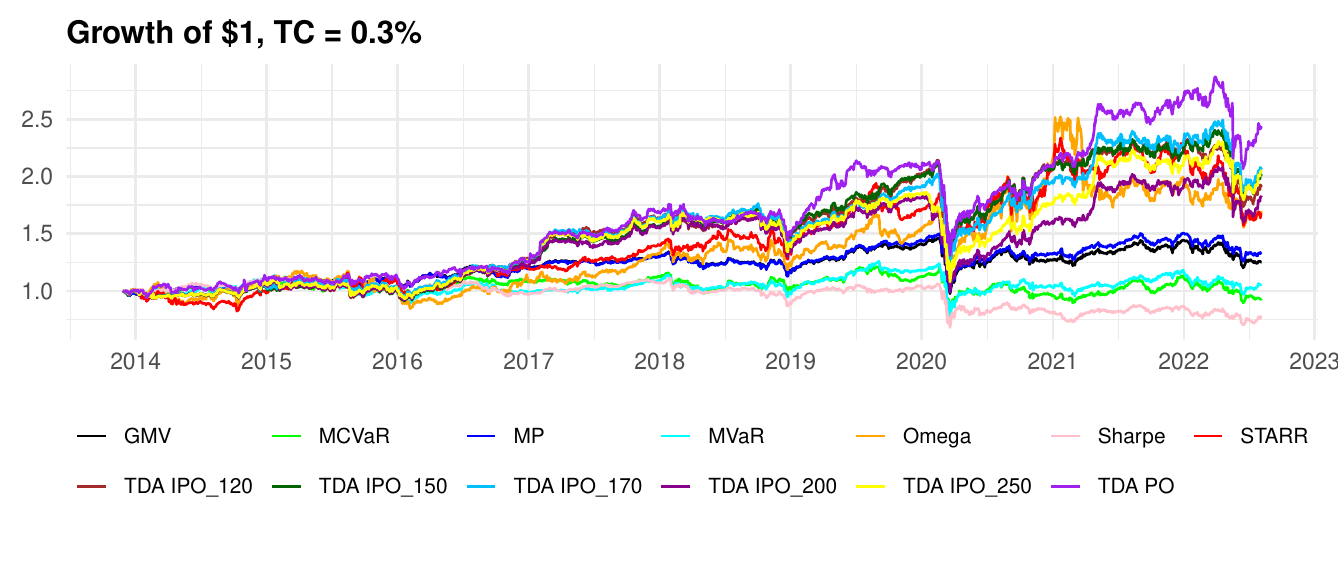}
        \label{fig:wealth_tc3}}
         \quad
    \subfigure[Transaction Cost = 0.4\%]{%
        \includegraphics[height=4.7cm,width=6cm]{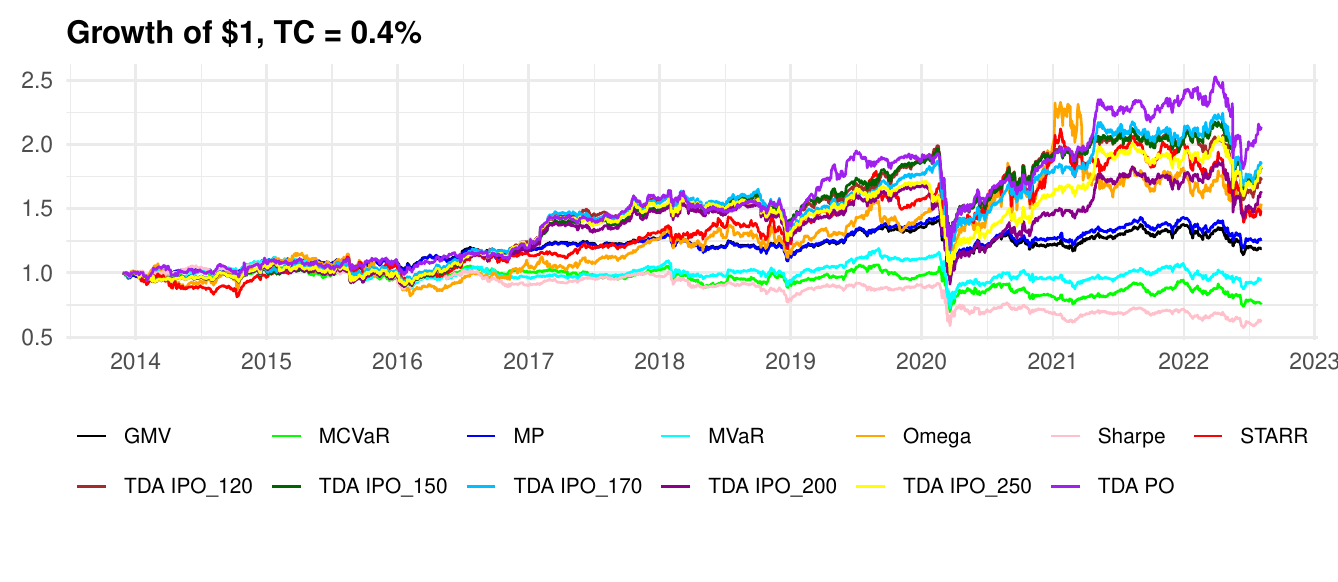}
        \label{fig:wealth_tc4}}
         \quad
    \subfigure[Transaction Cost = 0.5\%]{%
        \includegraphics[height=4.7cm,width=6cm]{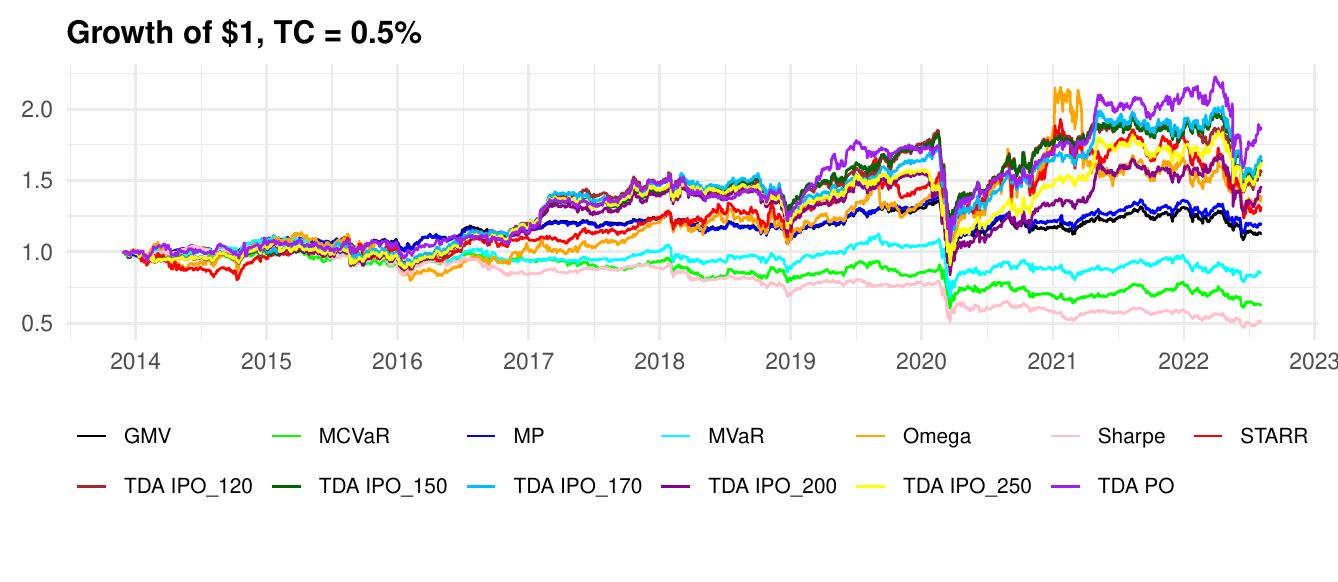}
        \label{fig:wealth_tc5}}
    \caption{%
        The wealth of the portfolio starting from \$1 for the TDA-PO strategy and the benchmark models. Returns have been adjusted for transaction costs at rates of 0.1\% to 0.5\%. Each plot shows the comparative performance of the index, standard models, and TDA-based portfolios over the investment horizon. The results demonstrate that while higher transaction costs reduce overall return levels, the TDA-based portfolio maintains consistent relative performance and robustness across all cost regimes.
    }
    \label{fig:wealth_tc_sensitivity}
\end{figure*}

To quantify the implications of this variation, we conduct a stress-test analysis using elevated transaction cost (TC) levels of 1\%, 2\%, 3\%, 4\%, and 5\%. These values simulate a range of trading frictions and allow us to assess the sensitivity of each strategy’s net returns to rebalancing activity and the transaction costs. The resulting cumulative wealth trajectories are reported in Figure~\ref{fig:wealth_tc_sensitivity}, where returns have been adjusted for proportional costs based on absolute portfolio weight changes at each step.

It is important to note that transaction costs are only incurred by strategies that involve active rebalancing. The naive and S\& P index portfolios, being static by construction, are unaffected and are thus excluded from these plots for clarity. %Their inclusion would result in flat trajectories that do not vary with transaction cost levels. 
The results reveal several key insights. First, while all actively rebalanced portfolios experience reduced wealth as TC increase, the magnitude of performance degradation is strongly aligned with portfolio turnover characteristics, as previously reported in Figure \ref{fig:turnover_distribution}. For instance, Sharpe portfolios, which exhibit the highest turnover, suffer the steepest decline in cumulative return as transaction costs rise.

In contrast, the TDA-PO strategy, which maintains a moderate and stable turnover profile, experiences a more gradual decline in performance. Even at high transaction cost levels, TDA-PO consistently outperforms all other optimized portfolios and maintains a clear edge over the static benchmark strategies. %This highlights its ability to balance structural adaptiveness with cost-aware execution.

Traditional models such as GMV and MP, which involve lower turnover, also exhibit more gradual performance degradation, though they remain consistently outperformed by TDA-PO across all transaction cost regimes. Notably, TDA-PO outperforms both (GMV and MP) even under the highest tested transcation cost of 5\%.

Importantly, the performance ranking across models is not uniformly preserved as transaction costs rise. Rather, high-turnover strategies are disproportionately penalized, causing models like Sharpe specially to fall below even the naive benchmark at higher TC levels. Overall, the transaction cost sensitivity analysis confirms that the advantages of the TDA-PO framework persist even under adverse implementation scenarios. 

Notably, it is further evident that the model TDA-PO, along with other TDA-IPO, outperform all other PO models by generating the best cumulative return (adjusted with turnover cost) pattern where the worst performance is generated by the Sharpe model followed by the performance of MCVaR and MVaR models. This remains valid for all the transaction cost values examined, confirming the consistent superior performance of the proposed scheme.

\subsubsection{Stability Analysis in terms of Out-of-Sample Versus In-Sample Findings}

\begin{figure*}[!h]
    \centering
    \includegraphics[width=1.1\textwidth]{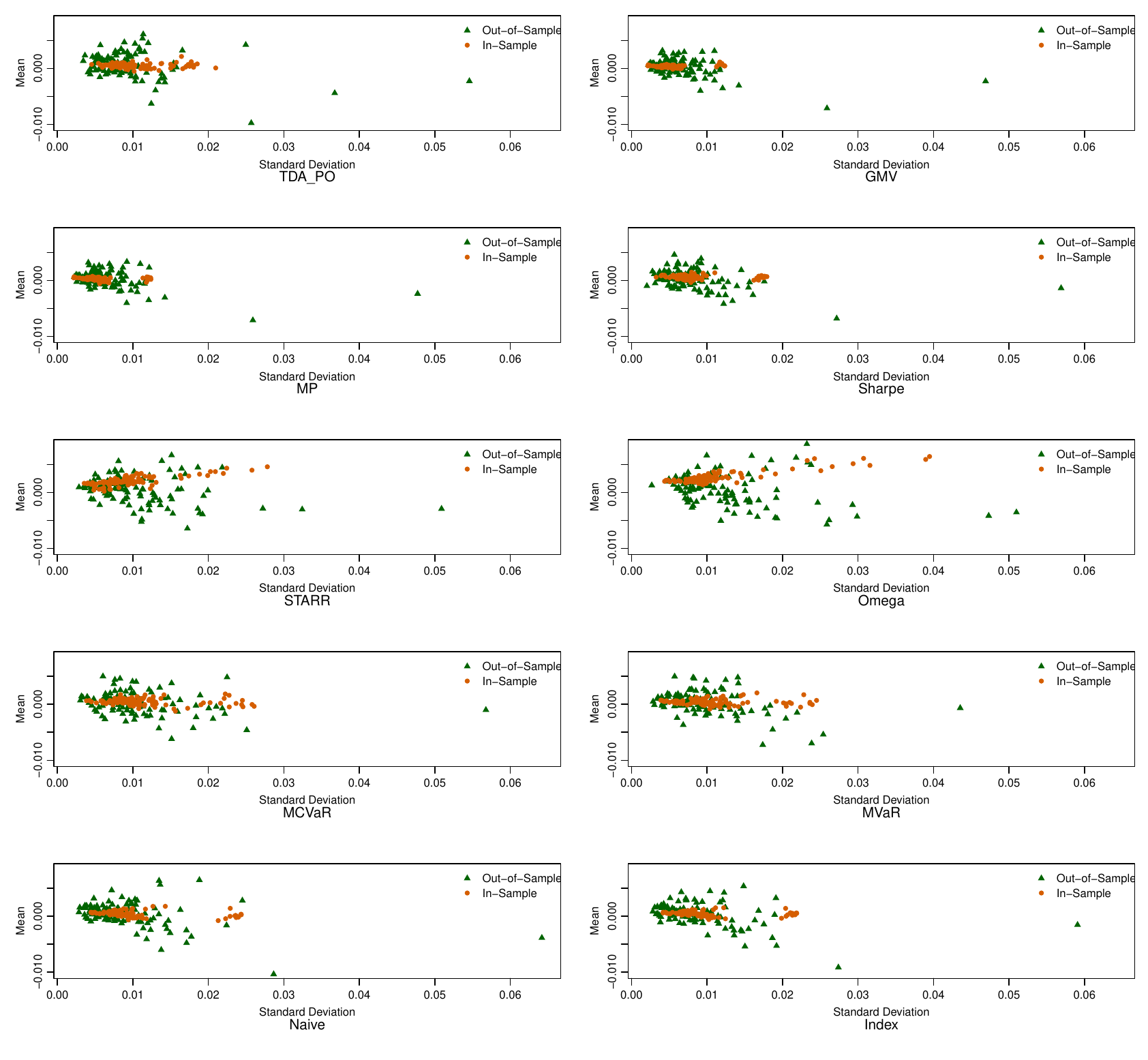}
    \caption{Scatter plots of In-sample and Out-of-Sample values of mean and standard deviation for all portfolios obtained from the TDA-PO model and the seven benchmark models.}
    \label{sharpe1}
\end{figure*}
To check the stability of results from the models over different windows, we compare their in-sample values of mean and standard deviation vis-a-vis their out-of-sample values (see Figure \ref{sharpe1}). The in-sample and out-of-sample coordinates for (mean, standard deviation) = ($\mu$, $\sigma$) are respectively, shown by the circular and triangle shapes. We notice from the Figure \ref{sharpe1} that the in-sample cluster of ($\mu$, $\sigma$) from the model TDA-PO  has comparatively closer connected points than any of the other models, with a similar pattern in its out-of-sample cluster, showing the robustness of the model with respect to the change of the data. 
Though the out-of-sample clusters of ($\mu$, $\sigma$) for variance-based PO models i.e. GMV, MP, and Sharpe, have closely connected points but not in their respective in-sample clusters. Lastly, the PO models STARR, Omega, MVaR and MVaR are found to be comparatively highly sensitive to the change of data, specifically for high values of $\sigma$.

So in short, we conclude the best performance of the proposed TDA-based optimization model than all the famous traditional PO models ranging from variance-based models to the reward-risk-based Omega model. This perhaps shows that capturing topological features of the assets presents a better feature selection than the traditional ways of portfolio allocation.   

\subsubsection{Statistical Evaluation of Risk Metrics}

Here, we aim to assess whether the proposed method TDA-PO and other models incur statistically higher downside risk or volatility when compared to the $1/N$ equal-weighted portfolio. For this purpose, we conduct formal hypothesis tests on three key dimensions of risk: VaR$_{0.95}$, CVaR$_{0.95}$, and variance ($\sigma^2$) (See Appendix A for details). 

The VaR$_{0.95}$ and CVaR$_{0.95}$ comparisons are implemented using large-sample $z$-tests based on empirical quantiles and tail averages Variance comparisons are performed using the standard $F$-test for equality of variances. Across all three tests, we evaluate whether given model's risk is statistically greater or worse than that of the $1/N$ naive strategy. The values worse than the naive are highlighted in red in all the relevant tables. 

The results indicate that, at the 95\% confidence level, \textbf{none} of the models—except for \textbf{Omega} and \textbf{STARR}—exhibit statistically higher risk than the $1/N$ naive benchmark. Both Omega and STARR portfolios are found to be significantly riskier than the naive allocation, not only at the 95\% level but also at the 90\% and  97\% statistic levels. These findings are consistent across the VaR$_{0.95}$, CVaR$_{0.95}$, and variance tests, reinforcing the conclusion that these two strategies carry systematically higher downside exposure. In contrast, the TDA-PO portfolio does not exhibit statistically greater risk than the naive benchmark under any of the risk measures considered. Its variants with cardinality constraints also show the same behavior. This suggests that, although the proposed strategy does not explicitly minimize classical risk metrics, it nevertheless maintains downside and volatility characteristics that are statistically comparable to the naive benchmark.

\section{Robust analysis}
%The robustness of portfolio performance depends not only on the underlying PO model but also on the length of the out-of-sample period.
In this section, to assess the robustness of the proposed TDA-PO model vis-a-vis benchmarked PO models and benchmarked portfolio strategies, we extend our analysis for varying window size in the rolling window scheme in the two directions: (1) Extension of out-of-sample (holding) period from 1 month to 3 months, 6 months, and 1 year (2) Extension of in-sample (training) period from 1 year to 2 years. This analysis provides further insights into the robustness of the proposed TDA-PO and TDA-IPO models, particularly in relation to risk-return dynamics, risk-adjusted efficiency, and stability of portfolio allocation with respect to the change in window size. 

\subsection{Impact of out-of-sample period size}
The results in Table \ref{tab:performance_metrics} provide insights into the effects of increasing the holding period for the cases of 3 months, 6 months, and 1 year on the performance metrics considered in the study.

\begin{table}[htbp]
  \centering
  \caption{Robustness check: Out-of-sample performance matrices obtained over rolling window scheme from the nine optimization models. The best values are highlighted in bold for the reader's convenience and second best are italicized. The $*$ is used to mark the significant values in the statistical tests for the Sharpe ratio at $90\%$ confidence level for the TDA-PO model vs others. Values highlighted in red represent performance metrics that are statistically inferior to those of the naive portfolio at the 95\% confidence level, with respect to the risk measures: variance, VaR$_{0.95}$, and CVaR$_{0.95}$. The assets are averages of total assets with non-zero weights across different windows.} 
  \label{tab:performance_metrics}
  \footnotesize
  \renewcommand{\arraystretch}{1}

\begin{tabular}{lrrrrrrrrr}
\toprule
Metric & TDA PO & TDA IPO & GMV & MP & Sharpe & STARR & Omega & MCVaR & MVaR \\
\midrule
\multicolumn{10}{c}{\textbf{In-sample period= 1 year, Out-of-sample period= 3 months}} \\
\midrule
EMR         & \textbf{5.90E-04}  & 5.09E-04  & 2.05E-04  & 2.19E-04  & 2.51E-04  & 5.04E-04  & \textit{5.14E-04} & 2.70E-04 & 2.97E-04 \\
Min         & -1.45E-01 & -1.05E-01 & \textbf{-8.94E-02} & \textit{-9.44E-02} & -1.65E-01 & -1.15E-01 & -1.33E-01 & -9.35E-02 & -1.04E-01 \\
Std Dev     & 1.15E-02  & 9.70E-03  & \textbf{8.42E-03}  & \textit{8.45E-03}  & 1.09E-02  & \textcolor{red}{1.25E-02}  & \textcolor{red}{1.46E-02} & 9.34E-03 & 9.74E-03 \\
DD          & 8.46E-03  & 7.15E-03  & \textbf{6.24E-03}  & \textit{6.27E-03}  & 8.06E-03  & 9.27E-03  & 1.07E-02 & 6.91E-03 & 7.21E-03 \\
$\text{VaR}_{0.95}$   & 1.61E-02  & 1.42E-02  & \textbf{1.14E-02}  & \textit{1.14E-02}  & 1.47E-02  & \textcolor{red}{1.80E-02}  & \textcolor{red}{2.25E-02} & 1.35E-02 & 1.41E-02 \\
$\text{CVaR}_{0.95}$  & 2.76E-02  & 2.41E-02  & \textit{2.04E-02}  & \textbf{2.03E-02}  & 2.57E-02  & \textcolor{red}{3.10E-02}  & \textcolor{red}{3.64E-02} & 2.26E-02 & 2.39E-02 \\
SR          & \textbf{5.25E-02}  & \textit{5.13E-02}  & 2.44E-02*  & 2.59E-02*  & 2.31E-02*  & 4.02E-02  & 3.53E-02* & 2.89E-02* & 3.05E-02* \\
$\text{SVR}_{0.95}$   & \textbf{3.65E-02}  & \textit{3.59E-02}  & 1.80E-02  & 1.92E-02  & 1.71E-02  & 2.80E-02  & 2.29E-02 & 2.00E-02 & 2.10E-02 \\
$\text{SCR}_{0.95}$   & \textbf{2.14E-02}  & \textit{2.11E-02}  & 1.01E-02  & 1.08E-02  & 9.78E-03  & 1.63E-02  & 1.41E-02 & 1.20E-02 & 1.24E-02 \\
Sortino     & \textbf{6.97E-02}  & \textit{5.72E-02}  & 3.28E-02  & 3.49E-02  & 3.12E-02  & 5.44E-02  & 4.80E-02 & 3.91E-02 & 4.12E-02 \\
$\text{Rachev}_{0.95}$ & \textbf{1.65E-03}  & \textit{1.49E-03}  & 3.61E-04  & 3.62E-04  & 5.76E-04  & 5.39E-04  & 1.16E-03 & 4.43E-04 & 4.89E-04 \\
PTR         & 3.75E-04 & 2.91E-04 & 2.46E-04 & 3.32E-04 & 1.55E-03 & 3.03E-04 & 4.14E-04 & 3.90E-04 & 4.00E-04 \\
\midrule
Assets      & 173 & 50 & 34 & 34 & 55 & 14 & 15 & 36 & 35 \\
Turnover    & 1.72 & 1.46 & \textbf{0.89} & \textit{0.91} & 1.86 & 1.82 & 1.75 & 1.26 & 1.43 \\
\midrule
\multicolumn{10}{c}{\textbf{In-sample period= 1 year, Out-of-sample period= 6 months}} \\
\midrule
EMR         & \textbf{5.65E-04}  & 5.18E-04  & 1.85E-04  & 1.84E-04  & 1.01E-04  & 4.86E-04  & \textit{5.39E-04} & 2.21E-04 & 2.89E-04 \\
Min         & -1.11E-01 & -1.05E-01 & \textbf{-8.94E-02} & \textit{-9.44E-02} & -1.37E-01 & -9.99E-02 & -1.33E-01 & -9.35E-02 & -1.04E-01 \\
Std Dev     & 1.13E-02  & 9.93E-03  & \textbf{8.52E-03}  & \textit{8.65E-03}  & 1.13E-02  & \textcolor{red}{1.18E-02}  & \textcolor{red}{1.39E-02} & 9.29E-03 & 9.80E-03 \\
DD          & 8.22E-03  & 7.39E-03  & \textbf{6.26E-03}  & \textit{6.41E-03}  & 8.40E-03  & 8.66E-03  & 1.02E-02 & 6.89E-03 & 7.26E-03 \\
$\text{VaR}_{0.95}$   & 1.70E-02  & 1.44E-02  & \textit{1.18E-02}  & \textbf{1.16E-02}  & 1.54E-02  & 1.69E-02  & \textcolor{red}{2.04E-02} & 1.33E-02 & 1.42E-02 \\
$\text{CVaR}_{0.95}$  & 2.79E-02  & 2.49E-02  & \textbf{2.05E-02}  & \textit{2.08E-02}  & 2.77E-02  & \textcolor{red}{2.93E-02}  & \textcolor{red}{3.45E-02} & 2.27E-02 & 2.40E-02 \\
SR          & \textbf{5.22E-02}  & \textit{4.99E-02}  & 2.17E-02*  & 2.12E-02*  & 8.96E-03*  & 4.13E-02  & 3.87E-02* & 2.38E-02* & 2.95E-02* \\
$\text{SVR}_{0.95}$   & \textbf{3.59E-02}  & \textit{3.32E-02}  & 1.57E-02  & 1.58E-02  & 6.55E-03  & 2.87E-02  & 2.64E-02 & 1.66E-02 & 2.04E-02 \\
$\text{SCR}_{0.95}$   & \textbf{2.08E-02}  & \textit{2.02E-02}  & 9.01E-03  & 8.83E-03  & 3.65E-03  & 1.66E-02  & 1.56E-02 & 9.75E-03 & 1.20E-02 \\
Sortino     & \textbf{5.76E-02}  & \textit{5.66E-02}  & 2.95E-02  & 2.86E-02  & 1.20E-02  & 5.61E-02  & 5.27E-02 & 3.21E-02 & 3.98E-02 \\
$\text{Rachev}_{0.95}$ & \textbf{1.97E-03}  & \textit{1.52E-03}  & 3.72E-04  & 3.82E-04  & 6.53E-04  & 7.47E-04  & 1.05E-03 & 4.39E-04 & 4.84E-04 \\
PTR         & 3.21E-04 & 2.66E-04 & 2.63E-04 & 3.23E-04 & 1.16E-03 & 3.11E-04 & 4.20E-04 & 2.72E-04 & 2.94E-04 \\
\midrule
Assets      & 191 & 50 & 32 & 32 & 57 & 15 & 15 & 45 & 35 \\
Turnover    & 1.51 & 1.54 & \textbf{1.23 } & \textit{1.26 } & 1.87 & 1.82 & 1.73 & 1.59 & 1.56 \\
\midrule
\multicolumn{10}{c}{\textbf{In-sample period = 1 year, Out-of-sample period= 1 year}} \\
\midrule
EMR         & \textbf{4.96E-04}  & 4.24E-04  & 1.39E-04  & 1.15E-04  & 1.63E-04  & \textit{4.79E-04}  & 4.72E-04 & 1.94E-04 & 2.15E-04 \\
Min         & -1.39E-01 & -1.45E-01 & \textit{-1.15E-01} & \textbf{-1.07E-01} & -1.39E-01 & -1.27E-01 & -1.33E-01 & -1.23E-01 & -1.30E-01 \\
Std Dev     & 1.19E-02  & 1.12E-02  & \textit{9.47E-03}  & \textbf{9.43E-03}  & 1.12E-02  & \textcolor{red}{1.30E-02}  & \textcolor{red}{1.48E-02} & 9.93E-03 & 1.06E-02 \\
DD          & 8.81E-03  & 8.34E-03  & \textbf{6.99E-03}  & \textit{7.00E-03}  & 8.40E-03  & 9.52E-03  & 1.07E-02 & 7.45E-03 & 7.84E-03 \\
$\text{VaR}_{0.95}$   & 1.80E-02  & 1.51E-02  & \textbf{1.27E-02}  & \textit{1.30E-02}  & 1.59E-02  & \textcolor{red}{1.85E-02}  & \textcolor{red}{2.19E-02} & 1.35E-02 & 1.51E-02 \\
$\text{CVaR}_{0.95}$  & 2.92E-02  & 2.77E-02  & \textbf{2.26E-02}  & \textit{2.26E-02}  & 2.80E-02  & \textcolor{red}{3.20E-02}  & \textcolor{red}{3.59E-02} & 2.41E-02 & 2.59E-02 \\
SR          & \textbf{4.15E-02}  & \textit{3.79E-02}  & 1.46E-02*  & 1.22E-02*  & 1.45E-02*  & 3.68E-02  & 3.18E-02* & 1.95E-02* & 2.03E-02* \\
$\text{SVR}_{0.95}$   & \textbf{2.80E-02}  & \textit{2.75E-02}  & 1.09E-02  & 8.82E-03  & 1.03E-02  & 2.60E-02  & 2.15E-02 & 1.43E-02 & 1.43E-02 \\
$\text{SCR}_{0.95}$   & \textbf{1.70E-02}  & \textit{1.53E-02}  & 6.15E-03  & 5.07E-03  & 5.83E-03  & 1.50E-02  & 1.31E-02 & 8.03E-03 & 8.31E-03 \\
Sortino     & \textbf{5.88E-02}  & \textit{5.36E-02}  & 1.98E-02  & 1.64E-02  & 1.94E-02  & 5.03E-02  & 4.40E-02 & 2.60E-02 & 2.74E-02 \\
$\text{Rachev}_{0.95}$ & \textbf{1.35E-03}  & \textit{1.25E-03}  & 4.47E-04  & 4.51E-04  & 6.59E-04  & 9.08E-04  & 1.17E-03 & 4.97E-04 & 5.74E-04 \\
PTR         & 3.68E-04 & 3.64E-04 & 3.77E-04 & 5.58E-04 & 1.61E-03 & 9.11E-04 & 9.82E-04 & 3.96E-04 & 4.56E-04 \\
\midrule
Assets      & 204 & 50 & 32 & 32 & 60 & 14 & 15 & 38 & 36 \\
Turnover    & 1.63 & \textbf{1.43} & \textit{1.66} & 1.65 & 1.93 & 1.95 & 1.97 & 1.87 & 1.87 \\
\bottomrule
\end{tabular}
\end{table}

\begin{itemize}
\item \textbf{Return performance in terms of EMR:} The results from the Table \ref{tab:performance_metrics} demonstrate that TDA-PO consistently achieves the highest EMR across all the out-of-sample periods, reinforcing its ability to generate persistent return outperformance relative to benchmarked PO models (GMV, MP, MCVaR, MVaR, Sharpe, Omega and STARR) and benchmarked portfolio strategies ($1/N$, and S\&P 500 market index). Interestingly, despite being constructed without directly optimizing mean return, TDA-PO consistently outperforms mean-return-based PO models, the result is consistent with the case of 1 month out-of-sample period (see Table \ref{tab:compare}). 

Similarly, TDA-IPO, which imposes an explicit constraint on the number of assets held ($k$ = 50)\footnote{In our previous analysis in the Section 5, where we considered asset constraints of $k=120, 150, 170, 200, 250$, we observed that the average number of assets with non-zero weights across rolling windows was approximately 50 for all cases of $k$. Additionally, the benchmark PO models, such as GMV, MP, MVaR, MCVaR, STARR, and Omega, also exhibited an average asset count close to or below this threshold. To ensure a fair and consistent comparison, we set $k=50$, aligning our model with the empirical characteristics of the competing strategies}, also delivers strong return performance. While the EMR for TDA-IPO is slightly lower than TDA-PO and Omega models due to diversification constraints, it remains competitive and significantly outperforms traditional mean-variance based models. %The constrained asset selection mechanism does not substantially hinder its ability to capture time-varying return opportunities, confirming that TDA-based models remain effective even under practical investment constraints.

However, as the out-of-sample period extends from 3 months to 1 year, we observe a gradual decline in EMR across all considered models. This phenomenon is expected in longer investment horizons, where market volatility, structural shifts, and macroeconomic factors erode short-term return advantages. While TDA-PO maintains its return dominance, the performance gap between TDA-PO and variance-based models slightly narrows over longer periods.  

%A key takeaway from these results is that TDA-PO’s structural advantage does not stem from chasing returns but rather from efficiently adapting to market dynamics. %B The same structural advantage is observed in TDA-IPO, where its adaptive allocation framework within a constrained asset universe ensures robust return generation without excessive reliance on historical expectations.

\item \textbf{Risk Analysis in terms of stdev, DD, VaR$_{0.95}$, and CVaR$_{0.95}$:}
Similar to the results reported in the Table \ref{tab:compare} for the case of out-of-sample of 1 month, risk levels at TDA-PO are only slightly higher than GMV, MP, MVaR, MCVaR and Sharpe while they remain significantly lower than those of STARR, Omega, and the naive $1/N$ portfolio. %This indicates that TDA-PO does not aggressively sacrifice risk control for return maximization - a key advantage in real-world applications, where excessive risk exposure can lead to substantial drawdowns during adverse market conditions. 
TDA-IPO exhibits a similar risk profile, but with slightly reduced risk compared to TDA-PO. The risk analysis further reveals that as the holding period increases, overall portfolio risk also increases for most of the models, reflecting natural amplification of the market volatility over extended horizons. This is observed across all risk (i.e. stdev, DD, VaR$_{0.95}$, and CVaR$_{0.95}$) metrics except PTR. Despite this upward trend in risk exposure, TDA-PO maintains a favorable risk profile. 
%\sout{ due to its asset constraints enforcing a more diversified structure. By capping the number of holdings, TDA-IPO avoids excessive concentration risk while maintaining the adaptability of TDA-PO.}
 %This balance ensures that investors seeking a more controlled exposure with predefined diversification limits can still achieve lower downside risk than Omega, STARR, and naive diversification strategies. 

A particularly notable observation is that VaR$_{0.95}$ and CVaR$_{0.95}$ increase at a slower rate for TDA-PO and TDA-IPO compared to STARR and Omega as the holding period extends. This suggests that both TDA-based models exhibit better downside risk containment over longer investment horizons, an essential characteristic for investors focused on capital preservation. We further conducted statistical tests and confirmed that the TDA-PO strategy does not exhibit significantly higher VaR, CVaR, or standard deviation compared to the naive benchmark at the 95\% confidence level. In contrast, the Omega and STARR models were found to be statistically worse than the naive portfolio across most cases.

In terms of topological risk i.e. PTR, the Sharpe model suffers with its highest values when second highest values given by the model Omega whereas TDA-IPO and GMV models (having similar values) delivered its lowest values. This is true for all the cases of out-of-sample period size. The values of PTR increased highly from the out-of-sample of 6 months to 1 year for the models MP, MCVaR, MVaR, STARR and Omega whereas it remains stable for the models, TDA-PO, TDA-IPO, GMV and Sharpe.

\item \textbf{Risk adjusted return performance:}  
One of the most striking results from Table \ref{tab:performance_metrics} is that TDA-PO consistently ranks first in all risk-adjusted return metrics when the second rank achieved by the TDA-IPO. The superior Sharpe, Sortino, Rachev, Sharpe-CVaR (SCR$_{0.95}$), and Sharpe-VaR (SVR$_{0.95}$) ratios of TDA-PO reaffirm its efficiency in delivering higher consistent returns without substantially increasing risk exposure. Similarly, TDA-IPO exhibits strong risk-adjusted performance despite the imposed constraints on asset selection. While it does not always surpass TDA-PO in risk-adjusted metrics, it remains superior to variance-based models and naive diversification approaches. The constrained selection of assets in TDA-IPO leads to a more stable return-risk tradeoff, reducing excess volatility while still achieving strong Sharpe and Sortino ratios.

%\sout{except for the Rachev ratio, where STARR occasionally outperforms it}
%Interestingly, while STARR occasionally leads in the Rachev ratio, this outperformance is achieved at the cost of significantly higher tail risk exposure. The performance of TDA-IPO remains highly competitive in this regard, demonstrating that even with asset constraints, it maintains a strong risk-adjusted return profile across multiple holding periods.

Further, as the holding period increases, risk-adjusted ratios decline across all models, reflecting the diminishing return-to-risk efficiency over extended horizons. %However, TDA-PO’s Sharpe and Sortino ratios remain 
\item \textbf{Turnover Effects:}
Turnover analysis reveals a clear inverse relationship between turnover and holding period. As expected, turnover rates decline as the holding period increases, reflecting lower trading frequency in longer investment windows. This aligns with the fundamental principle that shorter rebalancing periods require more frequent portfolio adjustments, whereas longer holding periods demand maintaining positions despite market fluctuations.

From a cost-efficiency standpoint, TDA-PO and TDA-IPO exhibit lower turnover than Sharpe, STARR, Omega, MVaR, and MCVaR (except to MVaR \& MCVaR over 3 months out-of-sample period) implying reduced transaction costs over time. However, turnover remains higher than GMV and MP, which are inherently designed for low trading activity. This suggests that TDA-based models strike a balance between flexibility and stability, ensuring responsiveness to market dynamics without excessive rebalancing costs. 

\item With the increased size of out-of-sample period to 3 months, 6 months and 1 year, the performance of all the PO models including the proposed TDA-PO and TDA-IPO in comparison to the benchmarked strategies, $1/N$ and S\&P 500, remain consistent with the case of out-of-sample period size of 1 month (see Table \ref{tab:compare}). That is, except the models, GMV, MP, MVaR, MCVaR, and Sharpe, all other PO models i.e. STARR, Omega, TDA- PO and TDA-IPO generate better outcomes in terms of EMR and all the financial ratios than the $1/N$ naive and S\&P 500 market index portfolios.  This confirms that systematic optimization provides tangible advantages over both of the benchmarked allocation strategies.  

\end{itemize}

\subsection{Impact of in-sample period size}
In this section, we examine the impact of training window length on portfolio performance by fixing its size to 2 years while varying the out-of-sample period to 1 month, 3 months, 6 months, and 1 year. The results, summarized in Table \ref{tab:performance_metrics2}, demonstrate key structural advantages of TDA-based models over conventional approaches.

\begin{table}[htbp]
\footnotesize
\setlength{\tabcolsep}{1.5pt}
  \centering
  \caption{Robustness check: Out-of-sample performance matrices obtained over rolling window scheme from the nine optimization models and two benchmark portfolio strategies. The best values are highlighted in bold for the reader's convenience and second best are italicized. The $*$ is used to mark the significant values in the statistical tests for the Sharpe ratio at $90\%$ confidence level for the TDA-PO model versus others. Values highlighted in red represent performance metrics that are statistically inferior to those of the naive portfolio at the 95\% confidence level, with respect to variance, VaR$_{0.95}$, and CVaR$_{0.95}$. The assets are averages of total assets with non-zero weights across different windows.}
  \label{tab:performance_metrics2}

\begin{tabular}{lrrrrrrrrrrrr}
\toprule
Metric & TDA PO & TDA IPO & GMV & MP & Sharpe & STARR & Omega &MCVaR & MVaR & Naive & Index \\
\midrule
\multicolumn{12}{c}{\textbf{In-sample period= 2 years, Out-of-Sample period= 1 month}} \\
\midrule
EMR         & 5.02E-04  & \textbf{8.22E-04}  & 2.23E-04  & 2.30E-04  & 1.65E-04  & 4.45E-04  & \textit{5.33E-04}  & 1.92E-04 & 1.29E-04 & 3.33E-04  & 3.51E-04 \\
Min         & -1.10E-01 & -1.37E-01 & \textit{-9.58E-02} & {-9.49E-02} & -1.45E-01 & -1.17E-01 & -1.16E-01 & \textbf{-7.60E-02} & -1.20E-01 & -1.39E-01 & -1.28E-01 \\
Std Dev     & 1.20E-02  & 1.39E-02  & \textbf{8.76E-03}  & {8.81E-03}  & 1.10E-02  & 1.29E-02  & \textcolor{red}{1.57E-02}  & \textit{8.79E-0}3 & 1.00E-02 & 1.20E-02  & 1.15E-02 \\
DD          & 9.07E-03  & 1.59E-02  & \textit{6.47E-03}  & {6.49E-03}  & 8.36E-03  & 9.54E-03  & 1.15E-02  & \textbf{6.37E-03} & 7.49E-03 & 8.92E-03  & 8.46E-03 \\
$\text{VaR}_{0.95}$   & 1.80E-02  & 3.02E-02  & \textbf{1.16E-02}  & \textit{1.19E-02}  & 1.50E-02  & 1.97E-02  & 2.60E-02  & 1.28E-02 & 1.35E-02 & 1.70E-02  & 1.74E-02 \\
$\text{CVaR}_{0.95}$  & 3.12E-02  & 3.55E-02  & \textbf{2.08E-02}  & \textit{2.10E-02}  & 2.76E-02  & 3.21E-02  & \textcolor{red}{3.95E-02}  & 2.13E-02 & 2.46E-02 & 3.01E-02  & 2.92E-02 \\
SR          & \textit{4.18E-02}  & \textbf{5.90E-02}  & 2.54E-02*  & 2.61E-02*  & 1.49E-02*  & 3.44E-02  & 3.39E-02  & 2.19E-02* & 1.29E-02* & 2.78E-02*  & 3.06E-02* \\
$\text{SVR}_{0.95}$   & \textbf{2.79E-02}  & \textit{2.72E-02}  & 1.92E-02  & 1.93E-02  & 1.10E-02  & 2.25E-02  & 2.05E-02  & 1.51E-02 & 9.54E-03 & 1.96E-02  & 2.01E-02 \\
$\text{SCR}_{0.95}$   & \textit{1.61E-02}  & \textbf{2.31E-02}  & 1.07E-02  & 1.10E-02  & 5.96E-03  & 1.39E-02  & 1.35E-02  & 9.03E-03 & 5.26E-03 & 1.11E-02  & 1.20E-02 \\
Sortino     & \textbf{4.73E-02}  & \textbf{5.18E-02}  & 3.44E-02  & 3.54E-02  & 1.97E-02  & 4.67E-02  & 4.65E-02  & 3.02E-02 & 1.73E-02 & 3.74E-02  & 4.15E-02 \\
$\text{Rachev}_{0.95}$ & \textit{2.79E-03}  & \textbf{3.41E-03}  & 3.84E-04  & 3.90E-04  & 6.16E-04  & 8.84E-04  & 1.34E-03  & 4.08E-04 & 5.05E-04 & 7.68E-04  & 7.10E-04 \\
PTR         & 4.19E-04 & 4.61E-04 & 3.83E-04 & 3.91E-04 & 1.06E-03 & 8.65E-04 & 7.01E-04 & 4.05E-04 & 5.64E-04 & 9.58E-04 & 6.17E-04 \\
\midrule
Assets      & 344 & 20 & 34 & 35 & 60 & 14 & 13 & 45 & 42 & 462 & 1 \\
Turnover    & 0.58 & 0.75 & \textbf{0.28} & \textit{0.30} & 1.87 & 0.96 & 0.68 & 0.57 & 1.01 & 0.00 & 0.00 \\
\midrule
\multicolumn{12}{c}{\textbf{In-sample period= 2 years, Out-of-Sample period= 3 months}} \\
\midrule
EMR         & \textit{6.82E-04}  & \textbf{8.11E-04}  & 2.25E-04  & 2.23E-04  & 3.69E-04  & 3.44E-04  & 5.95E-04  & 1.88E-04 & 1.33E-04 & 3.33E-04  & 3.51E-04 \\
Min         & -1.10E-01 & -1.48E-01 & \textit{-9.58E-02} & {-9.49E-02} & -1.22E-01 & -1.17E-01 & -1.16E-01 & \textbf{-7.66E-02} & -1.17E-01 & -1.39E-01 & -1.28E-01 \\
Std Dev     & 1.25E-02  & 1.51E-02  & \textbf{8.92E-03}  & {8.94E-03}  & 1.11E-02  & 1.27E-02  & \textcolor{red}{1.61E-02}  & \textit{8.93E-03} & 9.92E-03 & 1.20E-02  & 1.15E-02 \\
DD          & 9.05E-03  & 1.17E-02  & \textit{6.59E-03}  & {6.60E-03}  & 8.13E-03  & 9.31E-03  & 1.16E-02  & \textbf{6.53E-03} & 7.39E-03 & 8.92E-03  & 8.46E-03 \\
$\text{VaR}_{0.95}$   & 1.83E-02  & 2.63E-02  & \textbf{1.18E-02}  & \textit{1.19E-02}  & 1.52E-02  & 2.00E-02  & 2.67E-02  & 1.31E-02 & 1.43E-02 & 1.70E-02  & 1.74E-02 \\
$\text{CVaR}_{0.95}$  & 3.14E-02  & 4.03E-02  & \textbf{2.15E-02}  & \textit{2.15E-02}  & 2.72E-02  & 3.08E-02  & 3.99E-02  & 2.17E-02 & 2.43E-02 & 3.01E-02  & 2.92E-02 \\
SR          & \textbf{5.45E-02}  & \textit{5.37E-02}  & 2.52E-02*  & 2.49E-02*  & 3.33E-02*  & 2.70E-02*  & 3.70E-02*  & 2.11E-02* & 1.34E-02* & 2.78E-02*  & 3.06E-02* \\
$\text{SVR}_{0.95}$   & \textbf{3.73E-02}  & \textit{3.09E-02}  & 1.91E-02  & 1.87E-02  & 2.43E-02  & 1.72E-02  & 2.23E-02  & 1.43E-02 & 9.29E-03 & 1.96E-02  & 2.01E-02 \\
$\text{SCR}_{0.95}$   & \textbf{2.17E-02}  & \textit{2.01E-02}  & 1.05E-02  & 1.03E-02  & 1.36E-02  & 1.12E-02  & 1.49E-02  & 8.68E-03 & 5.47E-03 & 1.11E-02  & 1.20E-02 \\
Sortino     & \textbf{5.22E-02}  & \textit{4.76E-02}  & 3.41E-02  & 3.37E-02  & 4.54E-02  & 3.70E-02  & 5.11E-02  & 2.88E-02 & 1.80E-02 & 3.74E-02  & 4.15E-02 \\
$\text{Rachev}_{0.95}$ & 1.88E-03  & \textbf{3.85E-03}  & 4.04E-04  & 4.09E-04  & 6.40E-04  & 8.48E-04  & \textit{1.40E-03}  & 4.19E-04 & 4.97E-04 & 7.68E-04  & 7.10E-04 \\
PTR         & 4.17E-04 & 7.70E-04 & 4.83E-04 & 4.01E-04 & 8.97E-04 & 8.71E-04 & 6.98E-04 & 4.26E-04 & 5.65E-04 & 9.58E-04 & 6.17E-04 \\
\midrule
Assets      & 342 & 20 & 34 & 35 & 60 & 14 & 14 & 458 & 42 & 462 & 1 \\
Turnover    & 1.24 & 1.16 & \textbf{0.53} & \textit{0.56} & 1.90 & 1.33 & 1.08 & 0.86 & 1.25 & 0.00 & 0.00 \\
\midrule
\multicolumn{12}{c}{\textbf{In-sample period= 2 years, Out-of-Sample period= 6 months}} \\
\midrule
EMR         & \textit{6.90E-04} & \textbf{8.75E-04}  & 2.00E-04  & 1.81E-04  & 7.47E-05  & 3.82E-04  & 6.56E-04  & 2.02E-04 & 1.88E-04 & 3.33E-04  & 3.51E-04 \\
Min         & -1.10E-01 & -1.48E-01 & \textit{-9.58E-02} & {-9.49E-02} & -8.78E-02 & -1.17E-01 & -1.16E-01 & \textbf{-7.60E-02} & -1.10E-01 & -1.39E-01 & -1.28E-01 \\
Std Dev     & 1.30E-02  & 1.64E-02  & \textit{8.98E-03}  & {9.03E-03}  & 1.08E-02  & 1.28E-02  & \textcolor{red}{1.57E-02}  & \textbf{8.95E-03} & 9.94E-03 & 1.20E-02  & 1.15E-02 \\
DD          & 9.23E-03  & 1.51E-02  & \textit{6.64E-03}  & {6.66E-03}  & 8.03E-03  & 9.32E-03  & 1.12E-02  & \textbf{6.49E-03} & 7.39E-03 & 8.92E-03  & 8.46E-03 \\
$\text{VaR}_{0.95}$   & 1.91E-02  & 3.10E-02  & \textbf{1.19E-02}  & \textit{1.19E-02}  & 1.61E-02  & 2.00E-02  & \textcolor{red}{2.57E-02}  & 1.30E-02 & 1.42E-02 & 1.70E-02  & 1.74E-02 \\
$\text{CVaR}_{0.95}$  & 3.21E-02  & 3.18E-02  & \textbf{2.16E-02}  & \textit{2.16E-02}  & 2.76E-02  & 3.12E-02  & \textcolor{red}{3.82E-02}  & \textit{2.16E-02} & 2.44E-02 & 3.01E-02  & 2.92E-02 \\
SR          & \textbf{5.30E-02}  & \textit{5.33E-02}  & 2.23E-02*  & 2.01E-02*  & 6.92E-03*  & 2.98E-02*  & 4.17E-02  & 2.26E-02* & 1.89E-02* & 2.78E-02*  & 3.06E-02* \\
$\text{SVR}_{0.95}$   & \textbf{3.61E-02}  & \textit{2.82E-02}  & 1.69E-02  & 1.52E-02  & 4.62E-03  & 1.91E-02  & 2.56E-02  & 1.55E-02 & 1.33E-02 & 1.96E-02  & 2.01E-02 \\
$\text{SCR}_{0.95}$   & \textbf{2.15E-02}  & \textit{2.75E-02}  & 9.29E-03  & 8.37E-03  & 2.71E-03  & 1.22E-02  & 1.72E-02  & 9.36E-03 & 7.70E-03 & 1.11E-02  & 1.20E-02 \\
Sortino     & \textbf{5.93E-02}  & \textit{5.94E-02}  & 3.02E-02  & 2.72E-02  & 9.30E-03  & 4.09E-02  & 5.83E-02  & 3.11E-02 & 2.55E-02 & 3.74E-02  & 4.15E-02 \\
$\text{Rachev}_{0.95}$ & 1.99E-03  & \textbf{4.32E-03}  & 4.07E-04  & 4.13E-04  & 6.16E-04  & 8.74E-04  & \textit{1.35E-03}  & 4.19E-04 & 5.09E-04 & 7.68E-04  & 7.10E-04 \\
PTR         & 4.12E-04 & 4.88E-04 & 4.82E-04 & 4.01E-04 & 3.21E-04 & 8.70E-04 & 7.04E-04 & 4.02E-04 & 5.48E-04 & 9.58E-04 & 6.17E-04 \\
\midrule
Assets      & 339 & 16 & 35 & 35 & 60 & 14 & 14 & 47 & 42 & 462 & 1 \\
Turnover    & 1.00 & 1.37 & \textbf{0.78} & \textit{0.80} & 1.87 & 1.52 & 1.32 & 1.09 & 1.44 & 0.00 & 0.00 \\
\midrule
\multicolumn{12}{c}{\textbf{In-sample period= 2 years, Out-of-Sample period= 1 year}} \\
\midrule
EMR         & 6.12E-04  & \textbf{1.33E-03}  & 1.74E-04  & 1.69E-04  & 2.04E-04  & 4.35E-04  & \textit{6.46E-04}  & 8.66E-05 & 1.75E-04 & 3.33E-04  & 3.51E-04 \\
Min         & -1.40E-01 & -1.48E-01 & -1.19E-01 & -1.19E-01 & \textbf{-9.51E-02} & -1.65E-01 & \textit{-1.33E-01} & -1.20E-01 & -1.38E-01 & -1.39E-01 & -1.28E-01 \\
Std Dev     & 1.40E-02  & 1.83E-02  & \textbf{9.86E-03}  & \textit{9.85E-03}  & 1.11E-02  & 1.28E-02  & 1.54E-02  & 9.92E-03 & 1.04E-02 & 1.20E-02  & 1.15E-02 \\
DD          & 9.99E-03  & 1.26E-02  & \textbf{7.33E-03}  & \textit{7.32E-03}  & 8.08E-03  & 9.45E-03  & 1.11E-02  & 7.37E-03 & 7.83E-03 & 8.92E-03  & 8.46E-03 \\
$\text{VaR}_{0.95}$   & 2.11E-02  & 3.07E-02 & \textbf{1.28E-02}  & \textit{1.27E-02}  & 1.55E-02  & 1.81E-02  & 2.53E-02  & 1.33E-02 & 1.46E-02 & 1.70E-02  & 1.74E-02 \\
$\text{CVaR}_{0.95}$  & 3.43E-02  & 3.84E-02  & \textbf{2.37E-02}  & \textit{2.36E-02}  & 2.75E-02  & 3.05E-02  & 3.74E-02  & 2.41E-02 & 2.54E-02 & 3.01E-02  & 2.92E-02 \\
SR          & \textit{4.38E-02}  & \textbf{7.27E-02}  & 1.77E-02*  & 1.72E-02*  & 1.83E-02*  & 3.40E-02  & 4.19E-02  & 8.73E-03* & 1.67E-02* & 2.78E-02*  & 3.06E-02* \\
$\text{SVR}_{0.95}$   & \textit{2.90E-02}  & \textbf{4.33E-02}  & 1.36E-02  & 1.33E-02  & 1.32E-02  & 2.40E-02  & 2.56E-02  & 6.49E-03 & 1.19E-02 & 1.96E-02  & 2.01E-02 \\
$\text{SCR}_{0.95}$   & \textit{1.78E-02}  & \textbf{3.46E-02}  & 7.36E-03  & 7.18E-03  & 7.41E-03  & 1.43E-02  & 1.73E-02  & 3.59E-03 & 6.86E-03 & 1.11E-02  & 1.20E-02 \\
Sortino     & 5.41E-02  & \textbf{7.16E-02}  & 2.37E-02  & 2.31E-02  & 2.52E-02  & 4.60E-02  & \textit{5.82E-02}  & 1.17E-02 & 2.23E-02 & 3.74E-02  & 4.15E-02 \\
$\text{Rachev}_{0.95}$ & 1.12E-03  & \textbf{4.75E-03}  & 4.86E-04  & 4.85E-04  & 6.70E-04  & 8.32E-04  & \textit{1.28E-03}  & 5.03E-04 & 5.58E-04 & 7.68E-04  & 7.10E-04 \\
PTR         & 4.19E-04 & 7.53E-04 & 6.02E-04 & 6.02E-04 & 4.03E-04 & 1.42E-03 & 9.10E-04 & 6.02E-04 & 6.88E-04 & 9.58E-04 & 6.17E-04 \\
\midrule
Assets      & 341 & 16 & 34 & 34 & 60 & 13 & 15 & 58 & 41 & 462 & 1 \\
Turnover    & \textbf{1.21} & 1.43 & 1.23 & \textit{1.22} & 1.95 & 1.76 & 1.69 & 1.47 & 1.61 & 0.00 & 0.00 \\
\bottomrule
\end{tabular}

\end{table}

\begin{itemize}
   \item  The results indicate that TDA-PO and TDA-IPO maintain strong (EMR) performance across all out-of-sample periods (TDA-PO produced slightly lower return than Omega model only for the case of out-of-sample of 1 month and 1 year), reinforcing their ability to generate persistent return advantages over traditional PO models, specially the variance-based models i.e. GMV, MP, \& Sharpe;\, the tail risk measure based PO models, MVaR \& MCVaR, and the benchmark strategies ($1/N$ and S\&P 500 market index). 
   
%   \item As out-of-sample period time increases, we observe an overall decline in EMR across all PO models. This effect, driven by higher market uncertainty and diminishing short-term return differentials over longer horizons, is more pronounced in traditional variance-based models (GMV, MP, and Sharpe), whereas TDA-PO and TDA-IPO sustain a relatively stable performance trajectory.

   \item An interesting observation is that, unlike in the case of in-sample period of 1 year, TDA-IPO exhibits superior return stability compared to TDA-PO with the increased in-sample period size of 2 years, suggesting that the imposed asset constraints help mitigate overfitting to short-term fluctuations. The constrained asset selection in TDA-IPO leads to more persistent return efficiency, confirming its superiority over TDA-PO model.

   \item 
 The risk analysis (stdev, DD, VaR$_{0.95}$, and CVaR$_{0.95}$) follows an expected pattern, with most of the models exhibiting increased risk exposure as the holding period extends. %This increase is observed across 
 While variance-based and tail risk based models maintain the lowest absolute risk levels, they do so at the cost of significantly lower return potential as similar to all other cases. In contrast, TDA-PO and TDA-IPO achieve a more balanced risk-return tradeoff, controlling downside risk without overly restricting upside potential. 

\item In terms of topological risk, the $1/N$ strategy suffers with the highest values of PTR for all the cases when second highest values were generated by the model STARR (over out-of-sample period of 1 month and 6 months), Sharpe model (over out-of-sample period of 3 months), and Omega model (over out-of-sample period of 1 year). It can be easily seen that the PTR values remain very unstable for all the models over varied size of out-of-sample except for TDA-PO, Omega, MCVaR, MVaR, $1/N$ and the S\&P 500 index. However, among these models TDA-PO has the least values of PTR and it is true for all choices of the out-of-sample periods. 

   \item Examining risk-adjusted return performance, TDA-PO and TDA-IPO sustain consistently higher Sharpe, Sortino, Rachev, Sharpe-CVaR (SCR$_{0.95}$), and Sharpe-VaR (SVR$_{0.95}$) ratios compared to traditional models irrespective of window size, reaffirming their efficiency in translating risk into return over varying horizons. Despite an overall decline in risk-adjusted ratios as the out-of-sample period extends, TDA-based models retain their performance edge in terms of these financial ratios. %, particularly in downside-risk-adjusted measures such as Sortino and Sharpe-CVaR. Interestingly, while STARR occasionally leads in the Rachev ratio, its outperformance is accompanied by significantly higher tail risk exposure, making it less appealing for risk-averse investors. 
%   The out-performance of TDA based portfolios over the other benchmarked portfolio in terms of financial ratios, is consistent . 

   \item In terms of turnover ratio, %A crucial consideration in portfolio optimization is turnover behavior and asset concentration. 
it has been decreased comparatively for all the models when the in-sample size increases from 1 year (see Table \ref{tab:performance_metrics}) to 2 years (see Table \ref{tab:performance_metrics2}), indicating the effect of information gained during the longer training period time on adjusting the portfolio. It can be easily seen that the TDA based portfolios always exhibit lesser turnover ratios compared to the Sharpe, MCVaR and STARR (and many times from Omega also), %The model TDA-PO, while adaptive, exhibits a moderate turnover decline, 
suggesting that they remain responsive to market changes without excessive trading costs. Meanwhile, TDA-IPO achieves lower turnover than TDA-PO, reinforcing its stability and cost-efficiency, making it a particularly attractive model for investors seeking lower trading frictions. Similar to the case of in-sample period of 1 year, here also the portfolios from the model GMV and MP attain lower values of turnover ratios.  Further, the results show that turnover decreases as the out-of-sample period increases from 1 months to 1 year, reflecting a reduction in rebalancing frequency over longer investment horizons. 

\item Finally, in comparison to the $1/N$ and S\&P 500 market index, all the PO models except the models, GMV, MP, MVaR, MCVaR and Sharpe, generate better outcomes in terms of EMR and all the financial ratios. This conclusion is consistence through out our empirical analysis irrespective to the change in out-of-sample or in-sample time period length. 

 %In terms of portfolio concentration, TDA-IPO consistently exhibits lower Herfindahl-Hirschman Index (HHI) values than TDA-PO, confirming that the asset cap constraint enhances diversification over longer horizons. While TDA-PO remains slightly more concentrated, its diversification level is still significantly better than that of STARR, Omega, and Sharpe, which allocate excessive weight to fewer assets over time.

\end{itemize}

Overall, the results confirm that TDA-PO and TDA-IPO provide robust and adaptable investment solutions across different investment horizons. TDA-PO remains the preferred choice for investors prioritizing higher risk-adjusted returns and short-term adaptability, while TDA-IPO emerges as a strong alternative for investors seeking more stable, and cost-efficient portfolios over extended training horizons. The findings underscore the strength of TDA-based optimization in overcoming limitations of traditional PO models, including much celebrated variance-based models, offering superior performance and risk management across various market conditions.

\textbf{Note: Why TDA-PO and TDA-IPO are financially strong choices?}

The robustness analysis confirms that TDA-PO and TDA-IPO consistently outperform competing PO models and benchmarked 1/N and S\&P 500 index, across varying size of window size. Even as the holding period increases, TDA-PO maintains its superior EMR, risk-adjusted return profile, and downside risk control, positioning it as the most efficient investment strategy. Despite minor trade-offs in turnover, TDA-based models retain key advantages over reward-risk bases PO models, Sharpe, STARR, and Omega, striking an optimal balance between high returns, controlled risk exposure, and manageable trading costs. 

The financial strength of TDA-PO and TDA-IPO lies in their ability to generate superior returns without direct reliance on expected return estimates, allowing them to adapt dynamically to changing market conditions. This reduced estimation risk, combined with robust risk control mechanisms and competitive risk-adjusted returns, positions TDA-based models as a compelling choice for investors seeking stable, high-performing, and adaptable portfolio solutions.

Our findings align closely with previous studies that have employed TDA in financial applications. \cite{goel2020topological} demonstrated the effectiveness of TDA in enhanced indexing, showing that topological features extracted from persistence landscapes provide valuable insights into asset selection and portfolio construction. Similarly, \cite{goel2023sparse} introduced a TDA-based clustering framework for sparse portfolio selection, highlighting the superior performance of TDA-driven methodologies over traditional correlation-based approaches. Our results reinforce these conclusions, as we also observe that TDA-based methods capture essential structural patterns in financial time series. Specifically, our approach successfully identifies topological features reflecting market dynamics, enhancing decision-making in financial applications.

\section{Conclusion}
With the growing applications of TDA in various domains of data analysis, its usage in the area of portfolio optimization has recently been explored. In this paper, we aim to utilize persistence homology, a fundamental tool of TDA, to 
%The main hurdles of any traditional PO is the involvement of statistical measure based risk function which does not only suffer with the estimation error but unable to capture important qualitative features of assets such as the shape. To
%achieve reliable and stable outcomes, we 
construct an estimation error-free risk measure named as ``Topological Risk” to obtain robust, reliable, and stable outcomes. The topological risk of a portfolio combines the dynamic of topological properties for each asset, calling asset topological risk. The topological risk for each asset is quantified as the squared error of the asset’s time series of $L_p$ norms of persistence landscapes from the norm of the mean persistence landscape. An optimal portfolio is finally derived by minimizing the topological risk of a portfolio over an admissible set of portfolios.

Numerical results over the sample data of S\&P 500 (U.S) with the sample period of nearly 10 years from December 10, 2012, to August 11, 2022,  conclude overall best out-of-sample performance from the model TDA-PO in comparison to seven well-established PO models, namely  global minimum variance, mean-variance, mean-CVaR, mean-VaR, Sharpe, STARR, and Omega models, and the two benchmark portfolio strategies, the equally weighted portfolio $1/N$ and the benchmark market index S\&P 500. The TDA based portfolios achieve first rank in terms of EMR without generating high risk, resulting in the best risk-adjusted returns (financial ratios) performance. This finding remains consistent through out the empirical analysis considered for the varying size of holding and investment time horizon. We also check the effect of cardinality constraint in the proposed model (TDA-IPO) by considering the five different values for cardinality. We found that irrespective of the values for $k$ in TDA-IPO, it generates higher values of mean return and ratios in comparison to all other traditional PO models. These results confirm the supremacy of the proposed TDA-based portfolios in delivering high returns while managing risk, when applied in investment practice.

Having a young idea of using TDA in 
portfolio optimization, there is room for improvement and refinement in the proposed TDA-based risk measure. For instance, while utilizing the topological risk matrix represented by $Q$, we operate under the assumption of zero cross-interaction among distinct assets, a scenario that may not align with reality. Introducing cross-interaction into the model requires careful consideration, and various methods can be employed for its calculation. One possible approach could be, defining the covariance between the norms of the persistence landscapes associated with two assets, but it is not trivial. Alternatively, exploring the computation of persistence landscapes through point clouds derived from two-dimensional data for a pair of assets presents itself as a promising avenue for further investigation, representing the next challenge to address in our work.

Another important consideration for future work is the role of embedding parameters in shaping topological summaries. In our study, we fixed the embedding dimension and time delay to \( d = 3 \) and \( \tau = 1 \), respectively, a choice guided by theoretical motivation and consistency with prior TDA applications. While these values are not tuned, they are sufficient to preserve short-term dynamics in financial time series. Exploring how alternative embedding configurations affect the estimation of topological risk, particularly in models that account for interactions between assets, remains a promising avenue for further investigation.

\section{Declaration}
    
\begin{itemize}
\item  Ethics approval and consent to participate: NA

\item Consent for publication: All data contain in the paper have the consent of all the authors to be published.

    \item Availability of data and material: The datasets used are available from the corresponding author on reasonable request.

    \item Competing interests: No competing interests.

    \item Funding: The corresponding author gratefully acknowledges the financial support received under the Startup Research Grant (SRG) scheme, File Number: SRG/2022/001983, from the Anusandhan National Research Foundation (ANRF) (previously known as Science \& Engineering Research Board (SERB)), Department of Science and Technology (DST), Government of India. 

    \item Authors' contributions: Anubha Goel contributes in conceptualization and design of the study, investigation, analysis and interpretation of results, writing—original draft, writing—review \& editing. Amita Sharma endows in investigation, analysis, project administration, supervision, writing—original draft, writing—review \& editing. Juho Kanniainen helps in investigation, resources, software, supervision, writing—review \& editing. All author(s) read and approved the final manuscript

\item Acknowledgements: The authors sincerely thank the Editor and the anonymous reviewers for their valuable
 comments and suggestions, which have considerably improved the presentation and quality of the paper.
 
\end{itemize}

\backmatter

\begin{appendices}
\section{Performance Mesaures}\label{sec:Appendix_I}

We use the following performance measures to analyze the out-of-sample performance of the portfolios.
\begin{enumerate}
	\item \textbf{Excess Mean Return (EMR)}: Out-of-sample excess mean return measured by $\mu = \frac{\sum_{t =1}^{T} R_t - r_f}{T}$, where $R_t$ is the realization of portfolio return $R_w$ at time point $t$, $t=1,\ldots,T$, where $T$ is the total number of days in the out-of-sample period and $r_f$ is a risk free interest. Higher values of EMR are preferable. 
	\item \textbf{Standard deviation (stdev)}: Out-of-sample 
	%standard deviation 
	stdev of portfolio returns computed as 
	$$\sigma=\sqrt{\frac{\sum_{t=1}^T (E(R_w)-R_t)^2}{T}}.$$ 
 The lower values of stdev are preferable. We also test whether the out-of-sample Variance ($\sigma^2$) of portfolio $p1$ is statistically worse than portfolio $p_2$. We apply the one-sided F-test\footnote{Given two strategies $s_1$ and $s_2$, with sample variances $\hat{\sigma}_{s_1}^2$ and $\hat{\sigma}_{s_2}^2$ computed over a common sample size $n$, we test the equality of variances using the $F$-statistic:

\[
F_{\sigma^2} := \frac{\hat{\sigma}_{s_2}^2}{\hat{\sigma}_{s_1}^2}
\]

\noindent
The corresponding $p$-value is evaluated against the $F$ distribution with degrees of freedom $(n - 1, n - 1)$.} with the hypothesis $H_0:\, \sigma^2_{p_1} - \sigma^2_{p_2} = 0$ and $H_a: \sigma^2_{p_1} -  \sigma^2_{p_2} > 0$.

  \item \textbf{Downside Deviation (DD)}: It accounts for all the negative returns over out-of-sample period and is quantified as: 
    $$\sqrt{\frac{\sum_{t=1}^{T}  \left( \text{min} \{ R_{t}, 0 \} \right)^{2}}{T}}.$$
The lower values of DD are preferable. 

%\textcolor{red}{Sharpe CVaR = STARR?  } 

  \item \textbf{Value-at-risk (VaR$_\alpha$) and Conditional Value-at-Risk (CVaR$_{\alpha}$)}: 
 VaR$_\alpha$ is a popular quantile-based risk measure used to estimate the maximum potential loss in a portfolio over a specific time horizon, at a given confidence level $\alpha \in
(0, 1)$ whereas CVaR$_{\alpha}$ is a conditional  expectation of portfolio losses more than VaR$_\alpha$. 
  
Arranging the out-of-sample losses from portfolio $\mathbf{w}$ in ascending order as $L_{w1}, L_{w2}, \ldots, L_{wT}$, 
    %where $M$ is the total number of out-of-sample windows, 
    VaR and CVaR at $\alpha$ are calculated as:
    \begin{align*}
        \text{VaR}_{\alpha}(L_{w}) &=L_{wk},\\
        \text{CVaR}_{\alpha}(L_{w}) &=\frac{1}{T(1-\alpha)} \sum\limits_{i=k}^{T} L_{wi}
    \end{align*}
    where $k=\lfloor{T\alpha}\rfloor{}+1$ and $L_w\, = \, -R_w$ is portfolio loss. Here, $\lfloor{\cdot}\rfloor{}$ denotes the greatest integer function or the floor function. For a fixed value of $\alpha \in (0,1)$, lower values of $\text{VaR}_{\alpha}(L_w)$ and $\text{CVaR}_{\alpha}(L_w)$ are preferable. Furthermore, we test whether the out-of-sample VaR and CVaR from portfolio $p1$ are statistically worse than those of portfolio $p_2$. We apply the one-sided test\footnote{
Given a strategy $s_1$ and a target portfolio $s^*$, with $y_1, \dots, y_n$ as the return series of $s_1$ sorted from lowest to highest, $\widehat{\text{CVaR}}_\alpha$, $\widehat{\text{VaR}}_\alpha$ are their sample CVaR and VaR values over a sample period $n$ and $c$ denoting the $\widehat{\text{CVaR}}_p$ for the target portfolio. We evaluate the $p$-values by calculating the $z$-test statistic:

\[
z_{\text{CVaR}_\alpha} := \frac{ \sqrt{n(1 - \alpha)} \left( c - \widehat{\text{CVaR}}_\alpha \right) }
{ \sqrt{ \frac{1}{n(1 - \alpha)} \sum_{i = n\alpha + 1}^{n} \left( y_i - \widehat{\text{CVaR}}_\alpha \right)^2 + \alpha \left( \widehat{\text{CVaR}}_\alpha - \widehat{\text{VaR}}_\alpha \right)^2 } },
\quad \text{where} \quad \widehat{\text{VaR}}_\alpha := y_{n\alpha}
\]

\[
\widehat{\text{CVaR}}_\alpha := \frac{1}{n(1 - \alpha)} \sum_{i = n\alpha + 1}^{n} y_i
\]

Given a strategy $s_1$ and a target portfolio $s^*$, with $y_1, \dots, y_n$ as the return series of $s_1$, and $c$ denoting the $\widehat{\text{VaR}}_p$ for the target portfolio, we evaluate the $p$-value using the large-sample $z$-test statistic:

\[
z_{\text{VaR}_p} := \frac{ \#\{ y_i : y_i < c \} - np }{ \sqrt{np(1 - p)} }
\]

\noindent
where $p$ is the target confidence level, and $\#\{ y_i : y_i < c \}$ denotes the number of returns in the sample below the target threshold $c$.
}. with the hypothesis $H_0:\, VaR(CVaR)_{p_1} - VaR(CVaR)_{p_2} = 0$ and $H_a:  VaR(CVaR)_{p_1} -  VaR(CVaR)_{p_2} > 0$.
  	
	\item \textbf{Sharpe Ratio (SR)}: Sharpe ratio is defined as the ratio of excess mean return to its standard deviation, i.e., \begin{center}
		SR$=\frac{\mu - r_f}{\sigma};\; \mu > r_f$,
	\end{center}
%	where $E(R_x)$ and $\sigma_x$ are the respective mean and standard deviation of the portfolio return $R_x$. 
Furthermore, we test whether the out-of-sample Sharpe ratios of two portfolios $p_1$ and $p_2$ are statistically different.
We apply the one-sided $z_{SR}$ test with the hypothesis $H_0:\, SR_{p_1} - SR_{p_2} = 0$ and $H_a: SR_{p_1} - SR_{p_2} > 0$.\footnote{Given two portfolios $p_1$ and $p_2$, with $\mu_{p_1}$, $\mu_{p_2}$, $\sigma_{p_1}$, $\sigma_{p_2}$, $\sigma_{p_1,p_2}$ as their sample means, standard deviations, and the covariance of two strategies over a sample period $n$. The $z$-test statistic is~$$z_{SR} = \frac{\sigma_{p_2}\mu_{p_1} - \sigma_{p_1}\mu_{p_2}}{\sqrt{\Upsilon}},~~ \mbox{with} $$ 
\begin{eqnarray*}
\Upsilon = \frac{1}{n}\big(2\sigma_{p_1}^2\sigma_{p_2}^2 - 2\sigma_{p_1}\sigma_{p_2}\sigma_{p_1,p_2} + 0.5\mu_{p_1}\sigma_{p_2}^2 + 0.5\mu_{p_2}\sigma_{p_1}^2\\
- \frac{\mu_{p_1}\mu_{p_2}}{\sigma_{p_1}\sigma_{p_2}}\sigma_{p_1,p_2}^2\big).\end{eqnarray*}
}
	
	\item \textbf{Sortino Ratio (Sortino)}: Sortino ratio takes risk below to the mean return ($\bar{\sigma}$) instead of standard deviation in the Sharpe ratio and is given as, 
	\begin{center}
		Sortino$=\frac{\mu -r_f}{\bar{\sigma}}; \; \mu > r_f$,
	\end{center}
	where $$\bar{\sigma}=\sqrt{\frac{\sum_{t=1}^T (\min\{R_t-E(R_w),0\})^2}{T}}$$ is the semi-standard deviation of $R_w$.

	\item \textbf{Sharpe-VaR Ratio  (SVR$_{\alpha}$)}: It is a Sharpe ratio when the standard deviation is being replaced by VaR$_{\alpha}(L_w)$ and is defined as:
	\begin{center}
		
		SVR$_{\alpha}=\frac{\mu -r_f}{VaR_{\alpha}(L_w)}\;\, \mu > r_f,\; VaR_{\alpha}(L_w)>0$.
	\end{center}

	\item \textbf{Sharpe-CVaR Ratio (SCR$_{\alpha}$)}: It is a Sharpe ratio when the standard deviation is being replaced by CVaR$_{\alpha}(L_w)$ and is defined as:
	\begin{center}
		
		SCR$_{\alpha}=\frac{\mu -r_f}{CVaR_{\alpha}(L_w)}\;\, \mu > r_f,\; CVaR_{\alpha}(L_w)>0$.
	\end{center}
%	We select $\alpha=0.05$ for the calculation of STARR$_{\alpha}$ in the empirical analysis.

\item \textbf{Rachev ratio (Rachev$_{\alpha}$)}: It is the ratio of expected tail returns to the expected tail losses and is given as: 

$$\text{Rachev}_{\alpha}=\dfrac{\text{CVaR}_{\alpha}(R_{w})}{\text{CVaR}_{\alpha}(-R_{w})}.$$ Larger values are desirable of all the above-listed ratios.

\item \textbf{Turnover Ratio (TR):} It is the average of the absolute values of trades among the $n$ assets over the investment period. It is defined as  
    $$\textbf{Turnover}=\dfrac{1}{M-1} \sum\limits_{t=1}^{M-1}\sum\limits_{j=1}^{n}\vert w_{j, t+1}-w_{j,t},\rvert$$
 where $M$ is the total number of windows.  
    Smaller values of turnover ratio are beneficial as they imply lower transaction costs.

\end{enumerate}

\textbf{Note:} We report the values of VaR$_{\alpha}$, CVaR$_{\alpha}$, SVR$_{\alpha}$, SCR$_{\alpha}$, and Rachev$_{\alpha}$ ratios for $\alpha=0.95$. For simplicity, we choose $r_f=0$ in the out-of-sample analysis.

\section{Benchmarked portfolio optimization models for the comparative analysis}\label{sec:Appendix_II}

\textbf{Basic notation:}
%Consider %an investor whose aim is to select an optimal portfolio over a single investment horizon from the set of $n$ feasible assets, represented by $J = \{1,\ldots,n\}$ where each security $j \in J$ has a rate of return denoted by random variable $R_j$ with a mean value $\mu_j = \mathbb{E}(R_j)$. A vector $x= (x_1,\ldots,x_n) \in \mathbb{R}^n$ denotes a portfolio of $n$ assets where each $x_j$ presents a proportion of total budget to be invested in $j$th asset, $j=1,\dots,n$. Then 
Let the return of a portfolio $w$ is given by a random variable $R_{w} = \sum_{i=1}^{n} r_i w_i$ where $r_i$ represents the random rate of return from $i$th asset ; $i=1,\ldots,n$. Further, portfolio mean return and variance respectively, are denoted by $\mathbb{E}(R_{w}) = \sum_{i=1}^{n} \mu_i w_i$ and $w{'}\Sigma w =\displaystyle\sum_{i=1}^n\displaystyle\sum_{k=1}^{n}w_{i}w_{k}\sigma_{ik}$ where $\mu_i = \mathbb{E}(r_i)$; $i=1,\ldots,n$ and $\sigma_{ik} = cov(r_i, r_k); \; i,k =1,\ldots,n$. 

For STARR and Omega PO models, we assume a total of $T$ number of scenarios/time points under the discrete time setting %, , which %. These $T$ finite number of scenarios either  can be obtained using Monte Carlo simulations or using historical price data, 
with each realization $r_{it}$ of $r_i$ occurs with a uniform probability $p_t = 1/T; t = 1, \ldots, T$. Then $y_t = \sum_{i=1}^{n} r_{it} w_i$ becomes $t$th realization of the portfolio return $R_w$, $t=1,\ldots,T$.

\begin{itemize}
    \item \textbf{Mean-variance model:}
The famous mean-variance model by \cite{marko1952} is a quadratic program which quantifies the portfolio return and risk respectively, by mean and variance of the return distribution. For $\mu = (\mu_1, \ldots, \mu_n)$ and $\Sigma = [\sigma_{ik}]_{i,k=1}^n$ respectively, be the mean vector and covariance matrix, the mean-variance model is given as:

\begin{align}
& \min
\begin{aligned}[t]
\quad w^{'}\Sigma w - \mu^{'}w\\
\end{aligned} \notag \\
& \text{subject to} \notag\\
& w \in W \notag
\end{align}

The mean-variance model can be solved efficiently using standard quadratic programming algorithms.  As portfolio weights from mean-variance model tend to be highly sensitive to even very small changes in the mean returns, ignoring it completely can be one of the remedy. We next consider global minimum variance PO model which inspired from this wisdom.

\item \textbf{Global minimum variance model :} 
Global minimum variance model \cite{Clarke2010MinimumVP} gives us the portfolio with least variance, is free from mean return terms and therefore, believe to generate more reliable out-of-sample results than the mean-variance model. The model is described in below:

\begin{align}
& \min
\begin{aligned}[t]
\quad w^{'}\Sigma w\\
\end{aligned} \notag \\
& \text{subject to} \notag\\
& w \in W \notag
\end{align}

Just like the mean-variance model, this can also be solved efficiently using standard quadratic programming algorithms.

\item \textbf{Sharpe model:} Sharpe ratio \cite{sharpe94} is the classic reward-risk ratio, defined as the fraction of the mean return of a portfolio to its standard deviation. It is widely used due to its simplicity and intuitive appeal of considering the standard deviation as a risk function. An optimization model maximizing Sharpe ratio is given as: 

\begin{align}
& \max
\begin{aligned}[t]
\quad \frac{\mu^{'}w}{\sqrt{w^{'}\Sigma w}}\\
\end{aligned} \notag \\
& \text{subject to} \notag\\
& w \in W \notag
\end{align}

Since the numerator (mean return) of Sharpe ratio is linear and the denominator (standard deviation) is an increasing function of a quadratic form, the above program reduces to a quadratic programming problem which can be solved efficiently globally. 

\item \textbf{STARR ratio:} The STARR ratio \cite{starr_martin} is a reward-risk ratio which replaces the standard deviation in the Sharpe ratio with a conditional value-at-risk at a given level of confidence $\alpha \in (0,1)$, which is a coherent risk measure. An optimization model maximizing STARR ratio is given as: 

\begin{align}
& \max
\begin{aligned}[t]
\quad \frac{\mu^{'}w}{CVaR_{\alpha}(w)}\\
\end{aligned} \notag \\
& \text{subject to} \notag\\
& w \in W, \notag
\end{align}

where $CVaR_{\alpha}(w)$ is the conditional value-at-risk of a portfolio $w$ at a given level of confidence $\alpha \in (0,1)$. The above reward-risk PO model is a linear fractional program under the discrete time setting and can be translated into an equivalent linear program which therefore, can be solved efficiently using any LP solver.

\item \textbf{Omega ratio:} Omega ratio introduced by \cite{mausser2006}, is a fraction of upside deviation of a portfolio return relative to a constant threshold point to its downside deviation. Therefore, an optimal portfolio maximizing Omega ratio does not only exhibits minimal downside deviation but simultaneously maximal upside deviation from the given threshold point. 

For a given threshold $L \in \mathbb{R}$, a PO model maximizing the Omega ratio is given in below:    

\begin{align}
& \max
\begin{aligned}[t]
\quad \frac{E(R_w- L)^+}{E(L - R_w)^+},\\
\end{aligned} \notag \\
& \text{subject to} \notag\\
& w \in W, \notag
\end{align}

where $c^{+}=\max\{c,0\}$. Under the discrete structure of portfolio return, the above model can be translated into an equivalent linear program under the condition that $L < \max_{w \in W}E(R_w)$. The target value $L$ in the Omega model is set as the average value of Index return.

\item \textbf{Mean-CVaR model:} 
The mean-CVaR model \cite{Rockafellar2000} incorporates downside risk directly by penalizing the conditional value-at-risk (CVaR) of the portfolio loss distribution at a given confidence level $\alpha \in (0,1)$. The model under discrete scenarios seeks to maximize the expected return while penalizing CVaR as follows:
\begin{align}
\max_{w,\;\zeta,\;z_t} \quad 
& \mu^\top w - \left[ \zeta + \frac{1}{(1-\alpha)T}\sum_{t=1}^T z_t \right] \notag \\
\text{subject to} \quad 
& z_t \ge - y_t - \zeta,\quad t = 1,\dots,T, \notag\\
& z_t \ge 0,\quad t = 1,\dots,T, \notag\\
& w \in W. \notag
\end{align}
Here, $\zeta \in \mathbb{R}$ and $z_t \ge 0$ are auxiliary variables that linearize the CVaR term. %The parameter $\lambda \ge 0$ controls the trade-off between return and tail risk. 

\item \textbf{Mean-VaR model:} 
The mean-VaR model aims to optimize expected return while penalizing the value-at-risk (VaR) of the portfolio loss distribution at level $\alpha \in (0,1)$. VaR is defined as the $\alpha$-quantile of the loss distribution. The corresponding optimization model is formulated as a mixed-integer linear program \cite{lofti2016}:
\begin{align}
\max_{w,\;\eta,\;u_t,\;y_t} \quad 
& \mu^\top w - \eta \notag \\
\text{subject to} \quad 
& u_t \ge - y_t - \eta,\quad t = 1,\dots,T, \notag\\
& 0 \le u_t \le M\,y_t,\quad t = 1,\dots,T, \notag\\
& \sum_{t=1}^T y_t \le \lceil (1-\alpha)T \rceil,\quad y_t \in \{0,1\}, \notag\\
& w \in W. \notag
\end{align}
Here, $\eta \in \mathbb{R}$ represents the VaR level, $u_t \ge 0$ are exceedance variables,  $M > 0$ is a sufficiently large number and $y_t$ are binary variables indicating tail scenarios. We take $\lambda=1$.

For the computation purposes, we take value of the confidence level $\alpha$ = 0.95 in the models, Mean-CVaR, Mean-VaR, and STARR.

\end{itemize}

\textbf{Mathematical Optimization Techniques for Portfolio Models}
\begin{itemize}

    \item 
We compute the TDA-PO, global minimum variance, and mean-variance portfolios using a Quadratic Programming approach, implemented via the \texttt{quadprog} package in R. Specifically, we use the \texttt{solve.QP} function, which employs the dual method of \cite{goldfarb1983numerically} to efficiently solve the convex quadratic optimization problem.

    \item 
To construct the Omega-optimal and mean-CVaR portfolios, we employ a Linear Programming approach using the \texttt{ROI} package in R. The optimization problem is solved via the GLPK solver, which by default utilizes the simplex method to efficiently determine the optimal portfolio weights while ensuring feasibility under the given constraints.

\item For mean-VaR portfolios, we formulate the optimization as a Mixed-Integer Linear Program (MILP) and solve it using the \texttt{CVXR} interface with the Gurobi optimizer. The model incorporates binary variables and employs a big-$M$ formulation to linearize the Value-at-Risk constraints. To control computational complexity, a time limit of 30 minutes is specified within the Gurobi solver parameters.

    \item 
For the Sharpe Ratio maximizing portfolio, we adopt a stochastic search-based optimization approach using the \texttt{PortfolioAnalytics} package in R. The optimization process follows a random search heuristic, where 5,000 randomly generated portfolios are evaluated based on their Sharpe Ratios. The portfolio with the highest Sharpe Ratio is selected, ensuring an optimal risk-return trade-off. This heuristic-based approach allows for an extensive exploration of the solution space without requiring explicit derivatives, making it particularly useful for complex portfolio allocation problems.

    \item 
To construct the STARR-optimal portfolio, we employ a convex optimization framework, also implemented via the \texttt{PortfolioAnalytics} package in R. The optimization problem is solved using the \texttt{ROI} solver, leveraging Linear Programming and Conic Optimization techniques to efficiently identify the asset weights that maximize the STARR ratio while maintaining feasibility under the imposed constraints.

    \item 
 For TDA-IPO with integer constraint on the number of assets, we employ an Integer Programming approach. The problem is formulated as a Mixed-Integer Quadratic Program and solved using the Gurobi optimizer, which is well-suited for handling large-scale integer-constrained problems. Gurobi employs a Branch-and-Bound framework with Quadratic Programming Relaxation, efficiently navigating the solution space to find an optimal allocation while enforcing the desired cardinality constraints.
\end{itemize}

\end{appendices}

\bibliography{main.bib}


\begin{thebibliography}{93}
\ifx \bisbn   \undefined \def \bisbn  #1{ISBN #1}\fi
\ifx \binits  \undefined \def \binits#1{#1}\fi
\ifx \bauthor  \undefined \def \bauthor#1{#1}\fi
\ifx \batitle  \undefined \def \batitle#1{#1}\fi
\ifx \bjtitle  \undefined \def \bjtitle#1{#1}\fi
\ifx \bvolume  \undefined \def \bvolume#1{\textbf{#1}}\fi
\ifx \byear  \undefined \def \byear#1{#1}\fi
\ifx \bissue  \undefined \def \bissue#1{#1}\fi
\ifx \bfpage  \undefined \def \bfpage#1{#1}\fi
\ifx \blpage  \undefined \def \blpage #1{#1}\fi
\ifx \burl  \undefined \def \burl#1{\textsf{#1}}\fi
\ifx \doiurl  \undefined \def \doiurl#1{\url{https://doi.org/#1}}\fi
\ifx \betal  \undefined \def \betal{\textit{et al.}}\fi
\ifx \binstitute  \undefined \def \binstitute#1{#1}\fi
\ifx \binstitutionaled  \undefined \def \binstitutionaled#1{#1}\fi
\ifx \bctitle  \undefined \def \bctitle#1{#1}\fi
\ifx \beditor  \undefined \def \beditor#1{#1}\fi
\ifx \bpublisher  \undefined \def \bpublisher#1{#1}\fi
\ifx \bbtitle  \undefined \def \bbtitle#1{#1}\fi
\ifx \bedition  \undefined \def \bedition#1{#1}\fi
\ifx \bseriesno  \undefined \def \bseriesno#1{#1}\fi
\ifx \blocation  \undefined \def \blocation#1{#1}\fi
\ifx \bsertitle  \undefined \def \bsertitle#1{#1}\fi
\ifx \bsnm \undefined \def \bsnm#1{#1}\fi
\ifx \bsuffix \undefined \def \bsuffix#1{#1}\fi
\ifx \bparticle \undefined \def \bparticle#1{#1}\fi
\ifx \barticle \undefined \def \barticle#1{#1}\fi
\bibcommenthead
\ifx \bconfdate \undefined \def \bconfdate #1{#1}\fi
\ifx \botherref \undefined \def \botherref #1{#1}\fi
\ifx \url \undefined \def \url#1{\textsf{#1}}\fi
\ifx \bchapter \undefined \def \bchapter#1{#1}\fi
\ifx \bbook \undefined \def \bbook#1{#1}\fi
\ifx \bcomment \undefined \def \bcomment#1{#1}\fi
\ifx \oauthor \undefined \def \oauthor#1{#1}\fi
\ifx \citeauthoryear \undefined \def \citeauthoryear#1{#1}\fi
\ifx \endbibitem  \undefined \def \endbibitem {}\fi
\ifx \bconflocation  \undefined \def \bconflocation#1{#1}\fi
\ifx \arxivurl  \undefined \def \arxivurl#1{\textsf{#1}}\fi
\csname PreBibitemsHook\endcsname

%%% 1
\bibitem[\protect\citeauthoryear{Markowitz}{1952}]{marko1952}
\begin{barticle}
\bauthor{\bsnm{Markowitz}, \binits{H.}}:
\batitle{Portfolio selection}.
\bjtitle{Journal of Finance}
\bvolume{7},
\bfpage{77}--\blpage{91}
(\byear{1952})
\doiurl{10.2307/2975974}
\end{barticle}
\endbibitem

%%% 2
\bibitem[\protect\citeauthoryear{Bonnans and Shapiro}{2000}]{bonnans2000}
\begin{botherref}
\oauthor{\bsnm{Bonnans}, \binits{J.F.}},
\oauthor{\bsnm{Shapiro}, \binits{A.}}:
Perturbation analysis of optimization problems.
New York: Springer-Verlag
(2000)
\end{botherref}
\endbibitem

%%% 3
\bibitem[\protect\citeauthoryear{Jagannathan and Ma}{2003}]{JagannathanMa2003}
\begin{barticle}
\bauthor{\bsnm{Jagannathan}, \binits{R.}},
\bauthor{\bsnm{Ma}, \binits{T.}}:
\batitle{Risk reduction in large portfolios: Why imposing wrong constraints
  helps}.
\bjtitle{Journal of Finance}
\bvolume{58},
\bfpage{1651}--\blpage{1684}
(\byear{2003})
\doiurl{10.1111/1540-6261.00580}
\end{barticle}
\endbibitem

%%% 4
\bibitem[\protect\citeauthoryear{Ledoit and Wolf}{2003}]{ledoitwolf2003}
\begin{barticle}
\bauthor{\bsnm{Ledoit}, \binits{O.}},
\bauthor{\bsnm{Wolf}, \binits{M.}}:
\batitle{Improved estimation of the covariance matrix of stock returns with an
  application to portfolio selection}.
\bjtitle{Journal of Empirical Finance}
\bvolume{10},
\bfpage{603}--\blpage{621}
(\byear{2003})
\doiurl{10.1016/S0927-5398(03)00007-0}
\end{barticle}
\endbibitem

%%% 5
\bibitem[\protect\citeauthoryear{Ledoit and Wolf}{2004}]{ledoitwolf2004}
\begin{barticle}
\bauthor{\bsnm{Ledoit}, \binits{O.}},
\bauthor{\bsnm{Wolf}, \binits{M.}}:
\batitle{A well-conditioned estimator for large-dimensional covariance
  matrices}.
\bjtitle{Journal of Multivariate Analysis}
\bvolume{88}(\bissue{2}),
\bfpage{365}--\blpage{411}
(\byear{2004})
\doiurl{10.1016/S0047-259X(03)00096-4}
\end{barticle}
\endbibitem

%%% 6
\bibitem[\protect\citeauthoryear{DeMiguel et~al.}{2009}]{demigueletal2009}
\begin{barticle}
\bauthor{\bsnm{DeMiguel}, \binits{V.}},
\bauthor{\bsnm{Garlappi}, \binits{L.}},
\bauthor{\bsnm{Nogales}, \binits{F.G.}},
\bauthor{\bsnm{Uppal}, \binits{R.}}:
\batitle{A generalized approach to portfolio optimization: Improving
  performance by constraining portfolio norms}.
\bjtitle{Management Sciences}
\bvolume{55(5)},
\bfpage{798}--\blpage{812}
(\byear{2009})
\doiurl{10.1287/mnsc.1080.0986}
\end{barticle}
\endbibitem

%%% 7
\bibitem[\protect\citeauthoryear{Kolm et~al.}{2014}]{kolm201460}
\begin{barticle}
\bauthor{\bsnm{Kolm}, \binits{P.N.}},
\bauthor{\bsnm{T{\"u}t{\"u}nc{\"u}}, \binits{R.}},
\bauthor{\bsnm{Fabozzi}, \binits{F.J.}}:
\batitle{60 years of portfolio optimization: Practical challenges and current
  trends}.
\bjtitle{European Journal of Operational Research}
\bvolume{234}(\bissue{2}),
\bfpage{356}--\blpage{371}
(\byear{2014})
\end{barticle}
\endbibitem

%%% 8
\bibitem[\protect\citeauthoryear{DeMiguel et~al.}{2009}]{demiguel2009optimal}
\begin{barticle}
\bauthor{\bsnm{DeMiguel}, \binits{V.}},
\bauthor{\bsnm{Garlappi}, \binits{L.}},
\bauthor{\bsnm{Uppal}, \binits{R.}}:
\batitle{Optimal versus naive diversification: How inefficient is the 1/n
  portfolio strategy?}
\bjtitle{The review of Financial studies}
\bvolume{22}(\bissue{5}),
\bfpage{1915}--\blpage{1953}
(\byear{2009})
\end{barticle}
\endbibitem

%%% 9
\bibitem[\protect\citeauthoryear{Jobson and Korkie}{1981}]{Jobson1981}
\begin{barticle}
\bauthor{\bsnm{Jobson}, \binits{J.D.}},
\bauthor{\bsnm{Korkie}, \binits{R.M.}}:
\batitle{Putting markowitz theory to work}.
\bjtitle{The Journal of Portfolio Management}
\bvolume{7}(\bissue{4}),
\bfpage{70}--\blpage{74}
(\byear{1981})
\end{barticle}
\endbibitem

%%% 10
\bibitem[\protect\citeauthoryear{Jorion}{1991}]{Jorion1991}
\begin{barticle}
\bauthor{\bsnm{Jorion}, \binits{P.}}:
\batitle{Bayesian and capm estimators of the means: Implications for portfolio
  selection}.
\bjtitle{Journal of Banking and Finance}
\bvolume{15},
\bfpage{717}--\blpage{727}
(\byear{1991})
\end{barticle}
\endbibitem

%%% 11
\bibitem[\protect\citeauthoryear{Chopra and Ziemba}{1993}]{chopra1993}
\begin{barticle}
\bauthor{\bsnm{Chopra}, \binits{V.K.}},
\bauthor{\bsnm{Ziemba}, \binits{W.T.}}:
\batitle{The effect of errors in means, variances, and covariances on optimal
  portfolio choice}.
\bjtitle{The Journal of Portfolio Management}
\bvolume{19}(\bissue{2}),
\bfpage{6}--\blpage{11}
(\byear{1993})
\end{barticle}
\endbibitem

%%% 12
\bibitem[\protect\citeauthoryear{Husmann et~al.}{2022}]{sparsestable22}
\begin{barticle}
\bauthor{\bsnm{Husmann}, \binits{S.}},
\bauthor{\bsnm{Shivarova}, \binits{A.}},
\bauthor{\bsnm{Steinert}, \binits{R.}}:
\batitle{Sparsity and stability for minimum-variance portfolios}.
\bjtitle{Risk Management}
\bvolume{24},
\bfpage{214}--\blpage{235}
(\byear{2022})
\doiurl{10.1057/s41283-022-00091-0}
\end{barticle}
\endbibitem

%%% 13
\bibitem[\protect\citeauthoryear{Clarke et~al.}{2010}]{Clarke2010MinimumVP}
\begin{barticle}
\bauthor{\bsnm{Clarke}, \binits{R.}},
\bauthor{\bsnm{Silva}, \binits{H.D.}},
\bauthor{\bsnm{Thorley}, \binits{S.}}:
\batitle{Minimum variance portfolio composition}.
\bjtitle{Journal of Portfolio Management}
\bvolume{37},
\bfpage{31}--\blpage{45}
(\byear{2010})
\end{barticle}
\endbibitem

%%% 14
\bibitem[\protect\citeauthoryear{Khodamoradi et~al.}{2021}]{ccmv2020}
\begin{barticle}
\bauthor{\bsnm{Khodamoradi}, \binits{T.}},
\bauthor{\bsnm{Salahi}, \binits{M.}},
\bauthor{\bsnm{Najafi}, \binits{A.R.}}:
\batitle{Cardinality-constrained portfolio optimization with short selling and
  risk-neutral interest rate}.
\bjtitle{Decisions in Economics and Finance}
\bvolume{44},
\bfpage{197}--\blpage{214}
(\byear{2021})
\doiurl{10.1007/s10203-020-00293-9}
\end{barticle}
\endbibitem

%%% 15
\bibitem[\protect\citeauthoryear{Lee et~al.}{2020}]{leeetal_2020}
\begin{barticle}
\bauthor{\bsnm{Lee}, \binits{Y.}},
\bauthor{\bsnm{Kim}, \binits{M.J.}},
\bauthor{\bsnm{Kim}, \binits{J.H.}},
\bauthor{\bsnm{Jang}, \binits{J.R.}},
\bauthor{\bsnm{Kim}, \binits{W.C.}}:
\batitle{Sparse and robust portfolio selection via semi-definite relaxation}.
\bjtitle{Journal of the Operational Research Society}
\bvolume{71}(\bissue{5}),
\bfpage{687}--\blpage{699}
(\byear{2020})
\doiurl{10.1080/01605682.2019.1581408}
\end{barticle}
\endbibitem

%%% 16
\bibitem[\protect\citeauthoryear{Scherer}{2007}]{scherer2007can}
\begin{barticle}
\bauthor{\bsnm{Scherer}, \binits{B.}}:
\batitle{Can robust portfolio optimisation help to build better portfolios?}
\bjtitle{Journal of Asset Management}
\bvolume{7},
\bfpage{374}--\blpage{387}
(\byear{2007})
\end{barticle}
\endbibitem

%%% 17
\bibitem[\protect\citeauthoryear{Bubenik et~al.}{2015}]{bubenik2015statistical}
\begin{barticle}
\bauthor{\bsnm{Bubenik}, \binits{P.}}, \betal:
\batitle{Statistical topological data analysis using persistence landscapes.}
\bjtitle{J. Mach. Learn. Res.}
\bvolume{16}(\bissue{1}),
\bfpage{77}--\blpage{102}
(\byear{2015})
\end{barticle}
\endbibitem

%%% 18
\bibitem[\protect\citeauthoryear{Bubenik}{2018}]{bubenik2018persistence}
\begin{botherref}
\oauthor{\bsnm{Bubenik}, \binits{P.}}:
The persistence landscape and some of its properties.
arXiv preprint arXiv:1810.04963
(2018)
\end{botherref}
\endbibitem

%%% 19
\bibitem[\protect\citeauthoryear{Konno et~al.}{1993}]{konno93}
\begin{barticle}
\bauthor{\bsnm{Konno}, \binits{H.}},
\bauthor{\bsnm{Shirakawa}, \binits{H.}},
\bauthor{\bsnm{Yamazaki}, \binits{H.}}:
\batitle{A mean-absolute deviation-skewness portfolio optimization model}.
\bjtitle{Annals of Operations Research}
\bvolume{45},
\bfpage{205}--\blpage{220}
(\byear{1993})
\end{barticle}
\endbibitem

%%% 20
\bibitem[\protect\citeauthoryear{Bellini et~al.}{2014}]{bellklaretal14}
\begin{barticle}
\bauthor{\bsnm{Bellini}, \binits{F.}},
\bauthor{\bsnm{Klar}, \binits{B.}},
\bauthor{\bsnm{M\"uller}, \binits{A.}},
\bauthor{\bsnm{Rosazza~Gianin}, \binits{E.}}:
\batitle{Generalized quantiles as risk measures}.
\bjtitle{Insurance: Mathematics and Economics}
\bvolume{54},
\bfpage{41}--\blpage{48}
(\byear{2014})
\end{barticle}
\endbibitem

%%% 21
\bibitem[\protect\citeauthoryear{Ju and Pearson}{1998}]{pearson98}
\begin{barticle}
\bauthor{\bsnm{Ju}, \binits{X.}},
\bauthor{\bsnm{Pearson}, \binits{N.D.}}:
\batitle{Using value-at-risk to control risk taking: how wrong can you be?}
\bjtitle{Journal of risk}
\bvolume{1},
\bfpage{5}--\blpage{36}
(\byear{1998})
\end{barticle}
\endbibitem

%%% 22
\bibitem[\protect\citeauthoryear{Goldfarb and
  Iyengar}{2003}]{goldfarb2003robust}
\begin{barticle}
\bauthor{\bsnm{Goldfarb}, \binits{D.}},
\bauthor{\bsnm{Iyengar}, \binits{G.}}:
\batitle{Robust portfolio selection problems}.
\bjtitle{Mathematics of Operations Research}
\bvolume{28},
\bfpage{1}--\blpage{38}
(\byear{2003})
\end{barticle}
\endbibitem

%%% 23
\bibitem[\protect\citeauthoryear{Moon and Yao}{2011}]{moon2011}
\begin{barticle}
\bauthor{\bsnm{Moon}, \binits{Y.}},
\bauthor{\bsnm{Yao}, \binits{T.}}:
\batitle{A robust mean absolute deviation model for portfolio optimization}.
\bjtitle{Computers \& Operations Research}
\bvolume{38}(\bissue{9}),
\bfpage{1251}--\blpage{1258}
(\byear{2011})
\end{barticle}
\endbibitem

%%% 24
\bibitem[\protect\citeauthoryear{Lotfi and Zeniosn}{2016}]{lofti2016}
\begin{botherref}
\oauthor{\bsnm{Lotfi}, \binits{S.}},
\oauthor{\bsnm{Zeniosn}, \binits{S.A.}}:
Equivalence of robust var and cvar optimization.
Working papers,
University of Pennsylvania, Wharton School, Weiss Center
(2016).
\url{https://EconPapers.repec.org/RePEc:ecl:upafin:16-03}
\end{botherref}
\endbibitem

%%% 25
\bibitem[\protect\citeauthoryear{Rockafellar and
  Uryasev}{2000}]{Rockafellar2000}
\begin{barticle}
\bauthor{\bsnm{Rockafellar}, \binits{R.T.}},
\bauthor{\bsnm{Uryasev}, \binits{S.}}:
\batitle{Optimization of conditional value-at risk}.
\bjtitle{Journal of Risk}
\bvolume{3},
\bfpage{21}--\blpage{41}
(\byear{2000})
\end{barticle}
\endbibitem

%%% 26
\bibitem[\protect\citeauthoryear{Sharpe}{1994}]{sharpe94}
\begin{barticle}
\bauthor{\bsnm{Sharpe}, \binits{W.F.}}:
\batitle{The sharpe ratio}.
\bjtitle{Journal of Portfolio Management}
\bvolume{21}(\bissue{1}),
\bfpage{49}--\blpage{58}
(\byear{1994})
\end{barticle}
\endbibitem

%%% 27
\bibitem[\protect\citeauthoryear{Martin et~al.}{2003}]{starr_martin}
\begin{barticle}
\bauthor{\bsnm{Martin}, \binits{R.D.}},
\bauthor{\bsnm{Rachev}, \binits{S.T.}},
\bauthor{\bsnm{Siboulet}, \binits{F.}}:
\batitle{Phi-alpha optimal portfolios and extreme risk management}.
\bjtitle{Wilmott Journal}
\bvolume{November},
\bfpage{70}--\blpage{83}
(\byear{2003})
\end{barticle}
\endbibitem

%%% 28
\bibitem[\protect\citeauthoryear{Kapsos et~al.}{2014}]{kapsosetal14}
\begin{barticle}
\bauthor{\bsnm{Kapsos}, \binits{M.}},
\bauthor{\bsnm{Zymler}, \binits{S.}},
\bauthor{\bsnm{Christofides}, \binits{N.}},
\bauthor{\bsnm{Rustem}, \binits{B.}}:
\batitle{Optimizing the omega ratio using linear programming}.
\bjtitle{Journal of Computational Finance}
\bvolume{17},
\bfpage{49}--\blpage{57}
(\byear{2014})
\end{barticle}
\endbibitem

%%% 29
\bibitem[\protect\citeauthoryear{Yitzhaki}{1982}]{yitzhaki}
\begin{barticle}
\bauthor{\bsnm{Yitzhaki}, \binits{S.}}:
\batitle{Stochastic dominance, mean variance, and gini’s mean difference}.
\bjtitle{The American Economic Review}
\bvolume{72}(\bissue{1}),
\bfpage{178}--\blpage{185}
(\byear{1982})
\end{barticle}
\endbibitem

%%% 30
\bibitem[\protect\citeauthoryear{Ogryczak and Ruszczyński}{2001}]{ogryczak01}
\begin{barticle}
\bauthor{\bsnm{Ogryczak}, \binits{W.}},
\bauthor{\bsnm{Ruszczyński}, \binits{A.}}:
\batitle{On consistency of stochastic dominance and mean- semideviations
  models}.
\bjtitle{Mathematical Programming}
\bvolume{89}(\bissue{2}),
\bfpage{217}--\blpage{232}
(\byear{2001})
\end{barticle}
\endbibitem

%%% 31
\bibitem[\protect\citeauthoryear{Linsmeier and Pearson}{1996}]{linsmeier}
\begin{botherref}
\oauthor{\bsnm{Linsmeier}, \binits{T.J.}},
\oauthor{\bsnm{Pearson}, \binits{N.D.}}:
Risk measurement: An introduction to value at risk.
Technical report, Technical report 96-04, OFOR, University of Illinois, Urbana
  Champaign, IL
(1996)
\end{botherref}
\endbibitem

%%% 32
\bibitem[\protect\citeauthoryear{Rockafellar and Uryasev}{2002}]{rockafeller02}
\begin{barticle}
\bauthor{\bsnm{Rockafellar}, \binits{R.T.}},
\bauthor{\bsnm{Uryasev}, \binits{S.}}:
\batitle{Conditional value-at-risk for general loss distributions}.
\bjtitle{Journal of Banking and Finance}
\bvolume{26}(\bissue{7}),
\bfpage{1443}--\blpage{1471}
(\byear{2002})
\end{barticle}
\endbibitem

%%% 33
\bibitem[\protect\citeauthoryear{Roman and Gautam}{2009}]{roman09}
\begin{barticle}
\bauthor{\bsnm{Roman}, \binits{D.}},
\bauthor{\bsnm{Gautam}, \binits{M.}}:
\batitle{Portfolio selection models: a review and new directions}.
\bjtitle{Wilmott Journal}
\bvolume{1},
\bfpage{69}--\blpage{85}
(\byear{2009})
\end{barticle}
\endbibitem

%%% 34
\bibitem[\protect\citeauthoryear{Jobson and Korkie}{1981}]{jobson81}
\begin{barticle}
\bauthor{\bsnm{Jobson}, \binits{J.}},
\bauthor{\bsnm{Korkie}, \binits{B.}}:
\batitle{Putting markowitz theory to work}.
\bjtitle{Journal of Portfolio Management}
\bvolume{7},
\bfpage{70}--\blpage{74}
(\byear{1981})
\end{barticle}
\endbibitem

%%% 35
\bibitem[\protect\citeauthoryear{Kolma et~al.}{2014}]{kolma14}
\begin{barticle}
\bauthor{\bsnm{Kolma}, \binits{P.N.}},
\bauthor{\bsnm{Tütüncüb}, \binits{R.}},
\bauthor{\bsnm{Fabozzic}, \binits{F.J.}}:
\batitle{60 years of portfolio optimization: Practical challenges and current
  trends}.
\bjtitle{European Journal of Operational Research}
\bvolume{234},
\bfpage{356}--\blpage{371}
(\byear{2014})
\end{barticle}
\endbibitem

%%% 36
\bibitem[\protect\citeauthoryear{Michaud and Michaud}{2023}]{michaud}
\begin{bchapter}
\bauthor{\bsnm{Michaud}, \binits{R.O.}},
\bauthor{\bsnm{Michaud}, \binits{R.O.}}:
\bctitle{Efficient asset management: A practical guide to stock portfolio
  optimization and asset allocation},
pp. \bfpage{2143}--\blpage{2151}
(\byear{2023}).
\bcomment{New York, NY, 2008; online edn, Oxford Academic, 31 Oct. 2023,
  https://doi.org/10.1093/oso/9780195331912.001.0001, accessed 14 Mar. 2025}
\end{bchapter}
\endbibitem

%%% 37
\bibitem[\protect\citeauthoryear{Black and Litterman}{1991}]{black91}
\begin{barticle}
\bauthor{\bsnm{Black}, \binits{F.}},
\bauthor{\bsnm{Litterman}, \binits{R.B.}}:
\batitle{Asset equilibrium: Combining investor views with market equilibrium}.
\bjtitle{Journal of Fixed Income}
\bvolume{1},
\bfpage{7}--\blpage{18}
(\byear{1991})
\end{barticle}
\endbibitem

%%% 38
\bibitem[\protect\citeauthoryear{Best and Grauer}{2015}]{best15}
\begin{barticle}
\bauthor{\bsnm{Best}, \binits{M.J.}},
\bauthor{\bsnm{Grauer}, \binits{R.R.}}:
\batitle{On the sensitivity of mean-variance-efficient portfolios to changes in
  asset means: Some analytical and computational results}.
\bjtitle{The Review of Financial Studies}
\bvolume{4}(\bissue{2}),
\bfpage{315}--\blpage{342}
(\byear{2015})
\end{barticle}
\endbibitem

%%% 39
\bibitem[\protect\citeauthoryear{Martin}{2009}]{martin09}
\begin{botherref}
\oauthor{\bsnm{Martin}, \binits{H.}}:
Asset allocation and risk management.
Lecture notes: IEOR E4602: Quantitative Risk Management
(2009)
\end{botherref}
\endbibitem

%%% 40
\bibitem[\protect\citeauthoryear{Bodnar et~al.}{2019}]{bodnar19}
\begin{barticle}
\bauthor{\bsnm{Bodnar}, \binits{T.}},
\bauthor{\bsnm{Dette}, \binits{H.}},
\bauthor{\bsnm{Parolya}, \binits{N.}}:
\batitle{Testing for independence of large dimensional vectors}.
\bjtitle{The Annals of Statistics}
\bvolume{47}(\bissue{5}),
\bfpage{2977}--\blpage{3008}
(\byear{2019})
\end{barticle}
\endbibitem

%%% 41
\bibitem[\protect\citeauthoryear{Carlsson}{2009}]{carlsson2009topology}
\begin{barticle}
\bauthor{\bsnm{Carlsson}, \binits{G.}}:
\batitle{Topology and data}.
\bjtitle{Bulletin of the American Mathematical Society}
\bvolume{46}(\bissue{2}),
\bfpage{255}--\blpage{308}
(\byear{2009})
\end{barticle}
\endbibitem

%%% 42
\bibitem[\protect\citeauthoryear{Carlsson}{2020}]{carlsson2020topological}
\begin{barticle}
\bauthor{\bsnm{Carlsson}, \binits{G.}}:
\batitle{Topological methods for data modelling}.
\bjtitle{Nature Reviews Physics}
\bvolume{2}(\bissue{12}),
\bfpage{697}--\blpage{708}
(\byear{2020})
\end{barticle}
\endbibitem

%%% 43
\bibitem[\protect\citeauthoryear{Wasserman}{2018}]{wasserman2018topological}
\begin{barticle}
\bauthor{\bsnm{Wasserman}, \binits{L.}}:
\batitle{Topological data analysis}.
\bjtitle{Annual Review of Statistics and Its Application}
\bvolume{5},
\bfpage{501}--\blpage{532}
(\byear{2018})
\end{barticle}
\endbibitem

%%% 44
\bibitem[\protect\citeauthoryear{Chazal and
  Michel}{2017}]{chazal2017introduction}
\begin{botherref}
\oauthor{\bsnm{Chazal}, \binits{F.}},
\oauthor{\bsnm{Michel}, \binits{B.}}:
An introduction to topological data analysis: fundamental and practical aspects
  for data scientists.
arXiv preprint arXiv:1710.04019
(2017)
\end{botherref}
\endbibitem

%%% 45
\bibitem[\protect\citeauthoryear{Skaf and Laubenbacher}{2022}]{skaf20221}
\begin{barticle}
\bauthor{\bsnm{Skaf}, \binits{Y.}},
\bauthor{\bsnm{Laubenbacher}, \binits{R.}}:
\batitle{Topological data analysis in biomedicine: A review}.
\bjtitle{Journal of Biomedical Informatics}
\bvolume{130},
\bfpage{104082}
(\byear{2022})
\doiurl{10.1016/j.jbi.2022.104082}
\end{barticle}
\endbibitem

%%% 46
\bibitem[\protect\citeauthoryear{Wang et~al.}{2018}]{wang2018}
\begin{barticle}
\bauthor{\bsnm{Wang}, \binits{Y.}},
\bauthor{\bsnm{Ombao}, \binits{H.}},
\bauthor{\bsnm{Chung}, \binits{M.K.}}:
\batitle{Topological data analysis of single-trial electroencephalographic
  signals}.
\bjtitle{The annals of applied statistics}
\bvolume{12},
\bfpage{1506}--\blpage{1534}
(\byear{2018})
\end{barticle}
\endbibitem

%%% 47
\bibitem[\protect\citeauthoryear{Moraleda et~al.}{2020}]{moraleda2020}
\begin{barticle}
\bauthor{\bsnm{Moraleda}, \binits{R.R.}},
\bauthor{\bsnm{Xiong}, \binits{W.}},
\bauthor{\bsnm{Valous}, \binits{N.A.}},
\bauthor{\bsnm{Halama}, \binits{N.}}:
\batitle{Segmentation of biomedical images based on a computational topology
  framework}.
\bjtitle{Seminars in Immunology}
\bvolume{48},
\bfpage{101432}
(\byear{2020})
\doiurl{10.1016/j.smim.2020.101432} .
\bcomment{The Tumor Microenvironment: prognostic and theranostic impact. Recent
  advances and trends}
\end{barticle}
\endbibitem

%%% 48
\bibitem[\protect\citeauthoryear{Chung et~al.}{2021}]{chung2021}
\begin{botherref}
\oauthor{\bsnm{Chung}, \binits{M.K.}},
\oauthor{\bsnm{Smith}, \binits{A.D.}},
\oauthor{\bsnm{Shiu}, \binits{G.}}:
Reviews: Topological distances and losses for brain networks.
ArXiv
\textbf{abs/2102.08623}
(2021)
\end{botherref}
\endbibitem

%%% 49
\bibitem[\protect\citeauthoryear{Ravishanker and Chen}{2019}]{ravishankar2019}
\begin{botherref}
\oauthor{\bsnm{Ravishanker}, \binits{N.}},
\oauthor{\bsnm{Chen}, \binits{R.}}:
Topological data analysis (tda) for time series.
ArXiv
\textbf{arXiv:1909.10604}
(2019)
\end{botherref}
\endbibitem

%%% 50
\bibitem[\protect\citeauthoryear{Lum et~al.}{2013}]{lum.et.al.2013}
\begin{barticle}
\bauthor{\bsnm{Lum}, \binits{P.Y.}},
\bauthor{\bsnm{Singh}, \binits{G.}},
\bauthor{\bsnm{Lehman}, \binits{A.}},
\bauthor{\bsnm{Ishkanov}, \binits{T.}},
\bauthor{\bsnm{Vejdemo-Johansson}, \binits{M.}},
\bauthor{\bsnm{Alagappan}, \binits{M.}},
\bauthor{\bsnm{J.}, \binits{C.}},
\bauthor{\bsnm{Carlsson}, \binits{G.}}:
\batitle{Extracting insights from the shape of complex data using topology}.
\bjtitle{Scientific reports}
\bvolume{3},
\bfpage{1236}
(\byear{2013})
\end{barticle}
\endbibitem

%%% 51
\bibitem[\protect\citeauthoryear{Belchi et~al.}{2018}]{belchi2018}
\begin{barticle}
\bauthor{\bsnm{Belchi}, \binits{F.}},
\bauthor{\bsnm{Pirashvili}, \binits{M.}},
\bauthor{\bsnm{Conway}, \binits{J.}},
\bauthor{\bsnm{Bennett}, \binits{M.}},
\bauthor{\bsnm{Djukanovic}, \binits{R.}},
\bauthor{\bsnm{Brodzki}, \binits{J.}}:
\batitle{Lung topology characteristics in patients with chronic obstructive
  pulmonary disease}.
\bjtitle{Scientific reports}
\bvolume{8},
\bfpage{5341}
(\byear{2018})
\end{barticle}
\endbibitem

%%% 52
\bibitem[\protect\citeauthoryear{Ichinomiya et~al.}{2020}]{ichinomiya2020}
\begin{barticle}
\bauthor{\bsnm{Ichinomiya}, \binits{T.}},
\bauthor{\bsnm{Obayashi}, \binits{I.}},
\bauthor{\bsnm{Hiraoka}, \binits{Y.}}:
\batitle{Protein-folding analysis using features obtained by persistent
  homology}.
\bjtitle{Scientific reports}
\bvolume{118},
\bfpage{2926}--\blpage{2937}
(\byear{2020})
\end{barticle}
\endbibitem

%%% 53
\bibitem[\protect\citeauthoryear{Lo and Park}{2018}]{lo2018}
\begin{barticle}
\bauthor{\bsnm{Lo}, \binits{D.}},
\bauthor{\bsnm{Park}, \binits{B.}}:
\batitle{Modeling the spread of the zika virus using topological data
  analysis}.
\bjtitle{Scientific reports}
\bvolume{13},
\bfpage{1}--\blpage{12}
(\byear{2018})
\end{barticle}
\endbibitem

%%% 54
\bibitem[\protect\citeauthoryear{Carlsson et~al.}{2005}]{carlsson2005}
\begin{barticle}
\bauthor{\bsnm{Carlsson}, \binits{G.}},
\bauthor{\bsnm{Zomorodian}, \binits{A.}},
\bauthor{\bsnm{Collins}, \binits{A.}},
\bauthor{\bsnm{Guibas}, \binits{L.}}:
\batitle{Persistence barcodes for shapes}.
\bjtitle{International Journal of Shape Modeling}
\bvolume{11},
\bfpage{149}--\blpage{188}
(\byear{2005})
\doiurl{10.1145/1057432.1057449}
\end{barticle}
\endbibitem

%%% 55
\bibitem[\protect\citeauthoryear{Li et~al.}{2014}]{liet.al.2014}
\begin{botherref}
\oauthor{\bsnm{Li}, \binits{C.}},
\oauthor{\bsnm{Ovsjanikov}, \binits{M.}},
\oauthor{\bsnm{Chazal}, \binits{F.}}:
Persistence-based structural recognition,
2003--2010
(2014)
\end{botherref}
\endbibitem

%%% 56
\bibitem[\protect\citeauthoryear{Silva and Ghrist}{2007}]{silva2007}
\begin{botherref}
\oauthor{\bsnm{Silva}, \binits{V.D.}},
\oauthor{\bsnm{Ghrist}, \binits{R.}}:
Homological sensor networks.
Notices Amer. Math. Soc,
10--17
(2007)
\end{botherref}
\endbibitem

%%% 57
\bibitem[\protect\citeauthoryear{Adams and Carlsson}{2015}]{adams2015}
\begin{barticle}
\bauthor{\bsnm{Adams}, \binits{H.}},
\bauthor{\bsnm{Carlsson}, \binits{G.}}:
\batitle{Evasion paths in mobile sensor networks}.
\bjtitle{The International Journal of Robotics Research}
\bvolume{34},
\bfpage{90}--\blpage{104}
(\byear{2015})
\end{barticle}
\endbibitem

%%% 58
\bibitem[\protect\citeauthoryear{Perea and Harer}{2015}]{perea2015}
\begin{barticle}
\bauthor{\bsnm{Perea}, \binits{J.A.}},
\bauthor{\bsnm{Harer}, \binits{J.}}:
\batitle{Sliding windows and persistence: An application of topological methods
  to signal analysis}.
\bjtitle{Foundations of Computational Mathematics}
\bvolume{15},
\bfpage{799}--\blpage{838}
(\byear{2015})
\end{barticle}
\endbibitem

%%% 59
\bibitem[\protect\citeauthoryear{Pun et~al.}{2018}]{pun2018}
\begin{botherref}
\oauthor{\bsnm{Pun}, \binits{C.S.}},
\oauthor{\bsnm{Xia}, \binits{K.}},
\oauthor{\bsnm{Lee}, \binits{S.X.}}:
Persistent-Homology-based Machine Learning and Its Applications -- A Survey
\end{botherref}
\endbibitem

%%% 60
\bibitem[\protect\citeauthoryear{Berwald et~al.}{2014}]{berwald2014}
\begin{botherref}
\oauthor{\bsnm{Berwald}, \binits{J.}},
\oauthor{\bsnm{Gidea}, \binits{M.}},
\oauthor{\bsnm{Vejdemo-Johansson}, \binits{M.}}:
Automatic Recognition and Tagging of Topologically Different Regimes in
  Dynamical Systems
\end{botherref}
\endbibitem

%%% 61
\bibitem[\protect\citeauthoryear{Pereira and
  de~Mello}{2015}]{pereira2015persistent}
\begin{barticle}
\bauthor{\bsnm{Pereira}, \binits{C.M.M.}},
\bauthor{\bsnm{Mello}, \binits{R.F.}}:
\batitle{{Persistent homology for time series and spatial data clustering}}.
\bjtitle{Expert Systems with Applications}
\bvolume{42}(\bissue{15-16}),
\bfpage{6026}--\blpage{6038}
(\byear{2015})
\end{barticle}
\endbibitem

%%% 62
\bibitem[\protect\citeauthoryear{Wu and Hargreaves}{2022}]{wu2022topological}
\begin{barticle}
\bauthor{\bsnm{Wu}, \binits{C.}},
\bauthor{\bsnm{Hargreaves}, \binits{C.A.}}:
\batitle{Topological machine learning for multivariate time series}.
\bjtitle{Journal of Experimental \& Theoretical Artificial Intelligence}
\bvolume{34}(\bissue{2}),
\bfpage{311}--\blpage{326}
(\byear{2022})
\end{barticle}
\endbibitem

%%% 63
\bibitem[\protect\citeauthoryear{Perea et~al.}{2015}]{perea2015sw1pers}
\begin{barticle}
\bauthor{\bsnm{Perea}, \binits{J.A.}},
\bauthor{\bsnm{Deckard}, \binits{A.}},
\bauthor{\bsnm{Haase}, \binits{S.B.}},
\bauthor{\bsnm{Harer}, \binits{J.}}:
\batitle{Sw1pers: Sliding windows and 1-persistence scoring; discovering
  periodicity in gene expression time series data}.
\bjtitle{BMC bioinformatics}
\bvolume{16}(\bissue{1}),
\bfpage{1}--\blpage{12}
(\byear{2015})
\end{barticle}
\endbibitem

%%% 64
\bibitem[\protect\citeauthoryear{Masamichi}{2016}]{sato2016}
\begin{botherref}
\oauthor{\bsnm{Masamichi}, \binits{S.}}:
Can tda be a new risk measure? an application to finance of persistent homology
(2016)
\end{botherref}
\endbibitem

%%% 65
\bibitem[\protect\citeauthoryear{Gidea and Katz}{2018}]{gidea2018topological}
\begin{barticle}
\bauthor{\bsnm{Gidea}, \binits{M.}},
\bauthor{\bsnm{Katz}, \binits{Y.}}:
\batitle{Topological data analysis of financial time series: Landscapes of
  crashes}.
\bjtitle{Physica A: Statistical mechanics and its applications}
\bvolume{491},
\bfpage{820}--\blpage{834}
(\byear{2018})
\end{barticle}
\endbibitem

%%% 66
\bibitem[\protect\citeauthoryear{Saengduean
  et~al.}{2020}]{saengduean2020topological}
\begin{bchapter}
\bauthor{\bsnm{Saengduean}, \binits{P.}},
\bauthor{\bsnm{Noisagool}, \binits{S.}},
\bauthor{\bsnm{Chamchod}, \binits{F.}}:
\bctitle{Topological data analysis for identifying critical transitions in
  cryptocurrency time series}.
In: \bbtitle{2020 IEEE International Conference on Industrial Engineering and
  Engineering Management (IEEM)},
pp. \bfpage{933}--\blpage{938}
(\byear{2020}).
\bcomment{IEEE}
\end{bchapter}
\endbibitem

%%% 67
\bibitem[\protect\citeauthoryear{Gidea}{2017}]{gidea2017topological}
\begin{bchapter}
\bauthor{\bsnm{Gidea}, \binits{M.}}:
\bctitle{Topological data analysis of critical transitions in financial
  networks}.
In: \bbtitle{3rd International Winter School and Conference on Network Science:
  NetSci-X 2017 3},
pp. \bfpage{47}--\blpage{59}
(\byear{2017}).
\bcomment{Springer}
\end{bchapter}
\endbibitem

%%% 68
\bibitem[\protect\citeauthoryear{Vandewalle et~al.}{2001}]{vandewalle2001non}
\begin{barticle}
\bauthor{\bsnm{Vandewalle}, \binits{N.}},
\bauthor{\bsnm{Brisbois}, \binits{F.}},
\bauthor{\bsnm{Tordoir}, \binits{X.}}, \betal:
\batitle{Non-random topology of stock markets}.
\bjtitle{Quantitative Finance}
\bvolume{1}(\bissue{3}),
\bfpage{372}--\blpage{374}
(\byear{2001})
\end{barticle}
\endbibitem

%%% 69
\bibitem[\protect\citeauthoryear{Majumdar and Laha}{2020}]{majumdar2020}
\begin{barticle}
\bauthor{\bsnm{Majumdar}, \binits{S.}},
\bauthor{\bsnm{Laha}, \binits{A.K.}}:
\batitle{Clustering and classification of time series using topological data
  analysis with applications to finance}.
\bjtitle{Expert Systems with Applications}
\bvolume{162},
\bfpage{113868}
(\byear{2020})
\end{barticle}
\endbibitem

%%% 70
\bibitem[\protect\citeauthoryear{Gidea et~al.}{2020}]{gidea2020topological}
\begin{barticle}
\bauthor{\bsnm{Gidea}, \binits{M.}},
\bauthor{\bsnm{Goldsmith}, \binits{D.}},
\bauthor{\bsnm{Katz}, \binits{Y.}},
\bauthor{\bsnm{Roldan}, \binits{P.}},
\bauthor{\bsnm{Shmalo}, \binits{Y.}}:
\batitle{Topological recognition of critical transitions in time series of
  cryptocurrencies}.
\bjtitle{Physica A: Statistical Mechanics and its Applications}
\bvolume{548},
\bfpage{123843}
(\byear{2020})
\end{barticle}
\endbibitem

%%% 71
\bibitem[\protect\citeauthoryear{Aromi et~al.}{2021}]{aromi2021topological}
\begin{barticle}
\bauthor{\bsnm{Aromi}, \binits{L.L.}},
\bauthor{\bsnm{Katz}, \binits{Y.A.}},
\bauthor{\bsnm{Vives}, \binits{J.}}:
\batitle{Topological features of multivariate distributions: Dependency on the
  covariance matrix}.
\bjtitle{Communications in Nonlinear Science and Numerical Simulation}
\bvolume{103},
\bfpage{105996}
(\byear{2021})
\end{barticle}
\endbibitem

%%% 72
\bibitem[\protect\citeauthoryear{Akingbade et~al.}{2024}]{AKINGBADE2024107665}
\begin{barticle}
\bauthor{\bsnm{Akingbade}, \binits{S.W.}},
\bauthor{\bsnm{Gidea}, \binits{M.}},
\bauthor{\bsnm{Manzi}, \binits{M.}},
\bauthor{\bsnm{Nateghi}, \binits{V.}}:
\batitle{Why topological data analysis detects financial bubbles?}
\bjtitle{Communications in Nonlinear Science and Numerical Simulation}
\bvolume{128},
\bfpage{107665}
(\byear{2024})
\doiurl{10.1016/j.cnsns.2023.107665}
\end{barticle}
\endbibitem

%%% 73
\bibitem[\protect\citeauthoryear{Goel et~al.}{2023}]{goel2023sparse}
\begin{botherref}
\oauthor{\bsnm{Goel}, \binits{A.}},
\oauthor{\bsnm{Pasricha}, \binits{P.}},
\oauthor{\bsnm{Kanniainen}, \binits{J.}}:
Risk reduced sparse index tracking portfolio: A topological data analysis
  approach.
To appear in Omega
(2023)
\end{botherref}
\endbibitem

%%% 74
\bibitem[\protect\citeauthoryear{Goel et~al.}{2020}]{goel2020topological}
\begin{barticle}
\bauthor{\bsnm{Goel}, \binits{A.}},
\bauthor{\bsnm{Pasricha}, \binits{P.}},
\bauthor{\bsnm{Mehra}, \binits{A.}}:
\batitle{Topological data analysis in investment decisions}.
\bjtitle{Expert Systems with Applications}
\bvolume{147},
\bfpage{113222}
(\byear{2020})
\end{barticle}
\endbibitem

%%% 75
\bibitem[\protect\citeauthoryear{Rudkin et~al.}{2023}]{RUDKIN2023119894}
\begin{barticle}
\bauthor{\bsnm{Rudkin}, \binits{S.}},
\bauthor{\bsnm{Qiu}, \binits{W.}},
\bauthor{\bsnm{Dłotko}, \binits{P.}}:
\batitle{Uncertainty, volatility and the persistence norms of financial time
  series}.
\bjtitle{Expert Systems with Applications}
\bvolume{223},
\bfpage{119894}
(\byear{2023})
\doiurl{10.1016/j.eswa.2023.119894}
\end{barticle}
\endbibitem

%%% 76
\bibitem[\protect\citeauthoryear{Song and Li}{2025}]{SONG2025130194}
\begin{barticle}
\bauthor{\bsnm{Song}, \binits{S.}},
\bauthor{\bsnm{Li}, \binits{H.}}:
\batitle{Can topological transitions in cryptocurrency systems serve as early
  warning signals for extreme fluctuations in traditional markets?}
\bjtitle{Physica A: Statistical Mechanics and its Applications}
\bvolume{657},
\bfpage{130194}
(\byear{2025})
\doiurl{10.1016/j.physa.2024.130194}
\end{barticle}
\endbibitem

%%% 77
\bibitem[\protect\citeauthoryear{Baitinger and
  Flegel}{2021}]{baitinger2021better}
\begin{barticle}
\bauthor{\bsnm{Baitinger}, \binits{E.}},
\bauthor{\bsnm{Flegel}, \binits{S.}}:
\batitle{The better turbulence index? forecasting adverse financial markets
  regimes with persistent homology}.
\bjtitle{Financial Markets and Portfolio Management}
\bvolume{35}(\bissue{3}),
\bfpage{277}--\blpage{308}
(\byear{2021})
\end{barticle}
\endbibitem

%%% 78
\bibitem[\protect\citeauthoryear{Ruiz-Ortiz et~al.}{2022}]{ruiz2022tda}
\begin{botherref}
\oauthor{\bsnm{Ruiz-Ortiz}, \binits{M.A.}},
\oauthor{\bsnm{G{\'o}mez-Larra{\~n}aga}, \binits{J.C.}},
\oauthor{\bsnm{Rodr{\'\i}guez-Viorato}, \binits{J.}}:
A persistent-homology-based turbulence index \& some applications of tda on
  financial markets.
arXiv preprint arXiv:2203.05603
(2022)
\end{botherref}
\endbibitem

%%% 79
\bibitem[\protect\citeauthoryear{Takens}{1981}]{takens1981detecting}
\begin{barticle}
\bauthor{\bsnm{Takens}, \binits{F.}}:
\batitle{Detecting strange attractors in uid turbulence}.
\bjtitle{Dynamical Systems and Turbulence}
\bvolume{898},
\bfpage{366}
(\byear{1981})
\end{barticle}
\endbibitem

%%% 80
\bibitem[\protect\citeauthoryear{Horak et~al.}{2003}]{horak2003deterministicky}
\begin{bbook}
\bauthor{\bsnm{Horak}, \binits{J.}},
\bauthor{\bsnm{Krl{\'\i}n}, \binits{L.}},
\bauthor{\bsnm{Raidl}, \binits{A.}}:
\bbtitle{Deterministicky Chaos a Jeho Fyzikalni Aplikace}.
\bpublisher{Academia}, \blocation{???}
(\byear{2003})
\end{bbook}
\endbibitem

%%% 81
\bibitem[\protect\citeauthoryear{Khasawneh and
  Munch}{2017}]{khasawneh2017utilizing}
\begin{bchapter}
\bauthor{\bsnm{Khasawneh}, \binits{F.A.}},
\bauthor{\bsnm{Munch}, \binits{E.}}:
\bctitle{{Utilizing topological data analysis for studying signals of
  time-delay systems}}.
In: \bbtitle{Time Delay Systems},
pp. \bfpage{93}--\blpage{106}.
\bpublisher{Springer}, \blocation{???}
(\byear{2017})
\end{bchapter}
\endbibitem

%%% 82
\bibitem[\protect\citeauthoryear{Kim et~al.}{2019}]{kim2019}
\begin{botherref}
\oauthor{\bsnm{Kim}, \binits{K.}},
\oauthor{\bsnm{Kim}, \binits{J.}},
\oauthor{\bsnm{Rinaldo}, \binits{A.}}:
Time series featurization via topological data analysis.
arXiv preprint arXiv:1812.02987v2
(2019)
\end{botherref}
\endbibitem

%%% 83
\bibitem[\protect\citeauthoryear{Adams et~al.}{2017}]{adams2017}
\begin{barticle}
\bauthor{\bsnm{Adams}, \binits{H.}},
\bauthor{\bsnm{Emerson}, \binits{T.}},
\bauthor{\bsnm{Kirby}, \binits{M.}},
\bauthor{\bsnm{Neville}, \binits{R.}},
\bauthor{\bsnm{Peterson}, \binits{C.}},
\bauthor{\bsnm{Shipman}, \binits{P.}},
\bauthor{\bsnm{Chepushtanova}, \binits{S.}},
\bauthor{\bsnm{Hanson}, \binits{E.}},
\bauthor{\bsnm{Motta}, \binits{F.}},
\bauthor{\bsnm{Ziegelmeier}, \binits{L.}}:
\batitle{Persistence images: A stable vector representation of persistent
  homology}.
\bjtitle{The Journal of Machine Learning Research}
\bvolume{18}(\bissue{1}),
\bfpage{218}--\blpage{252}
(\byear{2017})
\end{barticle}
\endbibitem

%%% 84
\bibitem[\protect\citeauthoryear{Guo et~al.}{2020}]{guo2020empirical}
\begin{barticle}
\bauthor{\bsnm{Guo}, \binits{H.}},
\bauthor{\bsnm{Xia}, \binits{S.}},
\bauthor{\bsnm{An}, \binits{Q.}},
\bauthor{\bsnm{Zhang}, \binits{X.}},
\bauthor{\bsnm{Sun}, \binits{W.}},
\bauthor{\bsnm{Zhao}, \binits{X.}}:
\batitle{Empirical study of financial crises based on topological data
  analysis}.
\bjtitle{Physica A: Statistical Mechanics and its Applications}
\bvolume{558},
\bfpage{124956}
(\byear{2020})
\end{barticle}
\endbibitem

%%% 85
\bibitem[\protect\citeauthoryear{Tran et~al.}{2023}]{tran2023detecting}
\begin{bchapter}
\bauthor{\bsnm{Tran}, \binits{H.V.}},
\bauthor{\bsnm{McGregor}, \binits{C.}},
\bauthor{\bsnm{Kennedy}, \binits{P.J.}}:
\bctitle{Detecting stress from multivariate time series data using topological
  data analysis}.
In: \bbtitle{Australasian Joint Conference on Artificial Intelligence},
pp. \bfpage{341}--\blpage{353}
(\byear{2023}).
\bcomment{Springer}
\end{bchapter}
\endbibitem

%%% 86
\bibitem[\protect\citeauthoryear{Seversky et~al.}{2016}]{seversky2016time}
\begin{bchapter}
\bauthor{\bsnm{Seversky}, \binits{L.M.}},
\bauthor{\bsnm{Davis}, \binits{S.}},
\bauthor{\bsnm{Berger}, \binits{M.}}:
\bctitle{On time-series topological data analysis: New data and opportunities}.
In: \bbtitle{Proceedings of the IEEE Conference on Computer Vision and Pattern
  Recognition Workshops},
pp. \bfpage{59}--\blpage{67}
(\byear{2016})
\end{bchapter}
\endbibitem

%%% 87
\bibitem[\protect\citeauthoryear{Akingbade
  et~al.}{2024}]{akingbade2024topological}
\begin{barticle}
\bauthor{\bsnm{Akingbade}, \binits{S.W.}},
\bauthor{\bsnm{Gidea}, \binits{M.}},
\bauthor{\bsnm{Manzi}, \binits{M.}},
\bauthor{\bsnm{Nateghi}, \binits{V.}}:
\batitle{Why topological data analysis detects financial bubbles?}
\bjtitle{Communications in Nonlinear Science and Numerical Simulation}
\bvolume{128},
\bfpage{107665}
(\byear{2024})
\end{barticle}
\endbibitem

%%% 88
\bibitem[\protect\citeauthoryear{Fastrich
  et~al.}{2014}]{fastrich2014cardinality}
\begin{barticle}
\bauthor{\bsnm{Fastrich}, \binits{B.}},
\bauthor{\bsnm{Paterlini}, \binits{S.}},
\bauthor{\bsnm{Winker}, \binits{P.}}:
\batitle{Cardinality versus q-norm constraints for index tracking}.
\bjtitle{Quantitative Finance}
\bvolume{14}(\bissue{11}),
\bfpage{2019}--\blpage{2032}
(\byear{2014})
\end{barticle}
\endbibitem

%%% 89
\bibitem[\protect\citeauthoryear{Goel et~al.}{2025}]{goel2024sparse}
\begin{botherref}
\oauthor{\bsnm{Goel}, \binits{A.}},
\oauthor{\bsnm{Filipovi{\'c}}, \binits{D.}},
\oauthor{\bsnm{Pasricha}, \binits{P.}}:
Sparse portfolio selection via topological data analysis based clustering.
Quantitative Finance,
1--31
(2025)
\end{botherref}
\endbibitem

%%% 90
\bibitem[\protect\citeauthoryear{Mausser et~al.}{2006}]{mausser2006}
\begin{bchapter}
\bauthor{\bsnm{Mausser}, \binits{H.E.}},
\bauthor{\bsnm{Saunders}, \binits{D.}},
\bauthor{\bsnm{Seco}, \binits{L.A.}}:
\bctitle{Optimizing omega}.
(\byear{2006}).
\burl{https://api.semanticscholar.org/CorpusID:2235129}
\end{bchapter}
\endbibitem

%%% 91
\bibitem[\protect\citeauthoryear{Arnott and
  Wagner}{1990}]{arnott1990measurement}
\begin{barticle}
\bauthor{\bsnm{Arnott}, \binits{R.D.}},
\bauthor{\bsnm{Wagner}, \binits{W.H.}}:
\batitle{The measurement and control of trading costs}.
\bjtitle{Financial Analysts Journal}
\bvolume{46}(\bissue{6}),
\bfpage{73}--\blpage{80}
(\byear{1990})
\end{barticle}
\endbibitem

%%% 92
\bibitem[\protect\citeauthoryear{Yu et~al.}{2022}]{yu2022dynamic}
\begin{barticle}
\bauthor{\bsnm{Yu}, \binits{J.-R.}},
\bauthor{\bsnm{Chiou}, \binits{W.P.}},
\bauthor{\bsnm{Hung}, \binits{C.-H.}},
\bauthor{\bsnm{Dong}, \binits{W.-K.}},
\bauthor{\bsnm{Chang}, \binits{Y.-H.}}:
\batitle{Dynamic rebalancing portfolio models with analyses of investor
  sentiment}.
\bjtitle{International Review of Economics \& Finance}
\bvolume{77},
\bfpage{1}--\blpage{13}
(\byear{2022})
\end{barticle}
\endbibitem

%%% 93
\bibitem[\protect\citeauthoryear{Goldfarb and
  Idnani}{1983}]{goldfarb1983numerically}
\begin{barticle}
\bauthor{\bsnm{Goldfarb}, \binits{D.}},
\bauthor{\bsnm{Idnani}, \binits{A.}}:
\batitle{A numerically stable dual method for solving strictly convex quadratic
  programs}.
\bjtitle{Mathematical programming}
\bvolume{27}(\bissue{1}),
\bfpage{1}--\blpage{33}
(\byear{1983})
\end{barticle}
\endbibitem

\end{thebibliography}

\end{document}